\documentclass[12pt]{article}
\usepackage{amsmath}
\usepackage{graphicx}
\usepackage{enumerate}

\usepackage{url} 
\usepackage{tabularx} 
\DeclareMathOperator*{\argmax}{arg\,max}
\DeclareMathOperator*{\argmin}{arg\,min}
\usepackage{diagbox}
\newcommand{\yc}[1]{{{\color{black}  #1}}}

\usepackage[normalem]{ulem}

\usepackage[english]{babel}
\usepackage{amsthm}
\usepackage{amssymb}
\usepackage{multirow}
\usepackage{array,multirow}
\usepackage{booktabs}
\usepackage{subcaption}
\usepackage{mwe}
\usepackage[export]{adjustbox}
\usepackage{enumitem}
\usepackage{listings}
\usepackage{threeparttable}
\usepackage{romannum}
\usepackage{bm}
\usepackage{bbm}
\usepackage{rotating}
\usepackage[round]{natbib}
\usepackage{lscape} 
\usepackage{longtable}

\newtheorem{prop}{Proposition}[]

\newtheorem{condition}{Condition}
\usepackage[skip=1ex]{caption}
\newcommand\numberthis{\addtocounter{equation}{1}\tag{\theequation}}
\usepackage[ruled,vlined]{algorithm2e}
\usepackage{xcolor}

\usepackage{float}
\usepackage[utf8]{inputenc}
\usepackage{textcomp}
\usepackage{tabularx,ragged2e,booktabs,caption}
\newcolumntype{C}[1]{>{\Centering}m{#1}}

\usepackage{float}

\newtheorem{example}{Example}
\newtheorem{assumption}{Assumption}

\newtheorem{remark}{Remark}

\newcommand{\bc}{\bm{c}}

\newcommand{\bh}{\bm{h}}

\newcommand{\cM}{\mathcal{M}}
\newcommand{\cN}{\mathcal{N}}

\newcommand{\cS}{{\mathcal{S}}}
\newcommand{\cT}{{\mathcal{T}}}

\newcommand{\cZ}{\mathcal{Z}}

\newcommand{\II}{\mathbb{I}}

\newcommand{\bgamma}{\bm{\gamma}}

\newcommand{\bGamma}{\bm{\Gamma}}

\newcommand{\sign}{\mathop{\mathrm{sign}}}

\newcommand*{\zero}{{\bm 0}}

\title{\Large Accounting for Measurement Bias: A New Framework for Reliable Country Ranking in Large-Scale Educational Assessments}

\author{Jing Ouyang$^1$, Yunxiao Chen$^2$, Chengcheng Li$^3$, and Gongjun Xu$^4$\\
\small 1. Faculty of Business and Economics, University of Hong Kong\\
\small 2. Department of Statistics, London School of Economics and Political Science\\
\small 3. Microsoft\\
\small 4. Department of Statistics, University of Michigan}

\date{}
\begin{document}
\pagenumbering{arabic}

\def\spacingset#1{\renewcommand{\baselinestretch}%
{#1}\small\normalsize} \spacingset{1}

\maketitle

\vspace{1cm}

\begin{abstract}
{

International Large-scale Assessments (ILSAs), such as the Program for International Student Assessment (PISA) and the Trends in International Mathematics and Science Study (TIMSS), are cornerstone tools for global educational research and policy-making. By benchmarking educational quality and performance trends, these assessments enable countries to evaluate and share effective pedagogical structures. Specifically, ILSAs employ Item Response Theory (IRT) models to rank countries by students' performance on cognitive items. However, measurement bias—arising from linguistic, cultural, and curricular differences—poses a significant threat to the statistical inference of IRT models and, consequently, the validity of the resulting rankings. Neglecting this bias can lead to systematic errors in parameter estimation, ultimately distorting national standings. To address this, we propose a novel method that avoids the restrictive assumptions typical of existing approaches, such as the prior identification of unbiased ``anchor items" or designated reference groups. Our approach is computationally efficient and provides theoretical guarantees for the reliable recovery of group rankings. We apply this method to PISA 2022 data across the mathematics, science, and reading domains, yielding corrected performance rankings and insights into the survey's measurement-bias structures. 
}
 
\end{abstract}

\noindent%

 \spacingset{1.75} 
\section{Introduction}\label{sec:intro}

International large-scale assessments (ILSAs), including the Program for International Student Assessment (PISA), Program for the International Assessment of Adult Competencies (PIAAC),
Progress in International Reading Literacy Study (PIRLS) and Trends in International Mathematics and Science Study (TIMSS) play an important role in educational research and policy making. 
For example, PISA is a widely used ILSA conducted by the Organization for Economic Co-operation and Development (OECD) that evaluates educational systems by measuring 15-year-old students' performance in mathematics, science, and reading \citep{organisation2019pisa}.  
TIMSS and PIRLS are two other widely recognized ILSAs conducted by the International Association for the Evaluation of Educational Achievement (IEA). TIMSS monitors international trends in mathematics and science achievement, while PIRLS assesses reading achievement. They are administered every 4 years and target students in the 4th and 8th grades \citep{mullis2017timss}.
These ILSAs
collect valuable data on the educational quality and performance development across many education systems in the world, allowing countries to share techniques, organizational structures
and policies that have proven effective \citep{singer2018international,von2012role}. 

\begin{figure}[!h]
    \centering
    \includegraphics[height=13.5cm, width=14.5cm]{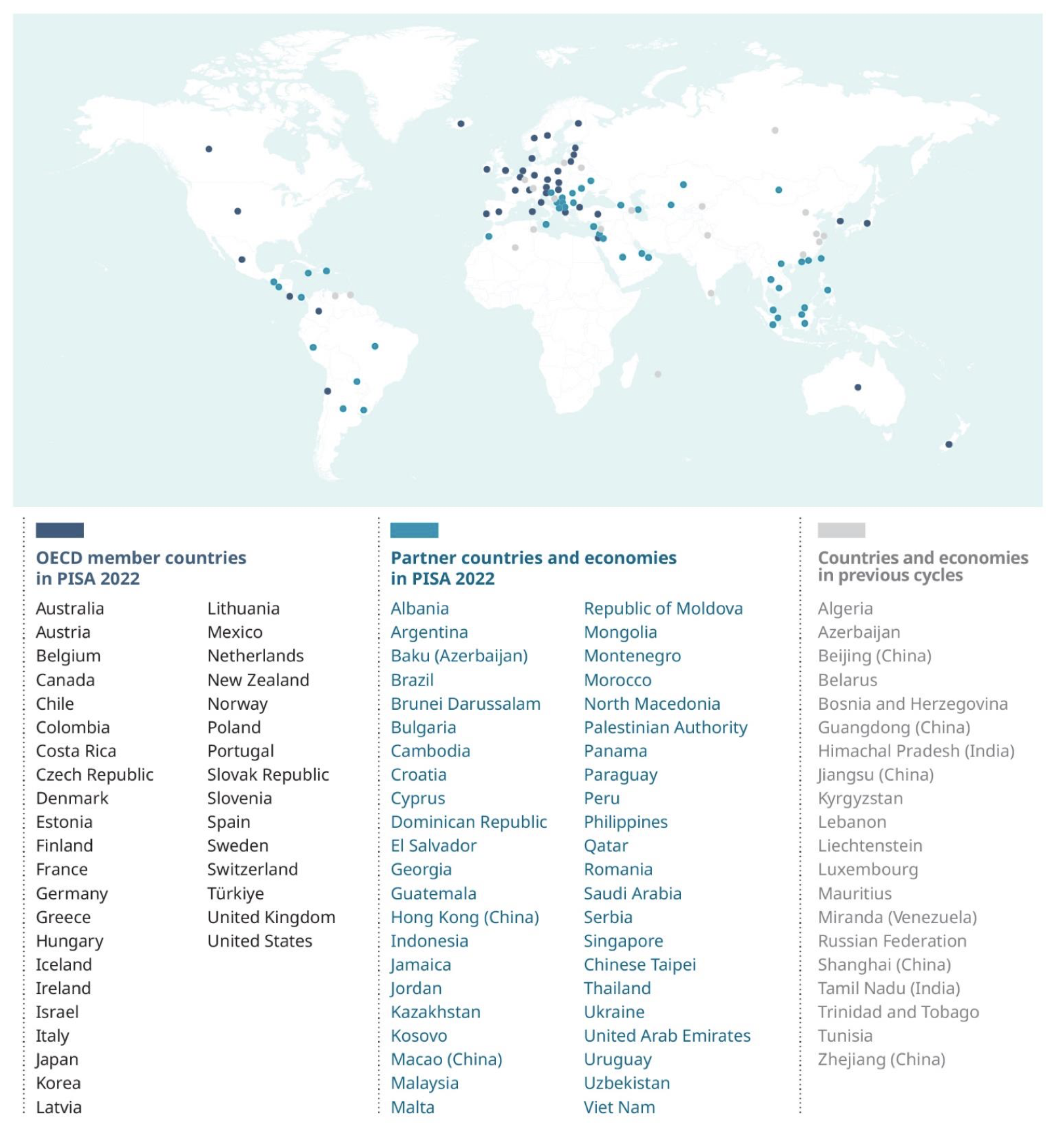}
    \caption{Geographical locations of the participating countries in PISA 2022.}
    \label{fig: country-pisa}
\end{figure}

An important task of an ILSA is to rank countries based on the overall performance of their students in measured subject domains (e.g., mathematics or science). For example, Figure~\ref{fig: country-pisa} illustrates the geographical locations of PISA 2022 participating countries/economies, spanning all five populated continents in the world\footnote{The figure is from PISA 2022 Results Volume II~\citep{pisa2022results2}.}.
The ranking problem is typically solved using a specific Item Response Theory (IRT) model \citep{embretson2013item,van2018handbook}, 
which uses a parameter to represent the overall performance of a country on a subject domain. The countries are then ranked according to the point estimates of their country-specific parameters.  
The specification of this IRT model is nontrivial because of potential measurement bias \citep{millsap2012statistical}, also known as measurement non-invariance,   arising from differences across countries, such as test languages, cultural contexts, and curriculum designs \citep{glas2014modeling}. That is, items may suffer from Differential Item Functioning (DIF), in the sense that different groups can have different probabilities of correctly answering the items after controlling for individual level of proficiency. Ignoring measurement non-invariance, i.e., using a standard IRT model with common item parameters shared across countries, can lead to biased estimates of country-specific parameters and inaccurate rankings of the countries. 

Many statistical procedures have been developed to account for measurement non-invariance in educational or psychological tests. These include methods based on anchor items \citep{mantel1959statistical, thissen1988use, zwick2000using, frick2015rasch, woods2013langer, tay2016item},
where anchor items are DIF-free items that have the same functioning across all groups, and methods without assuming anchor items, such as tree-based methods~\citep{tutz2016item,bollmann2018item}, regularization methods~\citep{tutz2015penalty,  belzak2020improving, bauer2020simplifying, schauberger2020regularization}, {alignment method~\citep{muthen2014irt, Alexander2020Lp, Alexander2023implementation}, {random effects models~\citep{muthen2018recent}}, anchor point selection~\citep{strobl2021anchor},} residual-based methods~\citep{candell1988iterative, clauser1993effects, fidalgo2000effects, glas2014modeling, kopf2015anchor, kopf2015framework, oliveri2011investigation,oliveri2014toward,wang2003effects, wang2004effects, wang2009mimic}, among others. 
However, these methods are all designed for settings with a small number of groups and are not suitable for ILSAs that involve many groups. More specifically, the measurement non-invariance problem that ILSAs face suffers from at least two challenges. First, traditional DIF methods require the specification of a reference group, and DIF effects are defined by contrasting it with one or more other group(s). For ILSAs, all countries are on an equal footing, so it is difficult to identify a single country as the reference group. Arbitrarily choosing the reference group can lead to undesirable results, as the analysis results, including the ranking of countries, may depend heavily on the choice of reference group. Second, many traditional DIF analysis methods require the specification of anchor items. Due to the high heterogeneity of ILSA data, some countries may always behave differently from the rest for each item, and thus, DIF-free items may not exist. Even if some DIF-free items exist, they are usually unknown prior to DIF analysis, as they are difficult to identify from the content. 

The mainstream ILSAs, including PISA, TIMSS, and PIRLS, use the root-mean-square deviation (RMSD) method~\citep{foy2019implementing, yin2021implementing, pisatechnicalreport2022} to address the challenges mentioned above. This method does not require specifying a reference group or anchor items.   
It starts by estimating a DIF-free IRT model using response data from all the countries. Based on the parameter estimates, RMSD fit statistics are computed for item-country pairs to assess how well the estimated model fits the response data for each country and item. For each item, countries with RMSD values above a certain threshold are flagged as deviating from the rest. Based on these flags, a new IRT model that accounts for measurement non-invariance is employed,  in which county-specific item parameters are added to address the misfit\footnote{In cases where large RMSD statistics remain for the new model, the IRT model may be further updated by adding additional country-specific item parameters. This process is typically repeated several times until there are no outlying RMSD statistics.}.

While the RMSD method seems sensible, it has two issues.  First, the result relies on a threshold value, but the choice of this threshold is often subjective. For example, the cut-off threshold was gradually reduced from 0.3 in PISA 2015 to 0.12 in PISA 2022, while a threshold of 0.1 was used in TIMSS 2019~\citep{pisatechnicalreport2015, pisatechnicalreport2022, foy2019implementing}, and the rationale behind these choices was not clearly explained in their technical reports. 
 Second, theoretically, the final model suggested by the RMSD method is not guaranteed to be the true model, even when the sample size goes to infinity and the data follow an IRT model with some group-specific parameters. As shown via a simulation study in \yc{Section A4 of the Supplementary Materials}, some DIF effects can be masked under the DIF-free model and thus cannot be identified using the RMSD fit statistics.   Consequently, the final model returned by the RMSD method can be far from the true model, and the resulting ranking of the countries may be inaccurate. 

This paper proposes a new method for addressing the measurement non-invariance problem in ILSAs. Like the RMSD method, the proposed method does not require the existence or knowledge of a reference group and anchor items. It makes an assumption about the deviation of the observed data from measurement invariance, 
characterized by the sparsity of item parameters for DIF effects. Under this sparsity assumption, we show that the true model can be identified among many equivalent models, and that the proposed method can consistently recover the true model parameters and further rank the countries. The proposed model is computationally efficient and requires no tuning.  We apply the proposed method to the mathematics, science, and reading domains of the PISA 2022 data. The results show that geographically proximate or culturally and linguistically similar countries tend to exhibit similar DIF patterns. Sensible country rankings are also obtained over the three subject domains.

The remainder of the article is organized as follows. \yc{Section \ref{sec-data-background} outlines ILSA data and the research questions through the lens of PISA 2022, alongside a review of the standard operational method for addressing measurement bias in ILSAs.} 
Section \ref{sec:meth} 
proposes a new IRT-based method to address measurement non-invariance in ILSAs. \yc{
Section \ref{Sec:Sim} summarizes the simulation studies conducted to evaluate the performance of the proposed method, with detailed settings and results provided in Section A4 of the Supplementary Materials.}
An application to the PISA 2022 data is given in Section \ref{sec:real-application}. 
We end with concluding remarks in Section \ref{sec:conc}. \yc{In addition,
proofs of the theoretical results, comparisons with competing methods, discussions of identifiability conditions, extensions to the non-uniform DIF setting, 
simulation studies, and additional real-data results are provided in the Supplementary Materials.} 
The code for simulation studies and real data application is available at \url{https://anonymous.4open.science/r/DIF_analysis_PISA_2022-EEBE}.

\section{Background}
\label{sec-data-background}

{

\subsection{Data and Research Questions}
\label{sec-data and research questions}
}

We consider the measurement non-invariance problem in International Large-scale Assessments (ILSAs).
We use the PISA 2022 cycle—the most recent data release\footnote{The data can be downloaded from:~\texttt{https://www.oecd.org/en/data/datasets/pisa-2022-database}.}—to motivate our research questions and to illustrate the proposed method as a case study. As ILSAs share a similar data structure and analysis tasks, the framework developed here is readily applicable to other assessments, such as TIMSS and PIRLS. The current analysis focuses on 37 OECD countries\footnote{Luxembourg, as an OECD country, did not participate in the 2022 cycle and is therefore excluded.}; see Figure~\ref{fig: country-pisa} for a full list of these countries.
Each PISA cycle designates either mathematics, science, or reading as its major domain, which receives more extensive assessment time and features a larger item pool. In 2022, mathematics was the major domain, while reading and science were the minor domains. We analyze all three domains using a dataset preprocessed to retain binary-scored items with at least 1,000 responses. This final dataset comprises 277,138 students and 169 items.

A primary objective of PISA is to rank participating countries by students' aggregate performance across the three assessed domains: mathematics, reading, and science. As reviewed in Section~\ref{subsec:IRTback}, an Item Response Theory (IRT) framework is employed by PISA and other ILSAs to derive these rankings, which  
transforms the ranking task into an estimation problem through a parametric model. 
However, the specification of the IRT model is susceptible to measurement noninvariance that arises from factors such as linguistic variation, cultural context, and divergent national curricula. Failing to account for these biases can lead to misaligned model parameters that substantially distort national rankings. Addressing measurement noninvariance is therefore essential to ensure a fair and reliable foundation for international comparisons and subsequent policymaking.

This paper introduces a new statistical framework to address the measurement noninvariance problem with ILSAs. It aims to answer the following research questions:

\begin{itemize}
    \item[(Q1)] How should country rankings be derived from ILSA latent proficiency estimates to ensure they are robust to the effects of measurement non-invariance? 
     \item[(Q2)] To what extent do estimated DIF effects reflect cross-country similarities? Specifically, do countries with shared geographic, linguistic, or cultural backgrounds exhibit similar DIF patterns?
     \item[(Q3)] In ILSAs that assess multiple subject domains, to what extent do the DIF patterns vary or remain consistent across these domains?
\end{itemize} 

In the context of the PISA 2022 case study, these research questions concern establishing robust rankings for the 37 OECD countries and analyzing their DIF patterns across the three assessment domains: mathematics, science, and reading. 
These research questions are central to both the design and the high-stakes reporting of ILSAs. The first question seeks to establish a more equitable and credible cross-national comparison by mitigating biases arising from linguistic variation, cultural context, and divergent national curricula. The subsequent questions extend beyond the ranking task to provide deeper insights into the structural patterns of measurement bias. Understanding these patterns is instrumental to the future development of assessment items, ensuring greater robustness and invariance across diverse international contexts.

\subsection{IRT and Analysis of Measurement Non-Invariance}\label{subsec:IRTback}

Suppose there are $N$ students from $p$ countries taking an ILSA. They answer $J$ cognitive items within a specific domain (e.g., mathematics) of the ILSA. 
Let $x_i \in \{1, ..., p\}$ indicate the country from which student $i$ takes the test. 
Further, let $Y_{ij}$ be student $i$'s score on item $j$. For simplicity, we consider binary responses in this work, where $Y_{ij} = 1$ indicates that the item is answered correctly and $Y_{ij} = 0$ otherwise.  In real-world ILSAs, the scoring of some items may allow partial credit, in which case the responses are ordinal. We note that, for data with ordinal responses, the discussions below on measurement non-invariance still apply, and the method we propose in this paper can be readily extended to accommodate ordinal response data. In the so-called matrix sampling design of ILSAs, each student is administered only a subset of the items in each domain. We use $z_{ij} \in \{0, 1\}$ to indicate whether item $j$ is administered to student $i$. If $z_{ij} = 1$, then student $i$ receives item $j$, in which case $Y_{ij}$ is observed. Otherwise, 
student $i$ does not receive item $j$, in which case $Y_{ij}$ is missing completely at random. Each individual student was sampled from the population with a weight $w_i$. Such sampling weights are used to control the proportional contribution from individuals to an overall population estimate~\citep{pisatechnicalreport2022}.   

An IRT model is used to scale students and rank countries within the ILSA domain considered. It
introduces a unidimensional latent variable $\theta_i \in \mathbb R$ to represent each student $i$'s proficiency for the domain. The model primarily assumes the following. 

\begin{enumerate}
    \item $\theta_i$'s are assumed to be independent and identically distributed for students from the same country. Typically, the conditional distribution of $\theta_i$ given $x_i = k$ is assumed to follow a normal distribution with mean $\mu_k$ and variance $\sigma_k^2$, for $k=1, ..., p$.
    \item For each student $i$, the observed item responses $Y_{ij}$ are conditionally independent given the latent variable $\theta_i$. 
    \item A distributional assumption is made regarding the conditional distribution of $Y_{ij}$ given $\theta_i$ and $x_i$. For an item that is invariant across countries, $Y_{ij}$ is conditionally independent of $x_i$ given $\theta_i$, in which case the probability of correctly answering the item given the student's proficiency level does not depend on the country he/she is from. Otherwise, we say an item is non-invariant. Typically, a two-parameter logistic (2PL) model is assumed for this conditional distribution, under which 
    \begin{equation}\label{eq:irf-dif}
P(Y_{ij} = 1\vert \theta_i = \theta, x_i = k) = \frac{\exp(a_{j}\theta + d_j +  \gamma_{jk} )}{1+\exp(a_{j}\theta + d_j +  \gamma_{jk})}. 
\end{equation}
Here, $a_j$ is a slope parameter associated with the latent variable shared across countries, and $d_j + \gamma_{jk}$ is the country-specific intercept. Further, $\gamma_{jk}$s are assumed to be sparse. {A brief discussion about this assumption is provided later, and the formal description of this assumption and its discussions are given in Section~\ref{sec:identifiability}.} Roughly speaking, this assumption requires that 
the true values of many $\gamma_{jk}$s are zero.  
This sparsity assumption is not only sensible for ILSAs where items are carefully designed to minimize their differential functioning across countries, but also essential for model identifiability. Under the sparsity assumption of $\gamma_{jk}$s, $d_j$ may be interpreted as the common intercept parameter shared by a set of countries to which the item functions the same, and $\gamma_{jk}$ may be interpreted as the DIF effect of country $k$ when compared with the baseline countries.  For each item $j$, the countries for which $\gamma_{jk} = 0$ are referred to as the baseline countries. 

\end{enumerate}
Additionally, constraints are imposed on certain model parameters to avoid model non-identifiability. Specifically, we set $\sum_{k=1}^K \mu_k = 0$ and $a_1 = 1$ to fix the location and scale of the latent variables.  
Below we provide some discussions about this IRT framework under measurement non-invariance. 

As mentioned in item~3 above, the sparsity assumption on the $\gamma_{jk}$s, which will be formally introduced in Section~\ref{sec:identifiability}, plays a crucial role in ensuring the identifiability of the true model parameters. 
  Let $a_j^*$, $d_j^*$, $\mu_k^*$ and $\gamma_{jk}^*$, $k=1, ..., p$, $j=1, ..., J$, be the true model parameters, where $\gamma_{jk}^*$s are sparse. 
   {Without the sparsity assumption, the model is unidentifiable.} That is,  we can simultaneously replace $\mu_k^*$ by $\mu_k^* + c_k$, replace $d_j^*$ by $d_j^* + h_j$, and replace $\gamma_{jk}^*$ by $\gamma_{jk}^* - a_j^*c_k - h_j$, for all $j = 1, ..., J$, $k = 1, ..., p$, without changing the model-implied distribution for observed data, for all $c_1$, ..., $c_p$ satisfying $\sum_{k=1}^p c_k = 0$ so that $\sum_{k=1}^K \mu_k = 0$ remains satisfied. {The sparsity assumption constrains the parameter space of interest to a subset of the full Euclidean space. As will be discussed in Section~\ref{sec:identifiability}, the model becomes identifiable within this constrained space, under the standard definition of identifiability (see, e.g., Chapter 1 of  \citealp{bickel2015mathematical}). We also note that this sparsity assumption is sensible for applications to large-scale assessments, such as PISA and TIMSS. That is, the test items in these large-scale assessments are carefully designed and reviewed to ensure cultural fairness and conceptual rigor, thereby largely controlling for DIF. As a result, it is reasonable to assume many DIF parameters are zero. 
We further note that we do not know which 
$\gamma_{jk}^*$s are zero, making it a challenge to estimate the model within this constrained parameter space.  A method will be proposed in Section \ref{sec:meth} to address this challenge.}



The concept of baseline countries introduced in the item~3 of model specification is closely related to the reference group in many traditional DIF analysis methods. However, we note that the set of baseline countries can vary across items, and there may not be a country that serves as a baseline for all items. Even if such a baseline country exists, it is almost impossible to identify it before the DIF analysis. Consequently, traditional DIF analysis methods that require specifying a reference group cannot be applied to this problem. 


In addition to reference group information, many traditional DIF analysis methods require knowledge of anchor items -- DIF-free items that function the same across all the countries in the comparison. However, with the current sparsity assumption, there does not necessarily exist an item $j$ for which $\gamma_{jk}^* = 0$ for all $k=1, ..., p$.  In other words, there may not exist an anchor item under the current setting for ILSAs. Even if there are some anchor items, it is almost impossible to identify them in advance for us to apply traditional DIF analysis methods that require the specification of some anchor items.  
 
Last but not least, the type of DIF in the IRT model \eqref{eq:irf-dif} is typically known as uniform DIF in the sense that country differences are only reflected by the intercept parameters $\gamma_{jk}$. This uniform DIF model can be extended by allowing the slope parameters to vary across countries. More specifically, we may replace \eqref{eq:irf-dif} by 
 \begin{equation}\label{eq:irf-dif-non-unif}
P(Y_{ij} = 1\vert \theta_i = \theta, x_i = k) = \frac{\exp(a_{j}\exp(\zeta_{jk} ) \theta + d_j +  \gamma_{jk} )}{1+\exp(a_{j}\exp(\zeta_{jk} )\theta + d_j +  \gamma_{jk})}, 
\end{equation} 
where the additional term $\exp(\zeta_{jk})$ is a multiplicative factor that captures the deviation of $k$th country's slope parameter from the common slope $a_j$. Similar to the sparsity assumption on $\gamma_{jk}$, there may be only a small number of item-country pairs for which the slope parameters deviate from the common ones. In other words, it may also be sensible to assume $\zeta_{jk}$s to be sparse, so that $a_j\exp(\zeta_{jk}) = a_j$ for many item-country pairs, which helps the identification of the extended DIF model. The method proposed in Section~\ref{sec:meth} can be extended to this non-uniform DIF setting; {see Section A2 of the Supplementary Materials.}


\subsection{The State of the Art: An RMSD Method for Treating Measurement Noninvariance in ILSAs}
\label{sec:RMSD}

The RMSD method has been employed as the default method to account for measurement noninvariance in ILSAs, including PISA, TIMSS, and PIRLS~\citep{ 
foy2019implementing,pisatechnicalreport2018,
yin2021implementing,
pisatechnicalreport2022}.
This method is performed under the same IRT framework described above and implicitly assumes the sparsity assumption about $\gamma_{jk}$s\footnote{As previously discussed, item responses can also be ordinal in practice. In such cases, our proposed model can be extended to accommodate ordinal responses, for example, by adapting the generalized partial credit model \citep{muraki1992generalized}, and the RMSD method is still applicable under this extension.}.  

More specifically, the RMSD method starts by estimating a baseline IRT model with $\gamma_{jk} = 0$ for all $ j = 1, ..., J$ and $ k = 1, ..., p$. The estimation is done by obtaining the marginal maximum likelihood (MML) estimator~\citep{bock1981marginal} 
using the Expectation-Maximization algorithm~\citep{dempster1977maximum}. 
The marginal loglikelihood function under model~\eqref{eq:irf-dif} can be written as 
\begin{equation*}
L(\Psi) = \sum_{i=1}^N w_i \log\left( \int \prod_{j: z_{ij}=1} \left[\frac{\exp\{Y_{ij}(a_j\theta_i + d_j + \gamma_{j, {x_i}})\}}{1+\exp(a_j\theta_i + d_j + \gamma_{j,x_i})}\right] \phi(\theta_i\vert \mu_{x_i}, \sigma^2_{x_i}) d\theta_i\Big\}\right),
\numberthis\label{eq:mml-rmsd}
\end{equation*}
where $\Psi = (a_1, ..., a_J, d_1, ..., d_J, \mu_1, ..., \mu_p, \sigma_1, ..., \sigma_p, \gamma_{11}, ..., \gamma_{Jp})^\top \in \mathbb R^{2J + 2p + Jp}$ combines all the model parameters together as a vector. The MML estimator is obtained by
\begin{equation}\label{eq:baseline}
    \check{\Psi} = \arg\max_{\Psi} L(\Psi), \quad s.t.~ a_1 = 1, \sum_{k=1}^p\mu_k=0,    \text{ and } \gamma_{jk}=0, j = 1, ..., J, k = 1, ..., p.
\end{equation}

The RMSD method \citep{pisatechnicalreport2022} computes a residual statistic, known as the root mean square
deviation, to measure the discrepancy between the observed data and the estimated constrained model \eqref{eq:baseline}. The definition of RMSD is based on the generalized residuals \citep{haberman2013generalized} for assessing the goodness of fit for models for contingency tables. 
For each student $i$, let 
$h_i(\theta\vert \check{\Psi})$ be the conditional density of the latent trait $\theta_i$ given their item responses and country membership, calculated under the estimated model with parameters $\check{\Psi}$. Define the pseudo-observed probability 
for item $j\in\{1,..., J\}$, country $k\in \{1, ..., p\}$ and response category $y \in \{0,1\}$ at latent trait level $\theta$ as 
$$o_{jk y}(\theta) = \frac{\sum_{i=1}^{N}  \II\{Y_{ij} = y, x_i = k, z_{ij} = 1 \} h_{i}(\theta\vert \check{\Psi})}{ \sum_{i=1}^{N}  \II\{x_i = k, z_{ij} = 1 \} h_{i}(\theta\vert \check{\Psi}) },$$  
where the numerator computes the frequency of the response category $y$ near $\theta$ weighted by the model-implied posterior densities $h_{i}(\theta\vert \check{\Psi})$, and the denominator computes the number of students from the $k$th country whose latent trait level is near $\theta$, weighted by the same posterior densities. 
Further, we define the corresponding model-implied probability as 
$$p_{jky}(\theta) = \frac{\exp\{y(\check a_j\theta + \check d_j + \check\gamma_{jk })\}}{1+\exp(\check a_j\theta + \check d_j + \check \gamma_{jk})},$$
where the parameters $\check a_j$, $\check d_j$ and $\check\gamma_{jk}$
are given by the estimated model \eqref{eq:baseline}. 
For each $j  \in \{1, ..., J\}$ and $k \in \{1, ..., p\}$, the RMSD statistic is defined as
\begin{align}
    \mathrm{RMSD}_{jk}=&\frac{1}{2}\Big\{ \sqrt{\int\left(o_{jk 0}(\theta)-p_{jk 0}(\theta)\right)^2 \phi(\theta|\check{\mu}_k, \check{\sigma}_k^2 ) d \theta} \nonumber \\
    &+ \sqrt{ \int\left(o_{jk 1}(\theta)-p_{jk 1}(\theta)\right)^2 \phi(\theta|\check{\mu}_k, \check{\sigma}_k^2 ) d \theta} \Big\}, \label{eq:RMSD}
\end{align}
which is the aggregated discrepancy between the pseudo-observed probabilities and the corresponding model-implied probabilities. The aggregation in \eqref{eq:RMSD} is based on the estimated marginal distribution for the latent trait of the $k$th country.

Given the RMSD statistics, the RMSD method then fits a relaxed IRT model, 
in which the parameter $\gamma_{jk}$ 
is freely estimated instead of being constrained  to zero when $\text{RMSD}_{jk}$ exceeds a pre-specified threshold. After estimating the updated model, $o_{jky}(\theta)$ and $p_{jky}(\theta)$ can be recalculated, based on which the RMSD statistics can be reevaluated and the IRT model may be further updated. 
We may continue this process of  updating the IRT model and calculating 
$\text{RMSD}$ statistics until all $\text{RMSD}_{jk}$s are less than the threshold \citep[see, e.g.,][]{jewsbury2020irt, sosa2024validity}.

The RMSD method identifies item-country pairs for which there is a mismatch between the observed item responses and the pattern implied by the constrained IRT model for which $\gamma_{jk} = 0$. This method implicitly assumes that
the DIF-free IRT model approximates the true model well so that a large value of $\text{RMSD}_{jk}$ indicates the violation of the constraint $\gamma_{jk} = 0$. However, this assumption may not hold in practice. 
In addition, the result of the RMSD method is often sensitive to the 
cut-off threshold, the choice of which, however, is often ad-hoc. Thus, the
RMSD method is not guaranteed to consistently learn the true model. 
In fact, in Section A4, simulation results show that the model selected by the 
RMSD method can deviate from the true model substantially. Moreover, the RMSD method typically requires estimating many IRT models. Given the large numbers of participants and items involved in ILSAs, implementing the RMSD method can be
time-consuming.

\section{Proposed Method} \label{sec:meth}

\subsection{An $L_1$ Assumption for Model Identification}
\label{sec:identifiability}

{To address the model identifiability issue mentioned in Section~\ref{subsec:IRTback}, 
we introduce the following assumption, which is based on the $L_1$ norm of $\gamma_{jk}^*$s.}

\begin{assumption}
     \label{cond:ML1} 
The true DIF parameters $\gamma_{jk}^*$ satisfy that 
     \begin{equation}\label{eq:spar2}
     \sum_{j=1}^J\sum_{k=1}^p\vert \gamma_{jk}^*\vert \leq \sum_{j=1}^J\sum_{k=1}^p \vert \gamma_{jk}^* - a_j^*c_k - h_j \vert,         
     \end{equation} 
     for all $c_1, ..., c_p$ and $h_1$, ..., $h_J$ satisfying $\sum_{k=1}^p c_k = 0$,  
     and the equality in \eqref{eq:spar2} only holds when $c_1 =\cdots = c_p =  0$ and $h_1 = \cdots = h_J = 0$. 
\end{assumption}

{This assumption requires the true $\gamma_{jk}^*$s to have the smallest $L_1$ norm, among all the equivalent models outlined in Section~\ref{subsec:IRTback},
Imposing this assumption leads to the model identifiability in a constrained space. Specifically, by constraining the parameter space to a subset of Euclidean space where the DIF effect $\gamma_{jk}$s has the smallest $L_1$ norm among equivalence models, the model indeterminacy 
is removed. Under the standard definition of identifiability (e.g., Chapter~1 of \citealp{bickel2015mathematical}), the model parameters are identifiable in this constrained space. As shown via Proposition~\ref{prop} and discussion paragraph 
below, this assumption holds when $\gamma_{jk}^*$s are sufficiently sparse. }

\begin{prop}\label{prop}
Suppose that $a_j^* > 0$ for all items $j \in \{1, ..., J\}$, which generally holds for IRT models for educational testing. Define $\Pi = \{\pi:\{1, \ldots, p\} \rightarrow\{1, \ldots, p\} ~\vert~ \pi \text { is a bijection}\}$ as all the permutations of $\{1, ..., p\}$.
Then Assumption~\ref{cond:ML1} holds if and only if the following conditions hold:

\begin{enumerate} 
          \item For each $j = 1, ..., J$,
         \begin{equation}\label{eq:rowspar}
\left\vert\left(\sum_{k=1}^p  \II_{\{\gamma_{jk}^* < 0\}}\right) -  \left(\sum_{k=1}^p \II_{\{\gamma_{jk}^* > 0\}}\right)\right\vert < \sum_{k=1}^p \II_{\{\gamma_{jk}^* = 0\}}. 
     \end{equation}

     \item For each $w = 1, ..., p-1$, and each $\pi \in \Pi$ 
     \begin{align}
     &    \sum_{j: w \geq k_j^*(\pi)} \sum_{k=w+1}^p a_j^*[ \II(\gamma_{j,\pi(k)}^* = 0) - \{ \II(\gamma_{j,\pi(k)}^* > 0) - \II(\gamma_{j,\pi(k)}^* < 0)\} ] \nonumber \\
    &+ \sum_{j: w < k_j^*(\pi)} \sum_{k=1}^w a_j^*[ \II(\gamma_{j,\pi(k)}^* = 0) + \{ \II(\gamma_{j,\pi(k)}^* > 0) - \II(\gamma_{j,\pi(k)}^* < 0)\} ] > 0,\label{eq:stronger sparsity}
     \end{align}
     where $k_j^*(\pi) \in \{1, ..., p\}$  depends only on the permutation $\pi$ and the signs of $(\gamma_{j1}^*, ..., \gamma_{jk}^*)$;  
     see equation (A2) in the Supplementary Materials for the definition.

\end{enumerate}
\end{prop}

Proposition~\ref{prop} provides a sufficient and necessary condition for Assumption~\ref{cond:ML1}, when the discrimination parameters $a_j^*$s are all positive. This result utilizes the fact that $\sum_{j=1}^J\sum_{k=1}^p $ $ \vert \gamma_{jk}^* - a_j^*c_k - h_j \vert$ as a function of $c_1$, ..., $c_p$, $h_1$, ..., $h_J$ is convex.  
The inequality in \eqref{eq:rowspar} requires that each row vector of the DIF matrix $(\gamma_{jk}^*)_{J \times p}$ is sufficiently sparse, in the sense that the number of zero entries in each row of $(\gamma_{jk}^*)_{J \times p}$ is required to dominate the difference between the numbers of positive and negative entries. 
This condition ensures that there does not exist a nonzero $h_j$ such that 
$\sum_{k=1}^p\vert \gamma_{jk}^*\vert \geq \sum_{k=1}^p \vert \gamma_{jk}^* - h_j \vert$, which is necessary for Assumption \ref{cond:ML1}  to hold. 
The condition in \eqref{eq:stronger sparsity}, roughly speaking,  requires each column of the DIF matrix $(\gamma_{jk}^*)_{J \times p}$ to be sufficiently sparse. This result 
 ensures that 
 $c_1 =\cdots = c_p =  0$  is the only 
minimizer of $\min_{h_1, ...h_J} \sum_{j=1}^J\sum_{k=1}^p \vert \gamma_{jk}^* - a_j^*c_k - h_j \vert$, subject to $\sum_{k=1}^p c_k = 0$. 
It is derived from verifying that all the directional derivatives of $\min_{h_1, ...h_J} \sum_{j=1}^J\sum_{k=1}^p \vert \gamma_{jk}^* - a_j^*c_k - h_j \vert$ at $c_1 =\cdots = c_p =  0$ are negative. As these directional derivatives are calculated separately for different orderings of $c_1$, ..., $c_p$, a permutation $\pi$ is involved. See Section~A4.1 for examples of DIF matrices $(\gamma_{jk}^*)_{J\times p}$ for which Assumption 1 holds.


We acknowledge that Assumption~\ref{cond:ML1}  may not always hold in practice. Nevertheless, the method we propose in Section~\ref{subsec:prop} is reasonably robust against mild violations of this assumption, when many $\gamma_{jk}^*$s are only approximately zero; see Section~A4.8 of the Supplementary Materials for a sensitivity analysis. We also note that this assumption tends to be weaker than what is required for the alignment method~\citep{muthen2014irt, Alexander2020Lp, Alexander2023implementation} to identify the true model parameters, 
when the loss function of the alignment method is also based on an  $L_1$ norm. As shown in Section A4.5.2 of the Supplementary Materials, there exist $(\gamma_{jk}^*)$ matrices that are identifiable under Assumption 1, but are unidentifiable under the loss function for the alignment method. 
Finally,
although it is possible to weaken Assumption~\ref{cond:ML1} for identifying the true model, the computation tends to be more challenging under weakened assumptions; see Section~\ref{sec:conc} for further discussions. 
 

\subsection{Proposed Estimator}\label{subsec:prop}

If the true model satisfies the $L_1$-norm-based sparsity condition, Assumption~\ref{cond:ML1}, then its parameter estimation becomes straightforward. In what follows, we propose an estimator that is statistically consistent and computationally efficient. 

Let 
\begin{equation*}
L(\Psi) = \sum_{i=1}^N w_i \log\left( \int \prod_{j: z_{ij}=1} \left[\frac{\exp\{Y_{ij}(a_j\theta_i + d_j + \gamma_{j, {x_i}})\}}{1+\exp(a_j\theta_i + d_j + \gamma_{j,x_i})}\right] \phi(\theta_i\vert \mu_{x_i}, \sigma^2_{x_i}) d\theta_i\Big\}\right)
\numberthis\label{eq:mml-dif}
\end{equation*}
be the weighted marginal log-likelihood function, where $\phi(\cdot|\mu, \sigma^2)$ is the density function of a normal distribution with mean $\mu$ and variance $\sigma^2$, and 
$$\Psi = (a_1, ..., a_J, d_1, ..., d_J, \mu_1, ..., \mu_p, \sigma_1, ..., \sigma_p, \gamma_{11}, ..., \gamma_{Jp})^\top \in \mathbb R^{2J + 2p + Jp}$$ is a generic notation that stacks all the model parameters together as a vector.  
Assumption~\ref{cond:ML1} motivates us to consider the following constrained parameter space for $\Psi$: 
\begin{equation*}
    \begin{aligned}
    \mathcal M = &\Big\{\Psi: 
\sum_{j=1}^J\sum_{k=1}^p\vert \gamma_{jk}\vert \leq \sum_{j=1}^J\sum_{k=1}^p \vert \gamma_{jk} - a_jc_k - h_j \vert, \mbox{~for all~} c_1, ..., c_p, h_1, ..., h_J \in \mathbb R\\
&\mbox{~satisfying~} \sum_{k=1}^p c_k = 0\Big\} \subset \mathbb R^{2J + 2p + Jp},
    \end{aligned}
\end{equation*}
in which $\gamma_{jk}$s have the smallest sum of absolute values among all the equivalent solutions.

We consider the constrained marginal maximum likelihood estimator 
\begin{equation}\label{eq:cmml}
    \begin{aligned}
\hat \Psi = & \argmax_{\Psi} L(\Psi)~\quad
s.t.  ~ \Psi \in \mathcal M,
~~ a_1 = 1, \mbox{~and~} \sum_{k=1}^p \mu_k = 0. 
    \end{aligned}
\end{equation}
Note that the last two constraints are introduced in Section~\ref{subsec:IRTback}, which are
needed for identifying the location and scale of the country-specific distributions of the latent variables. As shown in Proposition~\ref{prop:consistency} below, this estimator is statistically consistent. 

\begin{prop}\label{prop:consistency}
Suppose that Assumption~\ref{cond:ML1} and Condition~A1 in Section~A1.2.1 of the Supplementary Materials hold. 
Then, the constrained  marginal maximum likelihood estimator $\hat \Psi$ in \eqref{eq:cmml} satisfies that $\vert \hat \mu_k - \mu_k^*\vert=o_P(1), \vert \hat \sigma_k - \sigma_k^*\vert=o_P(1)$,  $\vert \hat a_j - a_j^*\vert =o_P(1), \vert \hat d_j - d_j^*\vert =o_P(1)$, $\vert \hat \gamma_{jk} - \gamma_{jk}^*\vert =o_P(1)$, as $N \rightarrow \infty$, 
for all $j=1,..,J$ and $k=1,...,p$. 

\end{prop}

Although the constraint space in \eqref{eq:cmml} seems complex, this optimization problem can be solved quite easily using a two-step procedure, as described in Algorithm~\ref{alg:ML1} below. 
Specifically, we establish the equivalence between the algorithmic output and the constrained maximum likelihood estimator in the following proposition.
\begin{algorithm}[!h]
\SetAlgoLined
 {\bf Step 1:}  Solve
\begin{align}
\tilde \Psi =& \arg\max_{\Psi}  L(\Psi), ~
 s.t.\;   a_1 = 1,\; \mu_k = 0,\; d_j = 0,   k  = 1, ..., p,  j =1, ..., J.  \label{eq:mml}
\end{align} 

 {\bf Step 2:} 
Solve  
\begin{align}\label{eq:lad}
 (\hat{c}_1, \dots, \hat{c}_p, \hat{h}_1, \dots, \hat{h}_J) =  \underset{c_1, \dots, c_p \atop h_1, \dots, h_J} {\argmin} &  \sum_{k=1}^p\sum_{j=1}^J \vert  \tilde\gamma_{jk}  - \tilde a_j c_k - h_j\vert \quad
 s.t.  \sum_{k=1}^p c_k = 0.  
\end{align}

 {\bf Output:} 
$\hat \gamma_{jk} = \tilde\gamma_{jk}  - \tilde a_j \hat c_k - \hat{h}_j,$
$\hat\mu_k = \hat c_k,$   $\hat a_j = \tilde a_j$, $\hat d_j =\hat h_j$ and $\hat\sigma_k=\tilde\sigma_k$. 

\caption{ }
\label{alg:ML1}
\end{algorithm}

\begin{prop}\label{prop:equivalence estimator}
    The output estimator in Algorithm~\ref{alg:ML1} is equivalent to the constrained marginal maximum likelihood estimator $\hat{\Psi}$ in~\eqref{eq:cmml}.
\end{prop}

We note that the first step of Algorithm~\ref{alg:ML1} finds a marginal maximum likelihood estimator under a set of minimum constraints different from those in the proposed estimator defined in \eqref{eq:cmml}. 
The constrained marginal maximum likelihood estimation problem in Step 1 is solved by EM algorithm~\citep{dempster1977maximum}, which is a common practice in PISA studies~\citep{pisatechnicalreport2015, pisatechnicalreport2018, pisatechnicalreport2022}.
In the second step, the estimator from the first step is transformed into the constrained space $\mathcal M$. This transformation is achieved by solving the convex optimization problem~\eqref{eq:lad}. 
More specifically, this optimization problem can be viewed as a Least Absolute Deviations objective function for median regression \citep{koenker_2005} and, thus, can be solved using standard software packages for quantile regression. In our implementation for the numerical studies, the R package \textit{quantreg} \citep{koenker2018package} is used to solve this optimization problem.

{
\section{Simulation Study}\label{Sec:Sim}
}

We summarize the simulation studies here; due to space constraints, the comprehensive settings and results are deferred to Section A4 of the Supplementary Materials.
We conducted simulation studies to evaluate the proposed DIF method across a diverse range of settings, varying the sample size ($n$), number of countries ($p$), number of items ($J$), and the sparsity patterns of the DIF structure. The performance of the proposed estimator was compared against several benchmarks: a baseline model that ignores DIF, RMSD-based methods with various thresholds, the alignment methods~\citep{Alexander2023implementation} and Lasso-regularization methods~\citep{belzak2020improving}. We evaluated these methods using Kendall rank correlation for country rankings, mean squared error (MSE) for parameter estimation, and model selection accuracy for identifying non-zero DIF effects. 

Across nearly all settings, the proposed method yields the most accurate country rankings—demonstrating the highest alignment with the ground truth—and the lowest estimation errors for model parameters. In contrast, the RMSD method, regardless of the chosen threshold, often fails to recover the true DIF structure. Overall, the simulation results indicate that the proposed method is more reliable and accurate than competing approaches, including the RMSD-based methodology currently employed in ILSA operations.

{
\section{Application to PISA 2022 Data}\label{sec:real-application}
}

\subsection{Overview}

In this section, we analyze the PISA 2022 data and answer the research questions raised in Section~\ref{sec-data and research questions}.  Specifically, we apply the proposed method in Algorithm~\ref{alg:ML1} to the mathematics, science, and reading domains of the PISA 2022 data. To answer Q1, we produce a new ranking of the 37 countries by their estimated $\hat \mu_k$'s for each domain, and compare the new ranking with the one produced by the RMSD method. To answer Q2, we examine the estimated DIF effects $\hat \gamma_{jk}$ and their cross-country similarity through multidimensional scaling representations. Specifically, we calculate the distance between two countries $i$ and $k$ as 
$\sqrt{\sum_{j=1}^J (\hat \gamma_{ji} - \hat \gamma_{jk})^2}$ and then apply metric Multidimensional Scaling~\citep[MDS;][]{borg2007modern} to the resulting $37\times 37$ distance matrix to embed the countries into a two-dimensional Euclidean space for visualization. Since the MDS procedure is designed to preserve relative distances between entities, the resulting embedding plots allow us to assess the similarity of DIF patterns visually—countries positioned closer together exhibit more similar measurement-bias patterns.
Finally, to answer Q3, we compare the distributions of the estimated DIF effects across the three domains using summary statistics, including representative quantiles and variances of $\hat{\gamma}_{jk}$, thereby assessing cross-domain heterogeneity. 

\subsection{Corrected Rankings}
\label{sec:corrected ranking}

\begin{figure}[!h]
    \centering
    \includegraphics[width=0.7\linewidth]{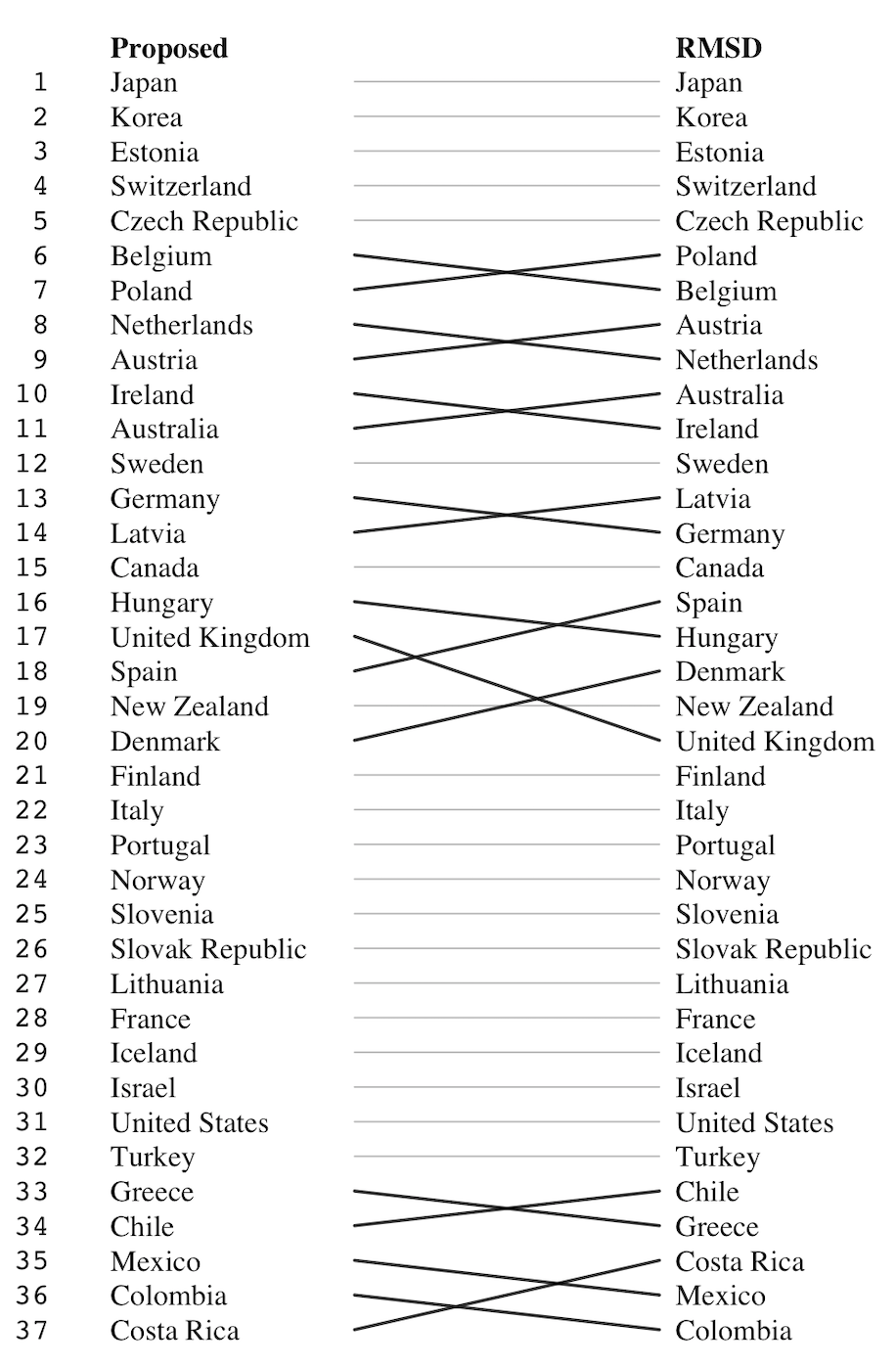}
    \caption{Comparison of rankings of country-wise average latent skill levels between proposed method and RMSD method for math domain.}
    \label{fig:math}
\end{figure}

\begin{figure}[!h]
    \centering
    \includegraphics[width=0.7\linewidth]{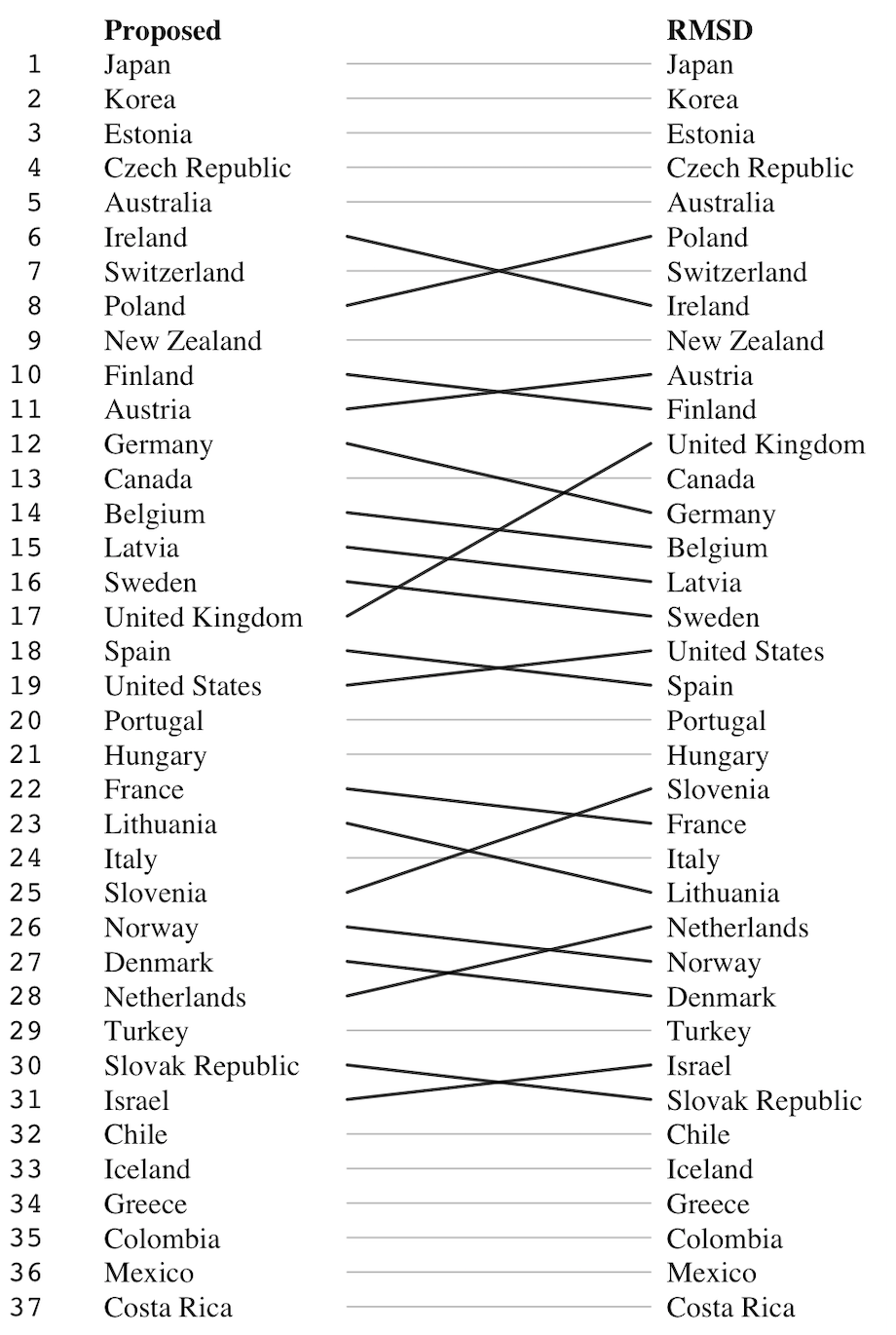}
    \caption{Comparison of rankings of country-wise average latent skill levels between proposed method and RMSD method for science domain.}
    \label{fig:science}
\end{figure}

\begin{figure}[!h]
    \centering
    \includegraphics[width=0.7\linewidth]{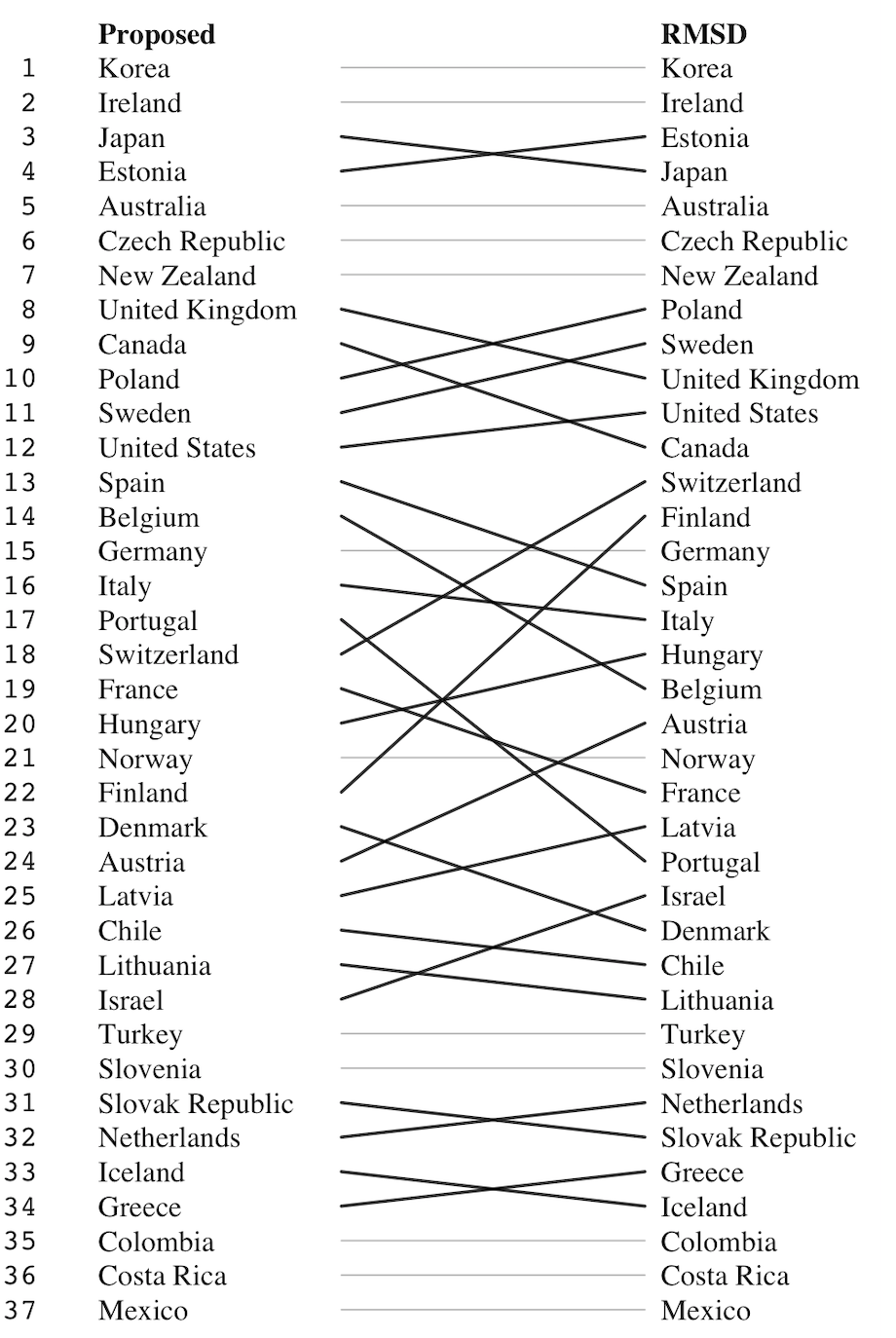}
    \caption{Comparison of rankings of country-wise average latent skill levels between proposed method and RMSD method for reading domain.}
    \label{fig:reading}
\end{figure}

Figures~\ref{fig:math}--\ref{fig:reading}
summarize the country rankings from the proposed method, for the mathematics, science, and reading domains of PISA 2022.
The estimates of the model parameters under the proposed method are provided in Sections~A5.1--A5.3 of the Supplementary Materials. For comparison, these figures also report the rankings obtained by the RMSD method with a threshold of 0.05.   
The rankings obtained by the RMSD method with thresholds of 0.10 and 0.15, together with those from the baseline method, are reported in Tables~A13, A18, and~A23 of the Supplementary Materials.
 In each figure, the same country under the proposed method and the RMSD method is connected by a line. Horizontal connections indicate that the two methods assign the same rank, whereas crossed lines indicate discrepancies in the rankings.
Therefore, the number and extent of line crossings provide a direct visual summary of the degree of disagreement between the two methods.

As shown in Figure~\ref{fig:math}, the math domain represents the highest level of agreement between the two methods, with a Kendall’s rank correlation of $0.964$. 
This strong concordance can be visualized as most connecting lines are nearly horizontal, with only a few crossed lines appearing in the middle and lower parts of the ranking.
In particular, the differences are almost exclusively ``micro-shifts" confined to adjacent entries. The top five countries (Japan, Korea, Estonia, Switzerland, and the Czech Republic) are perfectly identical across both models. Where discrepancies occur, the majority are limited to simple pairwise transpositions; for instance, the RMSD method merely ``flips" the relative ordering of Belgium and Poland (Ranks 6th–7th) and of the Netherlands and Austria (Ranks 8th–9th).  

The science domain in Figure \ref{fig:science} maintains a high level of concordance ($\tau = 0.945$), mirroring math’s stability at the extremes with an identical Top 5 and a perfectly consistent Bottom 6. 
Compared with mathematics, Figure~\ref{fig:science} shows more crossed lines in the upper-middle tier, indicating a moderate increase in disagreement between the two methods.
The most prominent difference is the United Kingdom, which the RMSD method ranks five places higher (12th) than the proposed method (17th). Furthermore, the two methods provide different views of the 6th through 8th positions: while both models identify Ireland, Switzerland, and Poland for these slots, the proposed method orders them Ireland-Switzerland-Poland, whereas the RMSD method produces the reverse sequence.

While all domains maintain high concordance at the extreme ends of the rankings, the reading domain (Figure \ref{fig:reading}) displays the larger number of crossed lines and exhibits notably greater middle-tier volatility than the math (Figure \ref{fig:math}) and science (Figure \ref{fig:science}) domains, with the Kendall's tau correlation dropping to $0.883$. This subject area moves beyond simple swaps and into ``macro-shifts" that suggest a fundamental difference in how the two methods interpret reading performance data. While the ``poles" remain relatively stable—with Korea and Ireland anchored at the top and Colombia, Costa Rica, and Mexico at the bottom—the middle of the ranking undergoes a substantial structural reordering. The most dramatic example is Finland, which climbs eight positions (from 22nd to 14th) under the RMSD method, while Switzerland rises five spots. Conversely, Portugal drops seven places and Belgium drops five in the RMSD ranking compared to the proposed model.

The increasing discrepancy—progressing from math to science and peaking in reading—is likely attributable to the varying severity of DIF across these subjects. As suggested in Section~\ref{subsec:domains}, the reading domain appears most affected by DIF, followed by science, and finally math. This trend likely reflects the increasing complexity and density of DIF patterns associated with higher language demands. As suggested by simulation results in Section~A4.8 of the Supplementary Materials, the RMSD method tends to lose efficacy as DIF patterns become denser, whereas the proposed method remains reasonably robust.

\subsection{Comparison across Countries}



\begin{figure}[!ht]
  \centering
  \begin{minipage}[v]{0.33\textwidth}
    \centering
    \includegraphics[width=\textwidth]{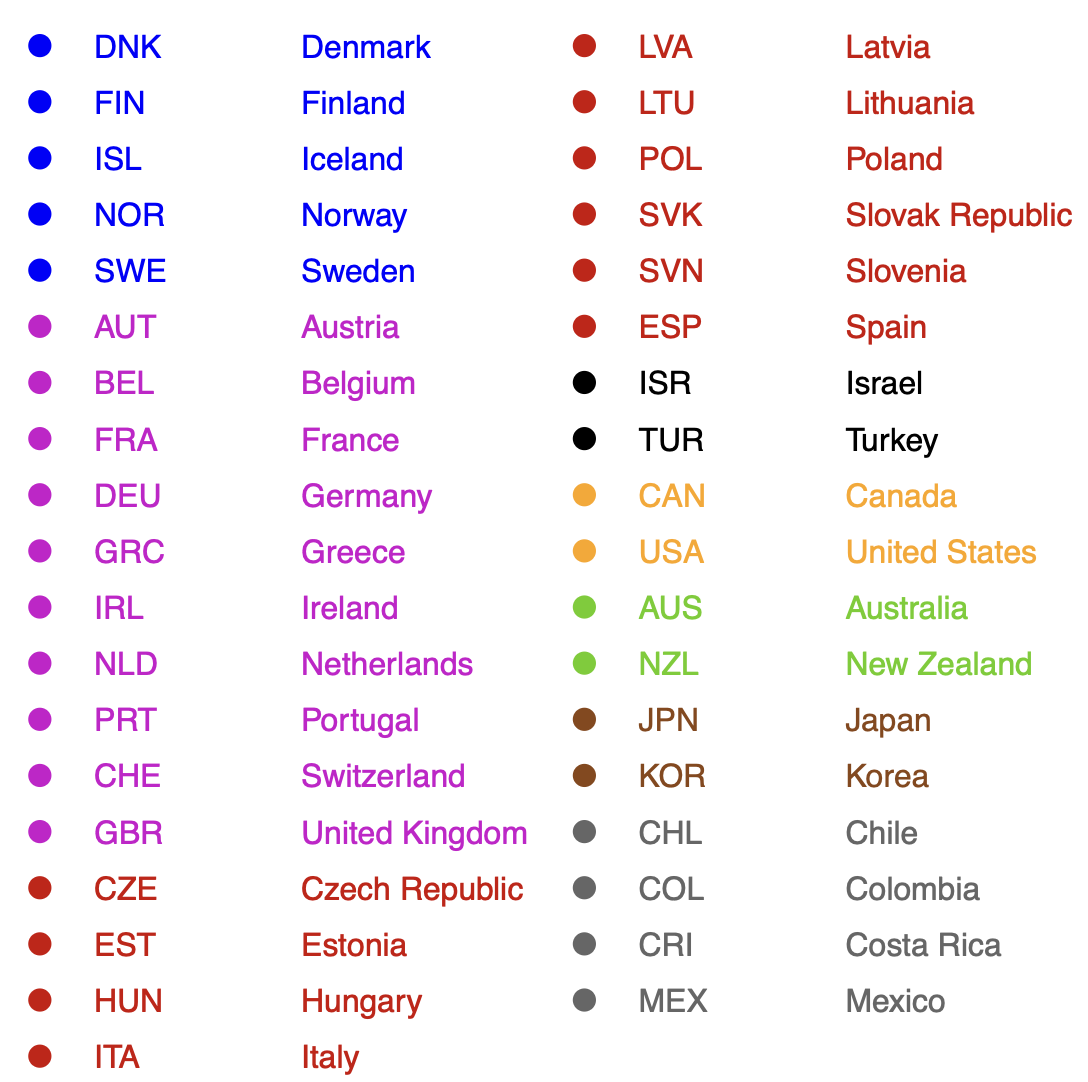}
  \end{minipage}~
  \begin{minipage}[c]{0.63\textwidth}
    \centering
    \includegraphics[width=\textwidth]{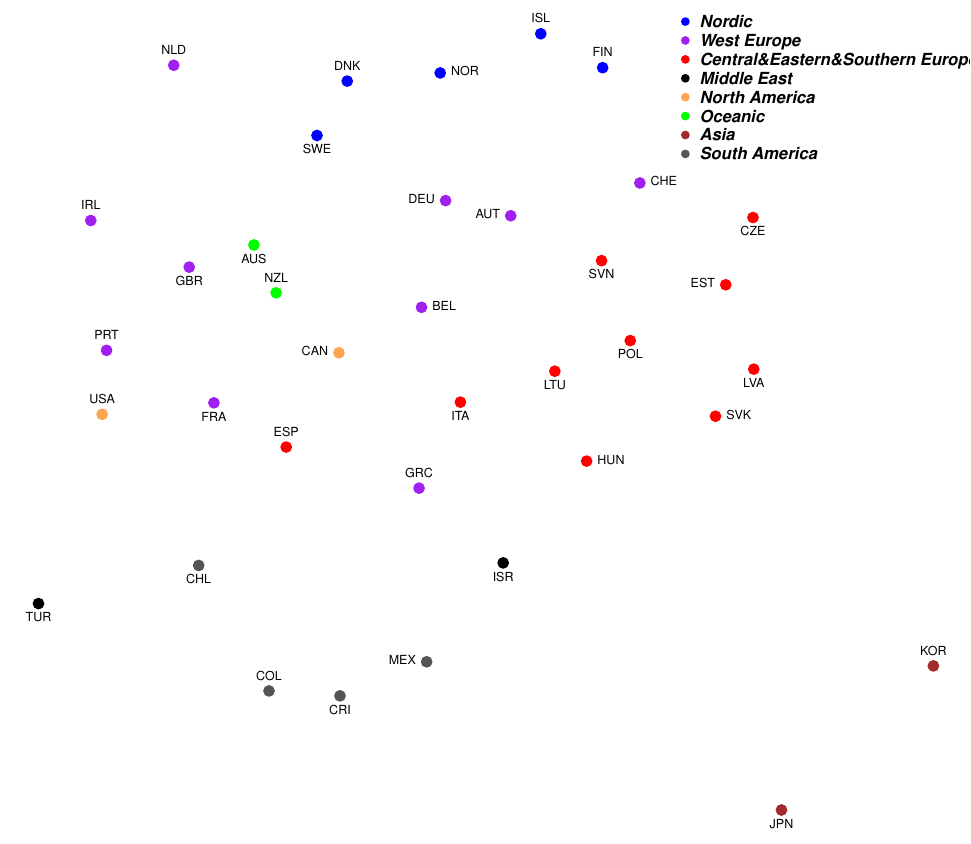}
  \end{minipage}
  \caption{
  Plot of two-dimensional representation of distance of $\hat{{\gamma}}_k$ between countries  through multi-dimensional scaling for the mathematics domain.}
    \label{fig: pca-cluster-math}

\end{figure}

\begin{figure}[!ht]
    \centering
   \begin{minipage}[c]{0.33\textwidth}
    \centering
    \includegraphics[width=\textwidth]{plots_da/da_legend.png}
  \end{minipage}~
  \begin{minipage}[c]{0.63\textwidth}
    \centering
    \includegraphics[width=\textwidth]{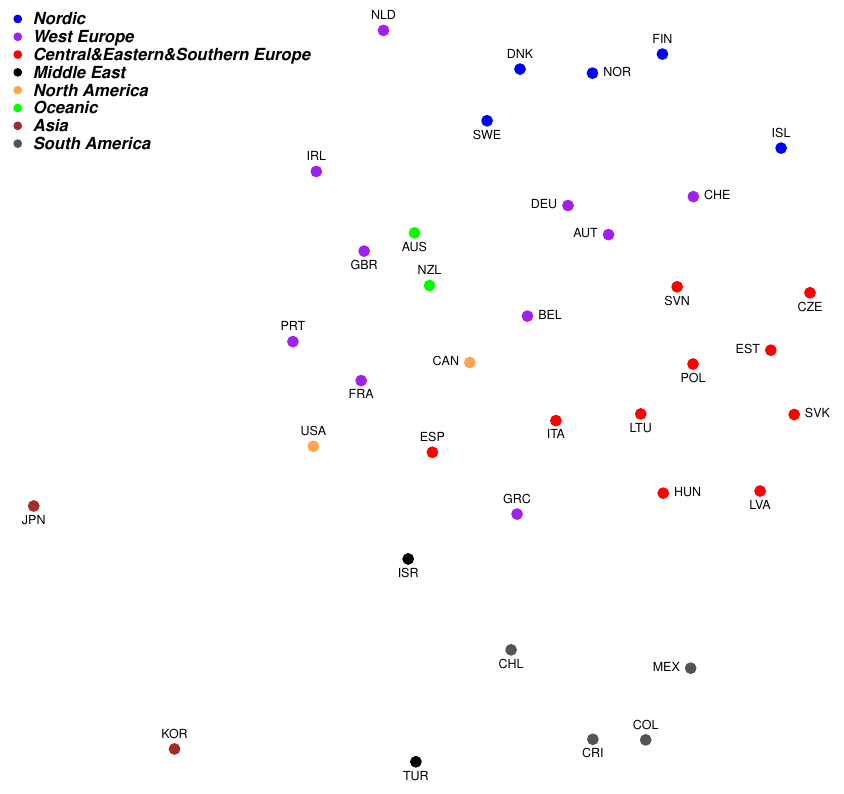}
  \end{minipage}
    \caption{
    Plot of two-dimensional representation of distance of $\hat{{\gamma}}_k$ between countries through multi-dimensional scaling for the science domain.}
    \label{fig: pca-cluster-sci}
\end{figure}

\begin{figure}[!ht]
    \centering
     \begin{minipage}[c]{0.33\textwidth}
    \centering
    \includegraphics[width=\textwidth]{plots_da/da_legend.png}
  \end{minipage}~
  \begin{minipage}[c]{0.63\textwidth}
    \centering
    \includegraphics[width=\textwidth]{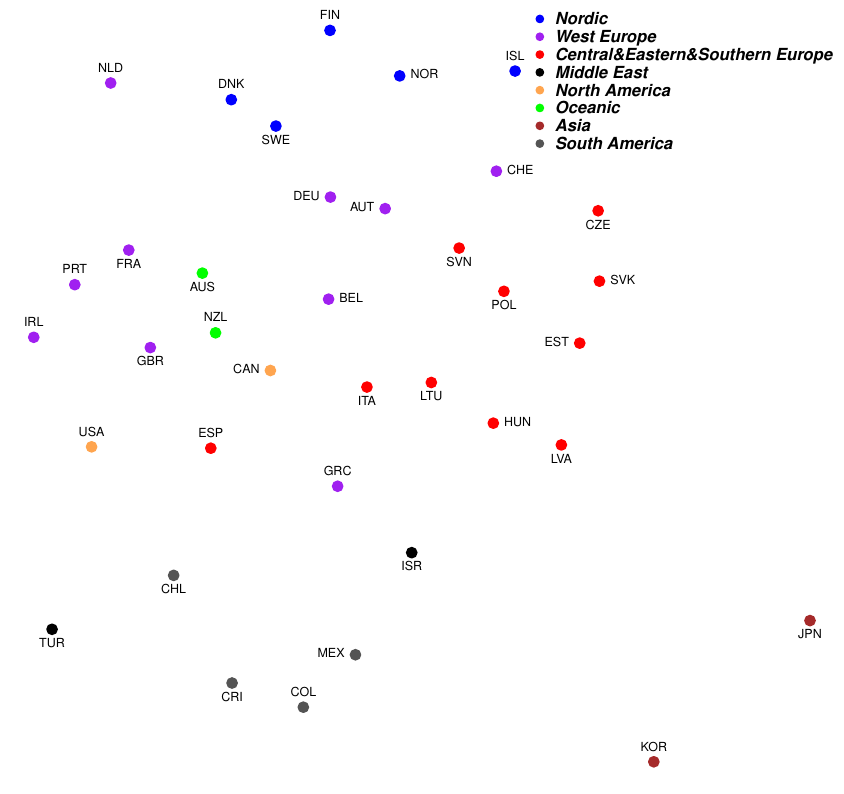}
  \end{minipage}
    \caption{Plot of two-dimensional representation of distance of $\hat{{\gamma}}_k$ between countries  through metric multi-dimensional scaling for the reading domain.}
    \label{fig: pca-cluster-reading}
\end{figure}

To address Q2 in Section~\ref{sec-data and research questions}, we examine the estimated DIF parameters $\hat{\gamma}_{jk}$ and investigate the cross-country similarity and heterogeneity of DIF patterns. For each domain, we apply metric MDS to the estimated DIF effects and embed the countries into a two-dimensional Euclidean space; the resulting plots are shown in Figures~\ref{fig: pca-cluster-math}--\ref{fig: pca-cluster-reading} for the math, science, and reading domains, respectively. Across all three domains, the MDS representations reveal broadly similar patterns: countries that are geographically close or share similar languages, cultures, or historical ties tend to cluster together. In particular, the Nordic countries, including Denmark, Finland, Iceland, Norway, and Sweden, tend to form a cluster, while the South American countries, such as Chile, Colombia, Costa Rica, and Mexico, also appear close to one another. In addition, the United States, Canada, Australia, and New Zealand are located near the United Kingdom and Western Europe, despite some of them being geographically distant, likely reflecting their historical, linguistic, and cultural connections. Finally, Japan and Korea remain relatively separated from most other OECD countries in all three domains, suggesting that these two East Asian countries have DIF patterns that are more distinct from the rest. Overall, the results indicate that the DIF effects in PISA 2022 are systematically related to geographic proximity as well as broader linguistic, cultural, and historical similarities across countries.

\subsection{Comparison across Domains}\label{subsec:domains}






We now address Q3. Table~\ref{tab:gamma_summary} summarizes the distribution of the estimated DIF effects, $\hat{\gamma}_{jk}$, for the mathematics, science, and reading domains. The table provides several representative quantiles and the variance of $\hat{\gamma}_{jk}$ for each subject area. As shown in Table~\ref{tab:gamma_summary}, the dispersion of these effects varies considerably across domains. Mathematics exhibits the most concentrated distribution, characterized by the lowest variance and the narrowest interquartile and 10th-to-90th percentile ranges. Science shows a moderate increase in spread, while reading displays the greatest overall dispersion. This monotone pattern suggests that country-related measurement non-invariance is least pronounced in mathematics, slightly more evident in science, and most significant in reading. As mentioned earlier, this result may explain the  the lower rank correlation between the proposed method and the RMSD method for reading reported in Section~\ref{sec:corrected ranking}.

The monotone pattern may be driven by the increasing amount of language-dependent content involved in the test items. 
Compared with mathematics items, science items often contain more contextual descriptions and information, and reading items are even more directly affected by language and interpretation. As a result, these items may be more sensitive to translation, cultural context, and country-specific interpretation, which could in turn lead to greater cross-country measurement bias.

\begin{table}
\centering
\resizebox{0.9\textwidth}{!}{
\begin{tabular}{lcccccc}
\toprule
Domain & 10th pct. & 25th pct. & Median & 75th pct. & 90th pct. & Variance \\
\midrule
Mathematics & -0.37 & -0.17 & 0.00 & 0.16 & 0.35 & 0.10 \\
Science & -0.43 & -0.19 & 0.00 & 0.18 & 0.37 & 0.11 \\
Reading & -0.48 & -0.21 & 0.00 & 0.22 & 0.46 & 0.16 \\
\bottomrule
\end{tabular}}
\caption{Summary of the estimated DIF effects $\hat{\gamma}_{jk}$ across the mathematics, science, and reading domains. For each domain, we report the 10th, 25th, 50th, 75th, and 90th percentiles, together with the variance of $\hat{\gamma}_{jk}$.}
\label{tab:gamma_summary}
\end{table}

\section{Concluding Remarks}\label{sec:conc}

This paper proposes the first statistically consistent method for the measurement non-invariance problem within the context of international large-scale assessments in education. 
This problem poses a unique challenge, diverging from standard DIF analysis because there is no prior information on anchor items or reference groups. Consequently, conventional DIF methodologies, which rely on such pre-specified elements, are rendered inapplicable in this critical setting. Furthermore, the state-of-the-art RMSD method, currently employed in the operational procedures of the popular international assessments, lacks theoretical guarantees and sometimes performs poorly, according to our simulation results.  The introduction of our method thus offers a more reliable and theoretically sound approach to ensuring the fairness and validity of these vital international educational comparisons.

\yc{We applied the proposed method to the PISA 2022 dataset, 
yielding a more robust ranking of the countries and identifying DIF effects related to countries and subject domains. These findings offer answers to the three research questions raised in Section~\ref{sec-data and research questions}. A natural next step for future research would be to conduct granular analyses of the specific items identified as exhibiting high DIF. By examining these items directly, researchers can validate the findings in the current paper and pinpoint underlying sources of bias, such as cultural nuances or linguistic barriers. Advancements in artificial intelligence may be particularly valuable in this regard, offering automated tools to detect subtle semantic shifts, explain the denser DIF patterns observed in the reading and science domains, and provide actionable insights for future item development.}

{We note that our sparsity condition, Assumption~\ref{cond:ML1}, plays a central role in the proposed method for identifying the true model parameters and consistent estimation. To our best knowledge, this is the first condition that gives 
such theoretical guarantees in the absence of prior information about anchor items and reference groups. We acknowledge that this assumption is relatively strong; however, as shown by a counterexample in Section A4.5.2 of the Supplementary Materials, it tends to be weaker than what is required for the $L_1$-based alignment method to identify the true DIF effects. A sensitivity analysis in Section A4.8 of the Supplementary Materials shows that the proposed method remains effective when Assumption 1 is mildly violated. However, it also shows that when the assumption is more severely violated, alternative methods, such as the Lasso-regularization method and the RMSD method with various thresholds, can outperform the proposed one. Moreover, other methods that are not included in the comparison, such as the alignment methods based on the $L_p$ ($0<p<1$) function~\citep{Alexander2020Lp, Alexander2023implementation} and the random effects method~\citep{muthen2018recent}, may also outperform the proposed method under severe violation of Assumption \ref{cond:ML1}.

 Assumption~\ref{cond:ML1} may be relaxed to identify models with a less sparse DIF effect matrix $(\gamma_{jk}^*)_{J\times p}$, by replacing the $L_1$ function in Assumption~\ref{cond:ML1} with a nonconvex sparsity-inducing function, such as the SCAD~\citep{fan2001variable}, MCP~\citep{zhang2010nearly} and $L_p$ ($0 < p<1$) functions. Compared with the $L_1$ function, these functions better approximate the $L_0$ function, and thus, may be better at identifying the DIF effect matrix when it is less sparse than what is required in Assumption \ref{cond:ML1}. However, these loss functions come at a computational cost. That is,  the 
objective function in the second step of Algorithm~\ref{alg:ML1} needs to be replaced with the corresponding nonconvex function. As a result,  the optimization problem is no longer convex, for which no polynomial-time algorithm  
is theoretically guaranteed to find the global solution. We leave the DIF analysis based on these nonconvex loss functions to future investigation. 
}

This work lays the foundation for several promising directions for future work. First, to apply in the operations of ILSAs, the proposed method needs to be extended to an IRT model allowing for ordinal items, such as the generalized partial credit model \citep{muraki1992generalized}, 
to accommodate items scored on a partial-credit basis. 
This can be achieved by generalizing Assumption~\ref{cond:ML1} for model identification and adapting Algorithm~\ref{alg:ML1} accordingly. Second, the current method only provides a point estimate of the country ranking. It may be beneficial to develop statistical inference procedures, such as bootstrap methods, to quantify the uncertainty of the estimated country ranking. 
Finally, we have extended the proposed framework to the non-uniform DIF setting; see Section~A2 of the Supplementary Materials for details. However, we have not yet empirically examined this extension. It would be valuable to evaluate its performance through simulation studies and real-data applications. Moreover, the extension enables model fit comparisons between uniform and non-uniform DIF models, which we plan to investigate further to determine which specification is better suited for different ILSAs.\\

\bigskip
\begin{center}
{\large\bf ACKNOWLEDGEMENT}
\end{center}
The authors are grateful to the Editor, Associate Editor, and two anonymous reviewers for their helpful and constructive comments. The authors also gratefully acknowledge the following sources of support: 
Ouyang is partially supported by the Hong Kong Early Career Scheme (ECS) Grant \#27308125 and the Seed Fund for Basic Research \#2401102046. Xu is partially supported by NSF SES-2150601.

\begin{center}
{\large\bf DISCLOSURE STATEMENT}
\end{center}
The authors report there are no conflicts of interest to declare.

\begin{center}
{\large\bf DATA AVAILABILITY}
\end{center}
The data that support the findings of this study are openly available at  
\url{https://www.oecd.org/en/data/datasets/pisa-2022-database.html}.

\begin{center}
{\large\bf SUPPLEMENTARY MATERIALS}
\end{center}
Supplementary Materials include the proof of propositions  and  additional results for the simulation studies and the  application to PISA 2022 data. The R code for simulation studies and data application can be found in \url{https://anonymous.4open.science/r/DIF_analysis_PISA_2022-EEBE}.

\bibliographystyle{apalike}
\spacingset{1.5}
{ \bibliography{Bibliography-MM-MC}}

\clearpage

\newpage 
\begin{appendix}

\renewcommand{\theequation}{A\arabic{equation}}
\renewcommand{\thelemma}{A\arabic{lemma}}
\renewcommand{\theassumption}{A\arabic{assumption}}
\renewcommand{\thecondition}{A\arabic{condition}}
\renewcommand{\thetable}{A\arabic{table}}
\renewcommand{\thesection}{A\arabic{section}}
\renewcommand{\theprop}{A\arabic{prop}}

\setcounter{page}{1}    

\begin{center}
    {\bf \large Supplementary Material for ``Accounting for Measurement Bias: A New Framework for Reliable Country Ranking in Large-Scale Educational Assessments''}

\end{center}


\spacingset{1}

The Supplementary Materials contain proofs of all propositions developed in the main article in Section~\ref{sec-proof}, additional discussions of non-uniform DIF and identifiability in Sections~A2 and A3, respectively, detailed simulation studies in Section~\ref{sec-simulation}, and additional results for the real-data application in Section~\ref{sec-real}.

\section{Proofs for the Propositions}\label{sec-proof}



\subsection{Proof of Proposition~\ref{prop}}

Consider the function
\[
g(\bc, \bh)
=
\sum_{j=1}^J \sum_{k=1}^p
\bigl|\gamma_{jk}^* - a_j^* c_k - h_j\bigr|.
\]
 The idea is to study the directional derivative of $g$ at $(\bc, \bh) = (\mathbf{0}, \mathbf{0})$ along all feasible directions. At $(\bc, \bh) = (\mathbf{0}, \mathbf{0})$, define
\[
s_{jk} \in
\begin{cases}
\{\sign(\gamma_{jk}^*)\}, & \gamma_{jk}^* \neq 0,\\[2pt]
(-1, 1), & \gamma_{jk}^* = 0,
\end{cases}
\]
which collects all possible subgradients of the absolute function $|\cdot|$ at $\gamma_{jk}^*$. For any direction $(\Delta\bc, \Delta\bh)$, the directional derivative of $g(\bc, \bh)$ is
\begin{align}
D g(\Delta\bc, \Delta\bh)
&:=
\lim_{t \downarrow 0}
\frac{g(t\Delta\bc, t\Delta\bh ) - g(\mathbf{0}, \mathbf{0})}{t}\nonumber\\
&=
\lim_{t \downarrow 0}
\sum_{j=1}^J \sum_{k=1}^p
\frac{\bigl|\gamma_{jk}^* - t(a_j^* \Delta c_k + \Delta h_j)\bigr|
      -|\gamma_{jk}^*|}{t}.\label{eq:Dg delta}
\end{align}

A standard case-by-case analysis of~\eqref{eq:Dg delta} yields
\[
D g(\Delta\bc, \Delta\bh)
=
\sum_{j=1}^J \sum_{k=1}^p
s_{jk}\bigl(-a_j^* \Delta c_k - \Delta h_j\bigr)
=
-\sum_{k=1}^p \Delta c_k
\Bigl(\sum_{j=1}^J a_j^* s_{jk} \Bigr)
- \sum_{j=1}^J \Delta h_j \Bigl( \sum_{k=1}^p s_{jk} \Bigr).
\]
In analogy with the univariate case (where left and right derivatives must be nonnegative at a minimum), in the current multivariate setting we require that $D g(\Delta\bc, \Delta\bh) \geq  0$ for all feasible nonzero directions $(\Delta\bc, \Delta\bh)$ in a neighborhood of $(\mathbf{0}, \mathbf{0})$ in order for $(\mathbf{0}, \mathbf{0})$ to be the minimizer. 
Thus we require 
\begin{align*}
    \sum_{j=1}^J a_j^* s_{jk} \leq 0, \quad \sum_{k=1}^p s_{jk} \leq 0,
\end{align*}
which are equivalently written as
\begin{enumerate}
    \item {Row sparsity:}\; for every $j$, at least one zero exists in row $j$ (i.e., some $k$ with $\gamma_{jk}^*=0$) and
\begin{align*}
    \left\vert\left(\sum_{k=1}^p  1_{\{\gamma_{jk}^* < 0\}}\right) -  \left(\sum_{k=1}^p 1_{\{\gamma_{jk}^* > 0\}}\right)\right\vert < \sum_{k=1}^p 1_{\{\gamma_{jk}^* = 0\}}.
\end{align*}
\item {Column sparsity:}\; for every $k$, at least one zero exists in column $k$ and
\begin{align*}
     \left\vert\left(\sum_{j=1}^J |a_j^*| 1_{\{\gamma_{jk}^* < 0\}}\right) -  \left(\sum_{j=1}^J |a_j^*| 1_{\{\gamma_{jk}^* > 0\}}\right)\right\vert < \sum_{j=1}^J |a_j^*| 1_{\{\gamma_{jk}^* = 0\}}.  
\end{align*}
\end{enumerate}

Nonetheless, the above conditions need further refinement.
 We next derive condition~\eqref{eq:stronger sparsity} in the main text to ensure that $(\zero,\zero)$ is the unique minimizer. Define
\[
\ell(\bc, \bh)
=
\sum_{j=1}^J \sum_{k=1}^p \bigl|\gamma_{jk}^* - a_j^* c_k - h_j\bigr|
\quad\text{and}\quad
g(\bc) := \min_{\bh} \ell(\bc, \bh)
=
\min_{\bh} \sum_{j=1}^J \sum_{k=1}^p \bigl|\gamma_{jk}^* - a_j^* c_k - h_j\bigr|.
\]
For each $j$, the minimization over $h_j$ decouples into $\min_{h_j} \sum_{k=1}^p \bigl|\gamma_{jk}^* - a_j^* c_k - h_j\bigr|$, whose solution is given by
\[
h_j
=
\text{median}\bigl\{\gamma_{jk}^* - a_j^* c_k : k = 1, \ldots, p\bigr\}.
\]

Let
\begin{align*}
p_{+,j} &= \#\{k : \gamma_{jk}^* > 0\},\\
p_{-,j} &= \#\{k : \gamma_{jk}^* < 0\},\\
p_{0,j} &= p - p_{+,j} - p_{-,j},\\
\cS_j &= \{k : \gamma_{jk}^* = 0\}.
\end{align*}

 We note that, since $c_k$'s are in the neighborhood of $\bc = \zero$, the sign of $\gamma_{jk}^* - a_j^* c_k$ is approximately the same as the sign of $\gamma_{jk}^*$, so  $p_{+,j} =  \# \{k: \gamma_{jk}^* - a_j^* c_k >0 \} = \# \{k: \gamma_{jk}^* >0 \}$. The same idea applies to $ p_{-,j}$. We also have $p_{0,j} = |\cS_j|$.

Define
\[
\Pi = \{\pi : \{1, \ldots, p\} \to \{1, \ldots, p\}\mid\pi \text{ is a bijection}\}.
\]
Fix a permutation $\pi \in \Pi$ and reorder $\bc$ so that
\[
0 = c_{\pi(1)} \le c_{\pi(2)} \le \cdots \le c_{\pi(p)}
\]
in a neighborhood of $\bc = \mathbf{0}$.

For each row $j$, we now locate the position of the overall median (which is zero under the row-sparsity condition~\eqref{eq:rowspar} in the main text) within the zero set~$\cS_j$. Define
\[
u_j(\bgamma_j^*)
=
\begin{cases}
\dfrac{p_{0,j} + p_{+,j} - p_{-,j}}{2}, & \text{if } p_{0,j} \text{ is even},\\[6pt]
\dfrac{p_{0,j} + p_{+,j} - p_{-,j} + 1}{2}, & \text{if } p_{0,j} \text{ is odd}.
\end{cases}
\]
For simplicity, write $u_j := u_j(\bgamma_j^*)$. Extract the indices of the zeros under $\pi$ via
\[
\cZ_j := \pi^{-1}(\cS_j) = \bigl(z_j(1), \ldots, z_j(p_{0,j})\bigr),
\]
arranged in ascending order. We then define
\begin{equation}\label{eq:k_j_star def}
k_j^*(\pi) := z_j(u_j) \in \cZ_j,
\end{equation}
so that the position in the full ordered sequence is $\pi(k_j^*)$. Because the median of $\{\gamma_{jk}^* - a_j^* c_k\}_{k=1}^p$ is the same as that of $\{\gamma_{jk}^*\}_{k=1}^p$, we obtain
\[
h_j = - a_j^* c_{\pi(k_j^*)}.
\]

\begin{example}
Consider an item $j$ with
\[
\bgamma_j^* = (-1, 0, 0, 0, 1, 4, 5).
\]
Then $\cS_j = \{2, 3, 4\}$, and the median of $\bgamma_j^*$ is $0$. One checks that
\[
u_j(\bgamma_j^*) = \frac{p_{0,j} + p_{+,j} - p_{-,j} + 1}{2}
= \frac{3 + 3 - 1 + 1}{2} = 3.
\]
Suppose the order of $\{c_k\}$ is
\[
c_7 \le c_5 \le c_3 \le c_1 \le c_6 \le c_4 \le c_2,
\]
which is represented by the permutation
\[
\pi : \{1, 2, 3, 4, 5, 6, 7\} \mapsto \{7, 5, 3, 1, 6, 4, 2\}.
\]
Restricting $\pi$ to $\cS_j = \{2, 3, 4\}$, we get
\[
\cZ_j = \pi^{-1}(\cS_j) = \{3, 6, 7\}.
\]
The $u_j$-th zero corresponds to $k_j^* = z_j(3) = 7$, and its position in the full sequence is
\[
\pi(k_j^*) = \pi(7) = 2.
\]
Hence $h_j = - a_j^* c_2$.
\end{example}

In a neighborhood of $\bc = \zero$, we have:
\begin{align}
& g(\bc) -g(\zero) \nonumber \\
     =& \sum_j\sum_k | \gamma_{jk}^* - a_j^*(c_k - c_{\pi(k_j^*)})| - |\gamma_{jk}^*|\nonumber \\
     =& \sum_j\sum_k | \gamma_{j,\pi(k)}^* - a_j^*(c_{\pi(k)} - c_{\pi(k_j^*)})|- |\gamma_{j,\pi(k)}^*|  \nonumber \\
    =& \sum_j\sum_k \II(k \geq k_j^*)\{-a_j^*\II(\gamma_{j,\pi(k)}^* > 0) + |a_j^*|  \II(\gamma_{j,\pi(k)}^* = 0) +a_j^* \II(\gamma_{j,\pi(k)}^* < 0)\}(c_{\pi(k)} - c_{\pi(k_j^*)}) \nonumber \\
   & +   \II(k< k_j^*)\{ a_j^*\II(\gamma_{j,\pi(k)}^* > 0)+ |a_j^*|  \II(\gamma_{j,\pi(k)}^* = 0) -a_j^*\II(\gamma_{j,\pi(k)}^* < 0)\}(c_{\pi(k_j^*)}- c_{\pi(k)})  \nonumber \\
    =& \sum_j \sum_k \II(k \geq k_j^*)[|a_j^*|  \II(\gamma_{j,\pi(k)}^* = 0) - a_j^*\{\II(\gamma_{j,\pi(k)}^* > 0) -  \II(\gamma_{j,\pi(k)}^* < 0)\}]\sum_{s=k_j^*}^{k-1} (c_{\pi(s+1)} - c_{\pi(s)}) \nonumber \\
    +& \II(k< k_j^*) [|a_j^*|  \II(\gamma_{j,\pi(k)}^* = 0)  +a_j^*\{\II(\gamma_{j,\pi(k)}^* > 0) -  \II(\gamma_{j,\pi(k)}^* < 0)\}]\sum_{s=k}^{k_j^*-1} (c_{\pi(s+1)} - c_{\pi(s)})  \nonumber \\
    =& \sum_{w=1}^{p-1} \Big\{\sum_{j=1}^J \II(w \geq k_j^*) \sum_{k=w+1}^p [|a_j^*| \II(\gamma_{j,\pi(k)}^* = 0) - a_j^*\{ \II(\gamma_{j,\pi(k)}^* > 0) - \II(\gamma_{j,\pi(k)}^* < 0)\} ] \nonumber \\
    &+ \II (w < k_j^*) \sum_{k=1}^w [|a_j^*| \II(\gamma_{j,\pi(k)}^* = 0) + a_j^*\{ \II(\gamma_{j,\pi(k)}^* > 0) - \II(\gamma_{j,\pi(k)}^* < 0)\} ]\Big\} (c_{\pi(w+1)} - c_{\pi(w)})   \nonumber \\
    =:& \sum_{w=1}^{p-1} t_w (\pi, \gamma_j^*) (c_{\pi(w+1)} - c_{\pi(w)})  \label{eq:gc - g0 permutation}
\end{align}

Note that the differences $c_{\pi(w+1)} - c_{\pi(w)}$ are nonnegative in a neighborhood of $\bc = \mathbf{0}$. Thus, $g(\bc) - g(\mathbf{0}) > 0$ for all nonzero $\bc$ if
$\forall w \in [p-1], \pi \in \Pi$
\begin{align}
    t_w(\pi, \bgamma_j^*) =& \sum_{j=1}^J \II(w \geq k_j^*) \sum_{k=w+1}^p [|a_j^*| \II(\gamma_{j,\pi(k)}^* = 0) - a_j^*\{ \II(\gamma_{j,\pi(k)}^* > 0) - \II(\gamma_{j,\pi(k)}^* < 0)\} ] \nonumber \\
    &+ \II (w < k_j^*) \sum_{k=1}^w [|a_j^*| \II(\gamma_{j,\pi(k)}^* = 0) + a_j^*\{ \II(\gamma_{j,\pi(k)}^* > 0) - \II(\gamma_{j,\pi(k)}^* < 0)\} ] > 0. \label{eq:tw condition}
\end{align}

This is exactly condition~\eqref{eq:stronger sparsity}. We next show the sufficiency and necessity of the condition below.

\begin{itemize}
    \item \textbf{Sufficiency.}  
    The argument above already establishes sufficiency. If
    \[
        t_w(\pi,\bgamma^*) > 0,\qquad \forall\, w\in[p-1],\ \forall\,\pi\in\Pi,
    \]
    then every coefficient of $(c_{\pi(w+1)} - c_{\pi(w)})$ is strictly positive. Consequently, for any nonzero $\bc$ in a neighborhood of $\zero$, we have $g(\bc) - g(\zero) > 0$, implying that $\bc=\zero$ is the unique minimizer.

    \item \textbf{Necessity.}  
    Suppose instead that there exist $w\in[p-1]$ and $\pi\in\Pi$ such that  
    \[
        t_w(\pi,\bgamma^*) \le 0,
    \]
    that is, condition~\eqref{eq:tw condition} does not hold. We show that in this case one can always find a nonzero $\bc'$ close to $\zero$ such that $g(\bc') \le g(\zero)$, and thus $\bc=\zero$ cannot be a unique minimum.

    Without loss of generality, assume $t_{k-1}(\pi,\bgamma^*) \le 0$ for some $k$.  
    In a sufficiently small neighborhood of $\bc=\zero$, construct a vector $\bc'$ satisfying
    \[
        c_{\pi(k)}' > 0,\qquad 
        c_{\pi(1)}'=\cdots=c_{\pi(k-1)}' = c_{\pi(k+1)}'=\cdots=c_{\pi(p)}' = 0.
    \]
    Then the only nonzero coordinate differences are
    \[
        c_{\pi(k)}' - c_{\pi(k-1)}' = c_{\pi(k)}',\qquad
        c_{\pi(k+1)}' - c_{\pi(k)}' = -c_{\pi(k)}'.
    \]
    Substituting into the expansion of $g(\bc') - g(\zero)$ yields
    \begin{align*}
        g(\bc') - g(\zero)
        &= \sum_{w=1}^{p-1} t_w(\pi,\bgamma^*)\,(c_{\pi(w+1)}' - c_{\pi(w)}') \\
        &= t_{k-1}(\pi,\bgamma^*)\, c_{\pi(k)}' 
           - t_k(\pi,\bgamma^*)\, c_{\pi(k)}' \\
        &= \bigl(t_{k-1}(\pi,\bgamma^*) - t_k(\pi,\bgamma^*)\bigr)\,c_{\pi(k)}'.
    \end{align*}
    For $c_{\pi(k)}'>0$, the right-hand side is negative because $t_{k-1}(\pi,\bgamma^*)\le 0$ and $t_k(\pi,\bgamma^*)> 0$.  
    Hence $g(\bc') < g(\zero)$.

    Therefore, if condition~\eqref{eq:tw condition} fails, then $\bc=\zero$ cannot be the unique minimizer, and Assumption~\ref{cond:ML1} does not hold.
\end{itemize}
\qedhere




\subsection{Proof of Proposition~\ref{prop:consistency}}
\label{sec:proof prof 2}
\subsubsection{Additional Regularity Conditions}
\label{sec:additional regularity conditions}
Let $\Psi^\dagger=\{\bm{\mu}^\dagger, \bm{\sigma}^\dagger, \bm{a}^\dagger, \bm{d}^\dagger, \bm{\gamma}^\dagger\}$ be the model parameters such that $a_1^\dagger=1, \bm{\mu}^{\dagger} = \bm{0}$, and $\bm{d}^{\dagger} = \bm{0}$.
Consider the following random function:
\begin{align*}
    L_N(\Psi) = \frac{1}{\sum_{i=1}^N w_i } \sum_{i=1}^N w_i \log\left( \int \prod_{j: z_{ij}=1} \left[\frac{\exp\{Y_{ij}(a_j\theta_i + d_j + \gamma_{j, {x_i}})\}}{1+\exp(a_j\theta_i + d_j + \gamma_{j,x_i})}\right] \phi(\theta_i\vert \mu_{x_i}, \sigma^2_{x_i}) d\theta_i\Big\}\right),
\end{align*}

The sequence of random functions converges to a fixed limit, which is denoted as $L^{\ast} (\Psi)$. That is, $L_N \overset{p}{\rightarrow} L^{\ast}(\Psi) $ for every parameter $\Psi$.
Let 
\begin{equation}\label{eq:step 1 estimation}
    \begin{aligned}
\tilde \Psi_N = & \argmax_{\Psi} L_N(\Psi)\\
s.t. & ~  a_1 = 1,\; \mu_k = 0,\; d_j = 0,   k  = 1, ..., p,  j =1, ..., J . 
    \end{aligned}
\end{equation}

\begin{condition}
    \label{condition:reg}
    Assume that \begin{enumerate}

    \item There exists $C$ such that $|w_i| \leq C$ for all $i = 1, ..., N$.

    \item $\Psi^{\dagger} \in [-M, M]^{2p + 2J+Jp}$ for sufficiently large $M >0 $.
    
    \item For every $\epsilon >0$,  \begin{align*}
        \sup_{\Psi: d(\Psi, \Psi^{\dagger}) \ge \epsilon } L^{\ast}(\Psi) \le L^{\ast}(\Psi^{\dagger}),
    \end{align*}
    where $d (\cdot, \cdot)$ is the Euclidean distance metric.
    \end{enumerate}
\end{condition}

\subsubsection{Proof}

Since model \eqref{eq:irf-dif} with constraint $a_1^\dagger=1, \bm{\mu}^{\dagger} = \bm{0}$, and $\bm{d}^{\dagger} = \bm{0}$ is identifiable, by the classical asymptotic theory for MLE \citep[see Theorem 5.14 in][]{van2000asymptotic}, under Condition~\ref{condition:reg}, $\tilde \Psi_N \overset{p}{\to} \Psi^\dagger$ as $N\to \infty$. 
Equivalently, as $N\to \infty$, for any $\epsilon>0$ and for all $j=1,...,J, k=1,...,p,$ we have with probability tending to 1 that
$\vert \tilde \mu_k - \mu_k^\dagger\vert\leq \epsilon$, 
$\vert \tilde \sigma_k - \sigma_k^\dagger\vert\leq \epsilon$,
$\vert \tilde a_j - a_j^\dagger\vert\leq \epsilon$,
$\vert \tilde d_j - d_j^\dagger\vert\leq \epsilon$,
$\vert \tilde \gamma_{jk} - \gamma_{jk}^\dagger\vert \leq \epsilon$. 


Define $g_{k,j} (c_k, h_j) = \vert \gamma_{jk}^\dagger  -a_{j}^\dagger c_k  - h_j\vert$
for any $k=1, ..., p$ and $j = 1, ..., J$. Furthermore, define $g_k(c_k, \bm{h}) = \sum_{j=1}^J  g_{k,j} (c_k, h_j)$, $g_j(\bm{c}, h_j) = \sum_{k=1}^p g_{k,j} (c_k, h_j)$, and $g(\bm c, \bm{h})=\sum_{k=1}^p\sum_{j=1}^J \vert \gamma_{jk}^\dagger - c_k a_j^\dagger - h_j\vert$ as function of $\bm c$ and $\bm h$ with model parameters defined at $\Psi^{\dagger}$.
In addition, we let $g_{N,kj} (c_k, h_j) = \vert \tilde{\gamma}_{jk}  -\tilde{a}_{j} c_k  - h_j\vert$, $g_{Nk}(c_k, \bm{h}) = \sum_{j=1}^J  g_{N,kj} (c_k, h_j)$, $g_{Nj}(\bm{c}, h_j) = \sum_{k=1}^p g_{N,kj} (c_k, h_j)$, and $g_N(\bm c, \bm{h})=\sum_{k=1}^p\sum_{j=1}^J \vert \tilde{\gamma}_{jk} - c_k \tilde{a}_j - h_j\vert$ with model parameters defined at $\tilde{\Psi}$. Denote their minimizers as $(c_1^{\dagger}, ..., c_p^{\dagger}, h_1^{\dagger}, ..., h_J^{\dagger}) = \argmin_{\bm c, \bm h} g(\bm c, \bm{h})$ and $(\hat{c}_1, ..., \hat{c}_p, \hat{h}_1, ..., \hat{h}_J) = \argmin_{\bm c, \bm h} g_N(\bm c, \bm{h})$ and we aim to establish that $\hat{c}_k \overset{p}{\rightarrow} c_k^\dagger$ and $\hat{h}_j \overset{p}{\rightarrow} h_j^{\dagger}$ as $N\to \infty$.

Note that the argument below would work for any $k=1,...,p$ and $j = 1, ..., J$.
By Condition~\ref{condition:reg}(2), there exists $M_{k1} <\infty$ such that $J, \vert \gamma_{jk}^\dagger\vert, \vert a_j^\dagger\vert \leq M_{k1}$.  
Then, there must exist $M_{k2}<\infty$ such that $\vert c_k^\dagger \vert \leq M_{k2}$ and $\vert h_j^{\dagger}\vert \leq M_{k2}$.
Furthermore, note $g_{N}$ is clearly continuous and jointly convex in $c_k$s and $h_j$s, so consistency will follow if $g_{N}$ can be shown to converge point-wise to $g^*$ that is uniquely minimized at the true value $c_k^\dagger$ (point-wise convergence of convex functions implies their uniform convergence on compact subsets).

{
Following the model identifiability definition and Assumption~\ref{cond:ML1} in the main text, $c_k^\dagger$s and $h_j^{\dagger}$s are unique. 
To see this, suppose for contradiction that there exist $c_{k1} \neq c_{k2}$ for some $k = 1, ..., p$ such that they are not equivalent but both minimizers for $c_k$, i.e., $c_{k1} \neq c_{k2} = \argmin_{c_k} g_k(c_k, \bm h^{\dagger})$. 
Based on Algorithm 1, we see the discrimination parameters $a_j$'s are not changed after transformation into the constrained space in Step 2, for which we have $a_j^\dagger=a_j^*$ for all $j=1,...,J$. 
As $\Psi^{\dagger}$ and $\Psi^*$ are in the same equivalence class which maximizing the likelihood function $L^*(\Psi)$, there exists transformation constant $c_{k3}$ such that $\gamma_{jk}^\dagger$ can be transformed from $\gamma_{jk}^*$ via $\gamma_{jk}^\dagger=\gamma_{jk}^*+c_{k3} a_j^* + h_j^{\dagger}$.
So we have
\begin{align}
    c_{k1}& =\arg\min_{c_k} \sum_{j=1}^J |\gamma_{jk}^{\dagger} - c_k a_j^* - h_j^{\dagger} | \nonumber \\
     & =\arg\min_{c_k} \sum_{j=1}^J |\gamma_{jk}^*+c_{k3} a_j^* + h_j^{\dagger} - c_k a_j^* - h_j^{\dagger} | \nonumber \\
      & =\arg\min_{c_k} \sum_{j=1}^J |\gamma_{jk}^* +(c_{k3}-c_k) a_j^*|, \label{eq:ck1}
\end{align}
and similarly,
\begin{align}
    c_{k2} = \arg\min_{c_k} \sum_{j=1}^J \vert \gamma_{jk}^* + (c_{k3}-c_k) a_j^* \vert. \label{eq:ck2}
\end{align}
Since $\sum_{j=1}^J |\gamma_{jk}^{\dagger}| \le \sum_{j=1}^J |\gamma_{jk}^{\dagger} - c_k a_j^{\dagger} - h_j^{\dagger} |$ where the equality is achieved at $c_k = c_{k1}$ and $c_k = c_{k2}$, together with~\eqref{eq:ck1} and~\eqref{eq:ck2}, we have $\gamma_{jk}^*=\gamma_{jk}^\dagger - (c_{k3}-c_{k1})a_j^* $ and $\gamma_{jk}^*=\gamma_{jk}^\dagger - (c_{k3}-c_{k2})a_j^*$. If Assumption~\ref{cond:ML1} in the main text holds, then $c_{k3}=c_{k1}$ and $c_{k3}=c_{k2}.$ This contradicts the assumption $c_{k1}\neq c_{k2}.$ Hence, $c_k^\dagger$ must be unique. 

For any $\vert c_k\vert \leq M_{k2},$\\
\begin{align*}
 & \vert g_{Nk}(c_k, \bh^{\dagger}) - g_k(c_k, \bh^{\dagger})\vert\\
& = \Big\vert \sum_{j=1}^J \Big(\vert \tilde\gamma_{jk} - c_k \tilde a_j - h_j^{\dagger} \vert-\vert \gamma_{jk}^\dagger - c_k a_j^\dagger - h_j^{\dagger}\vert\Big) \Big\vert \\
&\leq \Big\vert \sum_{j=1}^J \Big(\vert (\tilde\gamma_{jk} - c_k \tilde a_j - h_j^{\dagger})- ( \gamma_{jk}^\dagger - c_k a_j^\dagger - h_j^{\dagger})\vert\Big) \Big\vert \\
&=\Big\vert \sum_{j=1}^J \Big(\vert (\tilde\gamma_{jk} - \gamma_{jk}^\dagger)+ c_k( a_j^\dagger- \tilde a_j)\vert\Big) \Big\vert \\
&\leq \sum_{j=1}^J \Big(\vert \tilde\gamma_{jk} - \gamma_{jk}^\dagger\vert+ \vert c_k \vert \cdot \vert a_j^\dagger- \tilde a_j\vert\Big) \\
&\leq J \epsilon + \vert c_k \vert\epsilon.\\
&\leq (M_{k1}+M_{k2})\epsilon.\\
\end{align*}
Take $\epsilon_k=(M_{k1}+M_{k2})\epsilon$, it follows that for any fixed $\vert c_k\vert \leq M_{k2}$, $P\big(\vert g_{Nk}(c_k, \bh^{\dagger}) - g_k(c_k, \bh^{\dagger})\vert \leq \epsilon_{k}\big) \to 1$ as $N\to \infty$. Moreover, following from the uniqueness of $c_k^\dagger$ and the continuity and the convexity of $g_{Nk}(\cdot, \bh^{\dagger})$ in $c_k$, we must have $\vert \hat c_k - c_k^\dagger \vert = o_P(1)$ as $N \to \infty.$ Similarly we can show the uniqueness of $h_j^{\dagger}$ and by the continuity and convexity of $g_{Nj}(\bc^{\dagger}, \cdot)$ in $h_j$, we have $|\hat{h}_j - h_j^{\dagger}| = o_P(1)$.}

Note that $\hat \mu_k=\tilde\mu_k + \hat{c}_k$, $\hat{\sigma}_k=\tilde\sigma_k$, $\hat{\gamma}_{jk} = \tilde{\gamma}_{jk} - \hat{c}_k\tilde{a}_j -\hat{h}_j$, $\hat a_j=\tilde a_j$, $\hat d_j=\tilde d_j +\hat{h}_j$ for $j=1,...,J.$ From the model identifiability and Assumption~\ref{cond:ML1}, we know that $\mu_k^*=\mu_k^\dagger + c_k^\dagger$, $\sigma_k^*=\sigma_k^\dagger$, $\gamma_{jk}^* = \gamma_{jk}^\dagger - c_k^\dagger a_j^\dagger -h_j^{\dagger}$, $a_j^*=a_j^\dagger$, $d_j^*=d_j^\dagger +h_j^{\dagger}$ for all $j=1,...,J.$ 
Since $\vert \hat c_k - c_k^\dagger \vert = o_P(1), |\hat{h}_j - h_j^{\dagger}| = o_P(1), \vert \tilde \mu_k - \mu_k^\dagger\vert=o_P(1)$, $\vert \tilde \sigma_k - \sigma_k^\dagger\vert=o_P(1)$, $\vert \tilde \gamma_{jk} - \gamma_{jk}^\dagger\vert =o_P(1), \vert \tilde a_j - a_j^\dagger\vert =o_P(1), \vert \tilde d_j - d_j^\dagger\vert =o_P(1)$ as $N\to\infty$, it follows directly from the Slutsky's Theorem that $\vert \hat \mu_k - \mu_k^*\vert=o_P(1)$, $\vert \hat \sigma_k - \sigma_k^*\vert=o_P(1)$, $\vert \hat \gamma_{jk} - \gamma_{jk}^*\vert =o_P(1), \vert \hat a_j - a_j^*\vert =o_P(1)$, $\vert \hat d_j - d_j^*\vert =o_P(1)$ as $N\to \infty$.
Hence the proposition follows.

\subsection{Proof of Proposition~\ref{prop:equivalence estimator}}

Assume the contrary of the Proposition statement, i.e., the output estimator in Algorithm~\ref{alg:ML1} is {\em not equivalent} to the constrained maximum marginal likelihood estimator $\hat{\Psi}$ in~\eqref{eq:cmml}. Denote this algorithm output as $\hat{\Psi}^{\prime}$ and the intermediate estimator in Step 1 as $\tilde{\Psi}^{\prime}$. That is, $\tilde{\Psi}^{\prime}$ is the solution to the optimization problem~\eqref{eq:mml}:
\begin{align*}
    \tilde \Psi =& \arg\max_{\Psi}  L(\Psi), ~
 s.t.\;   a_1 = 1,\; \mu_k = 0,\; d_j = 0,   k  = 1, ..., p,  j =1, ..., J.
\end{align*}
 The uniqueness of $\hat{\Psi}$ has been proved in Section~\ref{sec:proof prof 2}. Therefore, we have
\begin{align}
    L(\hat{\Psi}) > L(\hat{\Psi}^{\prime}) = L(\tilde{\Psi}^{\prime}), \label{eq:L equivalence}
\end{align}
where the last equality is because the two estimators are from the same equivalence class, $\hat{\Psi}^{\prime} \sim \tilde{\Psi}^{\prime}$. Let the constrained parameter space for problem~\eqref{eq:cmml} to be $\cM_1 =\{\Psi: \Psi \in \cM, a_1 = 1, \sum_{k=1}^p \mu_k=0\}$ and the constrained parameter space for~\eqref{eq:mml} to be $\cM_2 = \{\Psi: a_1=1, \mu_1 = \cdots=\mu_p=0, d_1 =\cdots=d_J=0\}$. 

For any $\hat{\Psi} \in \cM_1$, there exists $\tilde{\Psi} \in \cM_2$ such that $\tilde{\Psi} \sim \hat{\Psi}$ and $\tilde{\Psi}$ can be transformed into $\hat{\Psi}$ via the following process. First, we solve for $\hat{c}_1, \dots, \hat{c}_p, \hat{h}_1, \dots, \hat{h}_J$ by the following:
\begin{align*}
     (\hat{c}_1, \dots, \hat{c}_p, \hat{h}_1, \dots, \hat{h}_J) =  \underset{c_1, \dots, c_p, h_1, \dots, h_J} {\argmin} &  \sum_{k=1}^p\sum_{j=1}^J \vert  \tilde\gamma_{jk}  - \tilde a_j c_k - h_j\vert \quad
 s.t.  \sum_{k=1}^p c_k = 0.  
\end{align*}
Then $\hat{\Psi}$ is obtained through the transformation: $\hat \gamma_{jk} = \tilde\gamma_{jk}  - \tilde a_j \hat c_k - \hat{h}_j,$
$\hat\mu_k = \hat c_k,$   $\hat a_j = \tilde a_j$, $\hat d_j =\hat h_j$ and $\hat\sigma_k=\tilde\sigma_k$. It can be verified that $\hat{\Psi} \in \cM_1$ since
\begin{align*}
    \sum_{j=1}^J\sum_{k=1}^p\vert \hat\gamma_{jk}\vert \leq \sum_{j=1}^J\sum_{k=1}^p \vert \hat\gamma_{jk} - \hat{a}_jc_k - h_j \vert, \forall c_k, h_j
\end{align*}
and $\hat{a}_1 = 1$ and $\sum_{k=1}^p\hat\mu_k = \sum_{k=1}^p \hat{c}_k = 0$. The equivalence of $\tilde{\Psi} \sim \hat{\Psi}$ implies $L(\tilde{\Psi}) = L (\hat{\Psi})$ and by~\eqref{eq:L equivalence}, we have 
\begin{align*}
    L(\tilde{\Psi}) > L(\tilde{\Psi}^{\prime}),
\end{align*}
which contradits to the fact that $\tilde{\Psi}^{\prime}$ maximizes $L(\cdot)$ in the parameter space $\cM_2$. Hence we conclude that the output estimator in Algorithm~\ref{alg:ML1} is equivalent to the constrained maximum marginal likelihood estimator $\hat{\Psi}$ in~\eqref{eq:cmml}.

{
\section{Additional Discussion on Non-Uniform DIF}
\label{sec:non uniform}
In this section, we provide more details on the non-uniform DIF model, where the measurement noninvariance is related to not only the intercept, but also the slope. 
As described in the last paragraph of Section 2.2 in the main text, we can allow the slope parameters to vary across countries through the following model setup:
 \begin{equation}\nonumber
P(Y_{ij} = 1\vert \theta_i = \theta, x_i = k) = \frac{\exp(a_{j}\exp(\zeta_{jk} ) \theta + d_j +  \gamma_{jk} )}{1+\exp(a_{j}\exp(\zeta_{jk} )\theta + d_j +  \gamma_{jk})}, j=1, ..., J, k=1, ...p, 
\end{equation} 
with the constraints 
$\sum \mu_k = 0$ and $\sum \log (\sigma_k) = 0$ to fix the location and scale of the latent variable.

In this setup, identifiability problems are more challenging, necessitating more sophisticated estimation methods. Even under the model constraints $\sum \mu_k = 0$ and $\sum \log (\sigma_k) = 0$, the model parameters remain unidentifiable without additional assumptions.
 In particular,
the model remains unchanged if we simultaneously replace:
\begin{align*}
&  a_j^* \text{  by  } a_j^* \exp (\eta_j);\quad  \mu_k^* \text{  by  }\mu_k^* + c_k; \quad \sigma_k^* \text{  by  }\sigma_k^*  \exp (m_k); \\
&  d_j^* \text{  by  } d_j^* + h_j;\quad \zeta_{jk}^* \text{  by  } \zeta_{jk}^* - m_k- \eta_j;\quad \gamma_{jk}^* \text{  by  } \gamma_{jk}^* - a_j^*\exp(\zeta_{jk}^*)c_k - h_j
\end{align*}
for all $j = 1, ..., J$ and $k = 1, ..., p$, satisfying $\sum_{k=1}^p c_k =0$ and $\sum_{k=1}^p m_k= 0$. These constraints on $c_k$'s and $m_k$'s correspond to the model constraints $\sum \mu_k = 0$ and $\sum \log (\sigma_k) = 0$. 
Here, the country-specific intercept is  $d_j + \gamma_{jk}$, while the parameter $a_j\exp(\zeta_{jk})$ is interpreted as country-specific slope parameter. The parameter $\zeta_{jk}$ captures the deviation of the slope parameters from the common slope parameters $a_j$ for each country $k = 1, ..., p$. If $\zeta_{jk}=0$ for all $k$, the slope parameters $a_j$ are common parameters shared across countries.
If $\gamma_{jk} = 0$ for all $k$, the intercept parameter $d_j + \gamma_{jk}$ are common across countries.
Therefore,
for any country $k = 1, ..., p$, an item $j$ is not a DIF item if $\zeta_{jk} = \gamma_{jk} = 0$.

In the same spirit as Assumption 1 in the main text, we may assume the true model parameters $\zeta_{jk}^*$ and $\gamma_{jk}^*$ satisfy a similar sparsity condition:
\begin{align}
    \sum_{j=1}^J\sum_{k=1}^p |\zeta_{jk}^*| \le  \sum_{j=1}^J\sum_{k=1}^p | \zeta_{jk}^* - m_k - \eta_j| \label{eq:nonuniform cond1}
\end{align}
and 
\begin{align}
     \sum_{j=1}^J\sum_{k=1}^p |\gamma_{jk}^*| \le \sum_{j=1}^J\sum_{k=1}^p |\gamma_{jk}^* - a_j^*\exp(\zeta_{jk}^*)c_k - h_j| \label{eq:nonuniform cond2}
\end{align}
 for all possible $c_1, ..., c_p$, $m_1, ..., m_p$, $h_1, ..., h_J$, and $\eta_1$, ..., $\eta_J$ satisfying constraints $\sum_{k=1}^p c_k =0$ and $\sum_{k=1}^p m_k= 0$. 
 The above inequalities tend to hold when both $(\zeta_{jk}^*)_{J\times p}$ and $(\gamma_{jk}^*)_{J\times p}$ are sufficiently sparse.

Under the sparsity assumptions described above, we briefly introduce the estimation method. The weighted marginal log-likelihood function is given by:
\begin{equation*}
L (\Psi) = \sum_{i=1}^N w_i \log\left( \int \prod_{j: z_{ij}=1} \left[\frac{\exp\{Y_{ij}(a_j\exp(\zeta_{j,x_i}) \theta_i + d_j + \gamma_{j, {x_i}})\}}{1+\exp(a_j\exp(\zeta_{j,x_i})\theta_i + d_j + \gamma_{j,x_i})}\right] \phi(\theta_i\vert \mu_{x_i}, \sigma^2_{x_i}) d\theta_i\Big\}\right)
\end{equation*}
where $\Psi = (a_1, ..., a_J, d_1, ..., d_J, \mu_1, ..., \mu_p, \sigma_1, ..., \sigma_p, \gamma_{11}, ..., \gamma_{Jp}, \zeta_{11}, ..., \zeta_{Jp})^\top \in \mathbb R^{2J + 2p + 2Jp}$ is a collection vector of all the parameters.

Similarly we aim to solve for the constrained marginal maximum  likelihood estimator 
\begin{equation}\label{eq:cmml}
    \begin{aligned}
\hat \Psi = & \argmax_{\Psi} L(\Psi)\\
s.t. & ~ \Psi \in \mathcal M,
~~  \sum_{k=1}^p \mu_k = 0 \mbox{~and~}  \sum_{k=1}^p \log(\sigma_k) = 0. 
    \end{aligned}
\end{equation}
where the constrained parameter space 
\begin{equation}
\begin{aligned}
\mathcal{M}
:= \biggl\{ \Psi  :\;
&\sum_{j=1}^J \sum_{k=1}^p \lvert \gamma_{jk} \rvert 
  \le 
  \sum_{j=1}^J \sum_{k=1}^p 
  \bigl\lvert \gamma_{jk} - a_j \exp({\zeta_{jk}}) c_k - h_j \bigr\rvert, \\
&\sum_{j=1}^J \sum_{k=1}^p \lvert \zeta_{jk} \rvert 
  \le 
  \sum_{j=1}^J \sum_{k=1}^p 
  \bigl\lvert \zeta_{jk} - m_k - \eta_j \bigr\rvert,
\\
& \text{for all } 
  c_1,\dots,c_p,\, 
  m_1,\dots,m_p,\, 
  h_1,\dots,h_J,\, 
  \eta_1,\dots,\eta_J \in \mathbb{R} \\
&\text{satisfying } 
  \sum_{k=1}^p c_k = 0
  \quad\text{and}\quad 
  \sum_{k=1}^p m_k = 0 \biggr\}\subset \mathbb R^{2J + 2p + 2Jp}
\end{aligned} \nonumber
\end{equation}
is set to pursue the sparsity conditions~\eqref{eq:nonuniform cond1}--\eqref{eq:nonuniform cond2} under the non-uniform DIF setting. Algorithm 1 in the main text is readily extended to solve this problem. 

\section{Additional Discussion on Identifiability}
\label{sec:discussion ID def}
 In the definition of model identifiability given in~\cite{bickel2015mathematical}, a model parameter $\Psi$ is identifiable when the corresponding distributions $P_{\Psi}=P_{\Psi^{\prime}}$ implies  $\Psi=\Psi^{\prime}$ for all $\Psi, \Psi^{\prime}$. In the main text, we introduce the sparsity assumption, Assumption~1, that constrains the parameter space into a subset of Euclidean space where the DIF effect $\gamma_{jk}$s has the smallest $L_1$ norm among equivalence models.
Within this constrained space implied by Assumption 1 and additional standard identifiability constraints to fix the location and scale of the latent variables (which is set as $a_1= 1$ and $\sum_{k=1}^p \mu_k = 0$ in the current paper), it holds that $P_{\Psi}=P_{\Psi^{\prime}}$ implies  $\Psi=\Psi^{\prime}$. 

To illustrate that the model is identifiable within the constrained space, we next show that after imposing the standard identifiability constraints to fix the location and scale of the latent variables, the linear transformation described in Section 2.2 of the main text is the only source of model indeterminacy for the identifiability definition given by~\cite{bickel2015mathematical} to be violated. 
To the end, suppose that there exist some transformation $\theta^{\prime} = T_k(\theta)$ about $\theta$ that gives the same response distribution as $\theta$: 
\begin{align*}
    P(Y \mid \theta, x_i) = P(Y \mid \theta^{\prime}, x_i).
\end{align*}
Here the transformation $T_k(\theta)$ would only be allowed to vary across country groups $k = 1,..., p$, but not vary across items $j = 1,..., p$ because of the structural assumption that $\theta_i$ given $x_i = k$ is assumed to follow a normal distribution $\cN(\mu_k, \sigma_k^2)$. Therefore, the latent variable $\theta$ is shared across all items. 

As a result,  we have 
\begin{align*}
    \frac{\exp(a_j \theta + d_j + \gamma_{jk})}{1+ \exp(a_j \theta + d_j + \gamma_{jk})} = \frac{\exp(a_j^{\prime} \theta^{\prime} + d_j^{\prime} + \gamma_{jk}^{\prime})}{1+\exp(a_j^{\prime} \theta^{\prime} + d_j^{\prime} + \gamma_{jk}^{\prime})}.
\end{align*}
As the logit function is strictly monotone, we must have
\begin{align*}
    a_j \theta + d_j + \gamma_{jk} = a_j^{\prime} T_k(\theta) + d_j^{\prime} + \gamma_{jk}^{\prime}, \quad \text{ for all }\theta
\end{align*}
So the transformation can only be linear:
\begin{align*}
    T_k(\theta) = \frac{a_j}{a_j^{\prime}} \theta + \frac{d_j + \gamma_{jk} - d_j^{\prime} - \gamma_{jk}^{\prime}}{a_j^{\prime}},
\end{align*}
and no nonlinear transformation $T_k$ can keep parameter within the equivalence class.

Given this, we assume the linear transformation of $T_k(\theta)$ takes a general form $T_k(\theta) = s_k \theta + t_k$. Substituting it into the distribution we have:
\begin{align*}
      a_j \theta + d_j + \gamma_{jk} = \frac{a_j}{s_k} \theta^{\prime} + d_j + \gamma_{jk} - \frac{a_j t_k}{s_k},
\end{align*}
where here the slope for $\theta^{\prime}$ denotes discrimination parameters and should be the same across countries $k = 1, ..., p$, which forces a single global scale that $s_1 = ... = s_p = : s$, so $T_k(\theta) = s \theta + t_k$ and $t_k$ is the only remaining freedom. The standard scale constraint \(a_1=1\) fixes the global scale \(s\) (so \(s=1\) after re-normalization),
leaving only group-specific translations,
\[
T_k(\theta)=\theta+t_k.
\]
Writing \(c_k:=t_k\), this induces the invariance
\[
\mu_k^{\prime}=\mu_k+c_k,
\qquad
d_j^{\prime}+\gamma_{jk}^{\prime}=d_j+\gamma_{jk}-a_jc_k.
\]
Moreover, the location constraint \(\sum_{k=1}^p \mu_k=0\) implies \(\sum_{k=1}^p c_k=0\).
Hence, after imposing the standard location/scale constraints, the only remaining likelihood-preserving freedom arising from transformation on latent factor $\theta$ is the group-specific shift \(\{c_k\}\). Combining this with the algebraic decomposition: \(d_j'=d_j+h_j\) and \(\gamma'_{jk}=\gamma_{jk}-h_j\)) gives the full
equivalence-class transformation stated in Section 2.2 of the main text and there is no other transformation than this that preserves the likelihood.

 This is why the model becomes identifiable within the constrained space under Assumption 1. In addition, when Assumption 1 is violated, Algorithm 1 no longer yields consistent results, as in this case the true model parameters are no longer identifiable within the constrained space $\mathcal M$.



}


\section{Simulation Study}\label{sec-simulation}


\subsection{Simulation Settings}

\label{sec-simulation settings}
For data generation, we vary the sample size $N\in \{10000, 20000\}$ and consider two settings with different numbers of countries: $p \in \{10, 20\}$. Let $x_i \in \{1, ..., p\}$ denote the country membership of each individual for $i = 1,...,N$ and each country $k$ have $N/p$ observations for $k = 1,...,p$. Each setting for the number of countries corresponds to a specific number of items: (i) $p = 10$ and $J = 15$; (ii) $p = 20$ and $J = 30$. 
The discrimination parameters $a_1, ..., a_J $  are set within the range $ [1,2]$ and the easiness parameter $d_1, ..., d_J $ are in the range $[-1, 1]$ {and $[-3, 3]$}.
 Detailed specifications of these parameter settings are provided in Tables~\ref{tab:a setting}--\ref{tab:d setting large}. 
 {The simulation results under the range $d_j^* \in [-3, 3]$ are provided in Section~\ref{sec:enlarge d}.} 
 For simplicity, we assume all the sampling weights are ones, i.e., $w_i = 1$, and there is no missing response, i.e., $z_{ij} = 1$ for all $i = 1,.., N$ and $j = 1, ..., J$.

\begin{table}[ht]
    \centering
    \begin{tabular}{c|ccccccccccccccc}
    \hline
     & \multicolumn{15}{c}{Discrimination parameters: $a_1^*, \dots, a_J^*$} \\
    \hline
 $J=15$  & 1.0 &  1.2 & 1.4 & 1.6 & 1.8 & 1.0 &  1.2 & 1.4 & 1.6 & 1.8 & 1.0 &  1.2 & 1.4 & 1.6 & 1.8 \\
       \hline
 \multirow{2}{*}{ $J=30$ }  &  1.0 &  1.2 & 1.4 & 1.6 & 1.8 & 1.0 &  1.2 & 1.4 & 1.6 & 1.8 & 1.0 &  1.2 & 1.4 & 1.6 & 1.8 \\
         & 1.0 &  1.2 & 1.4 & 1.6 & 1.8 & 1.0 &  1.2 & 1.4 & 1.6 & 1.8 & 1.0 &  1.2 & 1.4 & 1.6 & 1.8\\
       \hline
    \end{tabular}
    \caption{Settings of the discrimination parameter.}
    \label{tab:a setting}
\end{table}

\begin{table}[H]
    \centering
    \resizebox{\textwidth}{!}{
    \begin{tabular}{c|ccccccccccccccc}
    \hline
     & \multicolumn{15}{c}{Easiness parameters: $d_1^*, ..., d_J^*$} \\
    \hline
 $J=15$  & 0.8 &  0.2 & -0.4 & -1.0 & 1.0 & 0.8 &  0.2 & -0.4 & -1.0 & 1.0 & 0.8 &  0.2 & -0.4 & -1.0 & 1.0  \\
       \hline
 \multirow{2}{*}{$J=30$ }  & 0.8 &  0.2 & -0.4 & -1.0 & 1.0 & 0.8 &  0.2 & -0.4 & -1.0 & 1.0 & 0.8 &  0.2 & -0.4 & -1.0 & 1.0  \\
         & 0.8 &  0.2 & -0.4 & -1.0 & 1.0 & 0.8 &  0.2 & -0.4 & -1.0 & 1.0 & 0.8 &  0.2 & -0.4 & -1.0 & 1.0 \\
       \hline
    \end{tabular}}
    \caption{Settings of the easiness parameter under the range $d_j^* \in [-1, 1]$.}
    \label{tab:d setting}
\end{table}

\begin{table}[H]
    \centering
 {      \resizebox{\textwidth}{!}{
 \begin{tabular}{c|ccccccccccccccc}
    \hline
     & \multicolumn{15}{c}{Easiness parameters: $d_1^*, ..., d_J^*$} \\
    \hline
 $J=15$ & 2.0 & 1.0 & -0.5 & -3.0 & 3.0 & 2.0 & 1.0 & -0.5 & -3.0 & 3.0 & 2.0 & 1.0 & -0.5 & -3.0 & 3.0  \\
       \hline
 \multirow{2}{*}{$J=30$ }  & 2.0 & 1.0 & -0.5 & -3.0 & 3.0 & 2.0 & 1.0 & -0.5 & -3.0 & 3.0 & 2.0 & 1.0 & -0.5 & -3.0 & 3.0  \\
         & 2.0 & 1.0 & -0.5 & -3.0 & 3.0 & 2.0 & 1.0 & -0.5 & -3.0 & 3.0 & 2.0 & 1.0 & -0.5 & -3.0 & 3.0  \\
       \hline
    \end{tabular}}
    \caption{Settings of the easiness parameter under the range $d_j^* \in [-3, 3]$.}
    \label{tab:d setting large}}
\end{table}

For (i) $p = 10$ and $J = 15$, we consider 
two generation settings for the DIF parameters $ (\gamma_{jk}^*)_{J \times p}$. These two settings are denoted as $S1$ and $S2$ and shown in Table~\ref{tab: J15 dif}. In these settings, the 
values of the nonzero entries are randomly generated from Unif$[1,3]$. The locations of the nonzero entries are set to have a blockwise pattern in $S1$, while they are randomly dispersed in $S2$. Similarly, for (ii) with $p = 20$ and $J = 30$, we consider two analogous settings: $S3$, with a similar blockwise pattern for nonzero entries as in $S1$, and $S4$, with randomly dispersed nonzero entries. The specific values of the DIF parameters for $S3$ and $S4$ are given in Tables~\ref{tab: J30 dif setting 3} and \ref{tab: J30 dif setting 4} in Section~\ref{sec-sim}. For all the settings, 
Assumption~\ref{cond:ML1} is satisfied, and thus, the true model can be identified.  

Finally, for all the settings,  the latent traits $\theta_i$ are generated from country-specific Gaussian distributions with mean parameters in the range of $[-0.3, 0.3]$ and the variance parameters are in the range of $[0.4, 1.2]$. The specific values of these parameters are presented in Tables~\ref{tab:mean setting}-\ref{tab:var setting}. For each setting, 100 independent Monte Carlo simulations are conducted.

\begin{table}[!h]
\fontsize{8.5}{8.5}\selectfont 
\begin{center}
 \resizebox{0.45\textwidth}{!}{\begin{tabular}{cccccccccc}
\hline
\multicolumn{10}{c}{$S1$} \\
\hline 0.00 & 0.00 & 0.00 & 0.00 & 0.00 & 0.00 & 1.66 & 1.25 & 2.26 & 2.07 \\
 0.00 & 0.00 & 0.00 & 0.00 & 0.00 & 2.62 & 2.20 & 1.59 & 0.00 & 2.11 \\
 0.00 & 0.00 & 0.00 & 0.00 & 0.00 & 1.77 & 0.00 & 2.16 & 2.01 & 2.74 \\
 1.42 & 0.00 & 0.00 & 0.00 & 0.00 & 2.57 & 2.14 & 0.00 & 0.00 & 0.00 \\
 0.00 & 2.66 & 1.03 & 2.82 & 1.91 & 0.00 & 0.00 & 0.00 & 0.00 & 0.00 \\
 0.00 & 1.22 & 1.26 & 2.12 & 1.52 & 0.00 & 0.00 & 0.00 & 0.00 & 0.00 \\
 0.00 & 2.41 & 1.19 & 2.51 & 1.67 & 0.00 & 0.00 & 0.00 & 0.00 & 0.00 \\
 0.00 & 2.79 & 1.47 & 1.76 & 2.78 & 0.00 & 0.00 & 0.00 & 0.00 & 0.00 \\
 0.00 & 1.56 & 2.58 & 1.75 & 1.40 & 0.00 & 0.00 & 0.00 & 0.00 & 0.00 \\
 0.00 & 1.46 & 2.20 & 1.34 & 2.16 & 0.00 & 0.00 & 0.00 & 0.00 & 0.00 \\
 1.56 & 0.00 & 0.00 & 0.00 & 0.00 & 1.35 & 1.84 & 0.00 & 0.00 & 0.00 \\
 0.00 & 0.00 & 0.00 & 0.00 & 0.00 & 1.54 & 0.00 & 2.99 & 1.45 & 1.12 \\
 0.00 & 0.00 & 0.00 & 0.00 & 0.00 & 1.10 & 1.46 & 2.69 & 0.00 & 2.61 \\
 0.00 & 0.00 & 0.00 & 0.00 & 0.00 & 0.00 & 1.43 & 2.82 & 1.56 & 1.21 \\
 0.00 & 0.00 & 0.00 & 0.00 & 0.00 & 1.63 & 2.75 & 1.94 & 0.00 & 2.53 \\
\hline
\end{tabular}}~\;\;\;\;
 \resizebox{0.45\textwidth}{!}{\begin{tabular}{cccccccccc}
\hline
\multicolumn{10}{c}{$S2$} \\
\hline
 0.00 & 2.40 & 2.91 & 0.00 & 0.00 & 0.00 & 2.62 & 0.00 & 2.06 & 0.00 \\
 0.00 & 0.00 & 1.40 & 0.00 & 0.00 & 0.00 & 2.12 & 0.00 & 0.00 & 1.53 \\
 0.00 & 0.00 & 1.87 & 0.00 & 1.28 & 0.00 & 0.00 & 1.96 & 0.00 & 1.45 \\
 0.00 & 1.06 & 0.00 & 0.00 & 1.55 & 0.00 & 0.00 & 0.00 & 0.00 & 1.12 \\
 0.00 & 0.00 & 2.86 & 0.00 & 0.00 & 2.10 & 2.07 & 2.65 & 0.00 & 0.00 \\
 0.00 & 0.00 & 0.00 & 1.76 & 0.00 & 0.00 & 0.00 & 1.51 & 2.87 & 1.51 \\
 0.00 & 0.00 & 2.65 & 0.00 & 2.86 & 1.31 & 0.00 & 2.78 & 0.00 & 0.00 \\
 2.40 & 1.89 & 0.00 & 2.84 & 0.00 & 0.00 & 0.00 & 1.08 & 0.00 & 0.00 \\
 1.10 & 0.00 & 0.00 & 0.00 & 0.00 & 2.57 & 2.94 & 0.00 & 0.00 & 1.94 \\
 0.00 & 1.79 & 1.47 & 2.63 & 0.00 & 0.00 & 0.00 & 0.00 & 1.65 & 0.00 \\
 0.00 & 1.55 & 0.00 & 0.00 & 2.04 & 1.80 & 0.00 & 0.00 & 2.69 & 0.00 \\
 2.55 & 0.00 & 0.00 & 0.00 & 0.00 & 1.59 & 0.00 & 0.00 & 1.79 & 1.91 \\
 0.00 & 0.00 & 0.00 & 1.56 & 2.74 & 2.96 & 0.00 & 2.61 & 0.00 & 0.00 \\
 1.04 & 0.00 & 0.00 & 2.69 & 1.76 & 0.00 & 0.00 & 0.00 & 0.00 & 1.01 \\
 0.00 & 0.00 & 0.00 & 1.84 & 1.50 & 0.00 & 1.82 & 2.80 & 0.00 & 0.00 \\
\hline
\end{tabular}} \\
\caption{DIF effect parameters at (i)$~p = 10, J=15$ under $S1$ and $S2$. Each nonzero entry is randomly generated from Unif$[1,3]$.}
\label{tab: J15 dif}
\end{center}
\end{table}

\begin{table}[H]
    \centering
        \resizebox{\textwidth}{!}{
    \begin{tabular}{c|cccccccccc}
    \hline
    \multicolumn{1}{c}{} & \multicolumn{10}{c}{Mean proficiency: $\mu_1^*, ..., \mu_p^*$} \\
    \hline
 $p=10$  & 0.21 &  0.09 & 0.07 & 0.05 & 0.03 & -0.05 & -0.07 & -0.09 & -0.11 & -0.13 \\
       \hline
 \multirow{2}{*}{$p=20$}  &  0.252 &  0.122 & 0.102 & 0.092 & 0.082 & 0.062&  0.042  &0.032 & 0.022 & 0.002 \\
         & -0.018 & -0.028 & -0.038 & -0.058 & -0.078 & -0.098 & -0.108 & -0.118 & -0.128 & -0.138 \\
       \hline
    \end{tabular}}
    \caption{Settings of the mean proficiency parameter.}
    \label{tab:mean setting}
\end{table}

\begin{table}[H]
    \centering
    \begin{tabular}{c|cccccccccc}
    \hline
    \multicolumn{1}{c}{} & \multicolumn{10}{c}{ Variance parameter $\sigma_1^*, ..., \sigma_p^*$} \\
    \hline
 S1 $\&$ S2  & 0.8 & 0.4 & 0.5& $ $0.6& 0.7& 0.75& 0.8& 0.85& 0.9& 1.1 \\
       \hline
 \multirow{2}{*}{S3 $\&$ S4}  &  0.8 & 0.4 & 0.5& $ $0.6& 0.7& 0.75& 0.8& 0.85& 0.9& 1.1  \\
         & 0.8 & 0.4 & 0.5& $ $0.6& 0.7& 0.75& 0.8& 0.85& 0.9& 1.1  \\
       \hline
    \end{tabular}
    \caption{Settings of the variance parameter.}
    \label{tab:var setting}
\end{table}

\subsection{Competing Methods and Evaluation Criteria}

We compare the proposed method in Algorithm~\ref{alg:ML1} with the RMSD method with different threshold values and a baseline IRT model in \eqref{eq:baseline} that assumes common item parameters across countries. Specifically, we consider three RMSD thresholds: 0.05, 0.10, and 0.15, mimicking ILSA practice. 
{The RMSD method is applied with both one and multiple iterations; see a description in Section~\ref{sec:RMSD} of the main text. The results from the RMSD method with multiple iterations are deferred to Section~\ref{sec:iterRMSD}. In addition to the RMSD method, we also compare the proposed method with the alignment methods~\citep{muthen2014irt, Alexander2020Lp, Alexander2023implementation} and Lasso-regularization methods~\citep{belzak2020improving, bauer2020simplifying, schauberger2020regularization}, which are implemented using the ``\texttt{sirt}'' and ``\texttt{regDIF}'' R packages, respectively.   
We note that the Lasso-regularization method requires specifying a reference group, which does not exist in our simulation settings. We thus randomly assign a group as the reference group in each simulation when implementing the Lasso-regularization method. 
A discussion of the alignment and Lasso-regularization methods is provided in Section~\ref{sec:alignment lasso}.}

To compare different methods, we consider the following evaluation criteria.

\begin{enumerate}
    \item Quality of ranking. An important task of an ILSA is to rank countries based on the estimated mean parameters $\mu_1$, ..., $\mu_p$. To evaluate the ranking quality, we use Kendall's rank correlation,   defined as $$ \text{Rank-Kendall} =  2p^{-1}(p-1)^{-1} \sum_{l< m \in \{ 1,...,p\}} \operatorname{sign}\left(\hat\mu_l-\hat\mu_m\right)\operatorname{sign}\left(\mu_l^*-\mu_m^*\right).$$ A higher Rank-Kendall value indicates a greater concordance between the estimated and true rankings. 
    \item Accuracy of parameter estimation.
   It is also important to estimate the parameters of the IRT model accurately, as they will be used in subsequent analyses, such as scaling of the students and linking different cycles of the ILSA. 
    We evaluate the accuracy in estimating the unknown parameters of the true IRT model using the mean squared errors (MSEs). 
    
    \item  Accuracy of model selection. We check how the RMSD method performs in terms of the selection of zero elements among the DIF parameter matrix $(\gamma_{jk}^*)_{J\times p}$. We calculate the number of zero elements selected by the RMSD statistics and the false positive rates (FPRs) and false negative rates (FNRs), where a false positive error occurs when a zero $\gamma_{jk}^*$ is selected as nonzero, and a false negative error occurs when a nonzero $\gamma_{jk}^*$ is not selected. {For comparison, we also apply a hard-thresholding rule for the proposed method, classifying a $\gamma_{jk}$ entry to be zero when $|\hat{\gamma}_{jk}|$ is below a given threshold and to be nonzero when $|\hat{\gamma}_{jk}|$ is above the threshold. Based on the classifications, FPR and FNR are calculated. 
    Detailed results across multiple choices of hard thresholds are reported in Section~\ref{sec:DIF detection}.}
    



\end{enumerate}

\subsection{Main Results}
\begin{figure}[!htbp]
\centering    
        \includegraphics[width=2.7in]{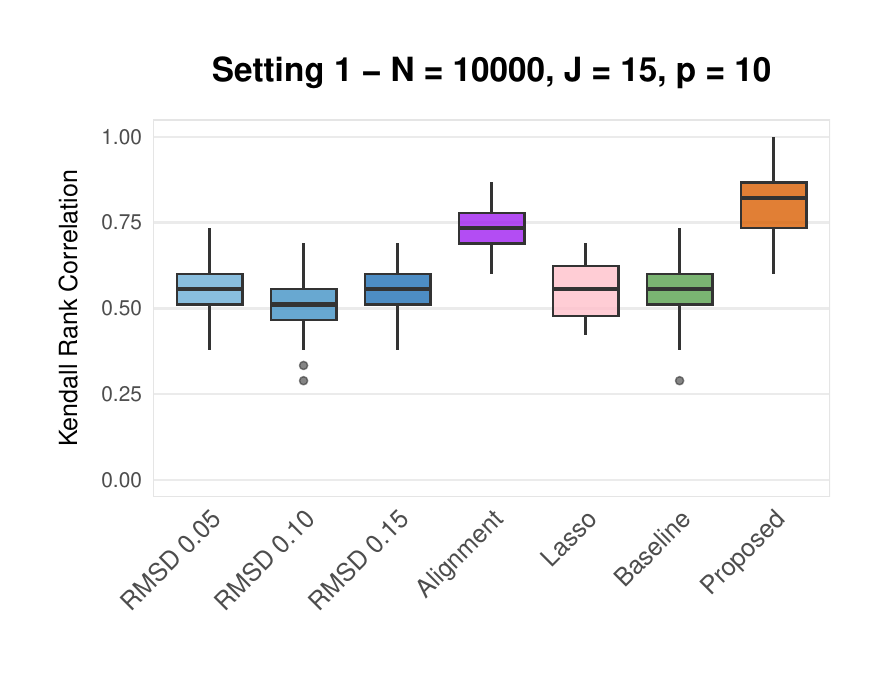}~
       \includegraphics[width=2.7in]{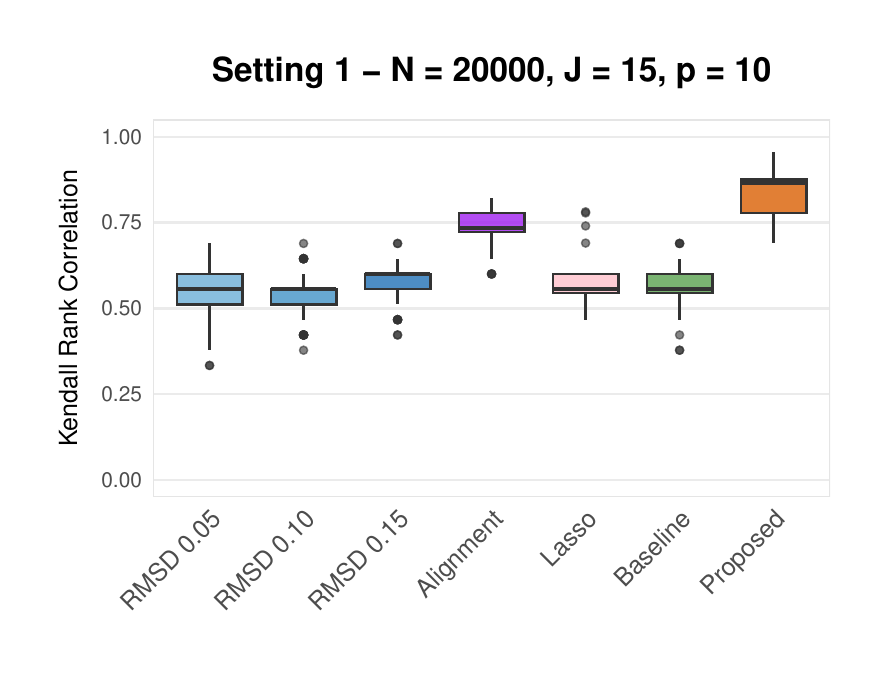}\\
       \includegraphics[width=2.7in]{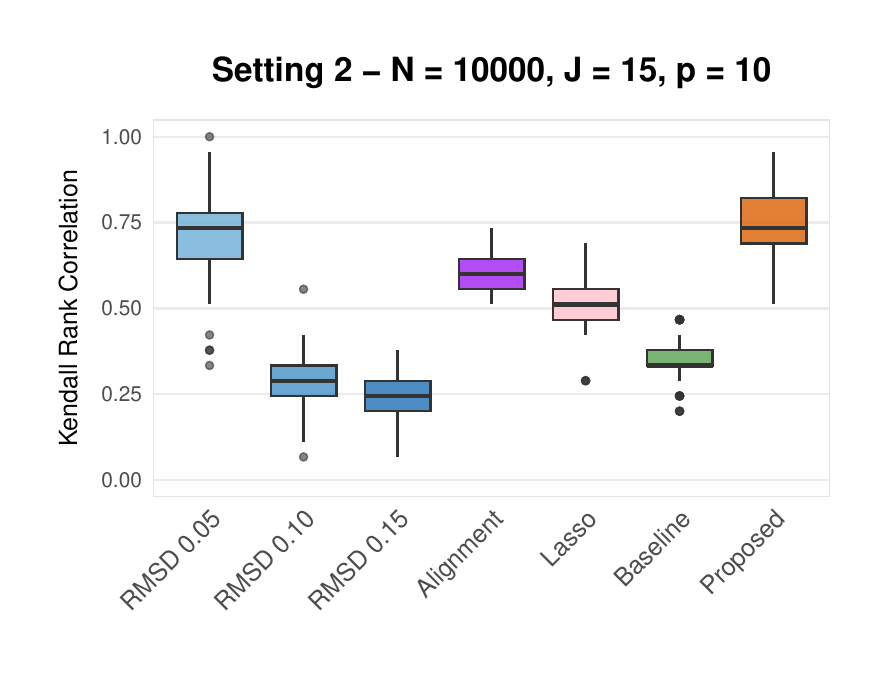}~
       \includegraphics[width=2.7in]{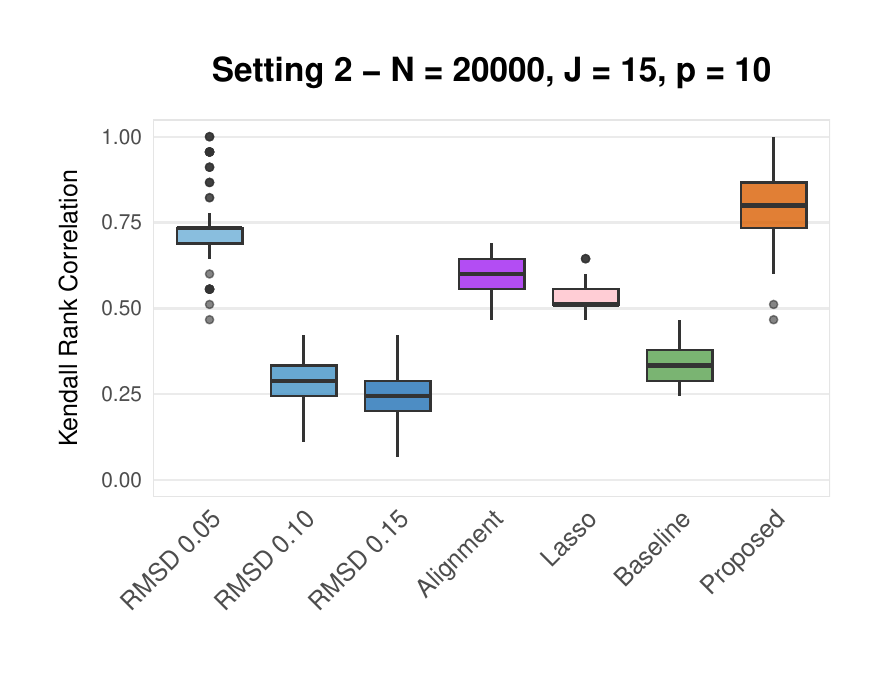} \\
       \includegraphics[width=2.7in]{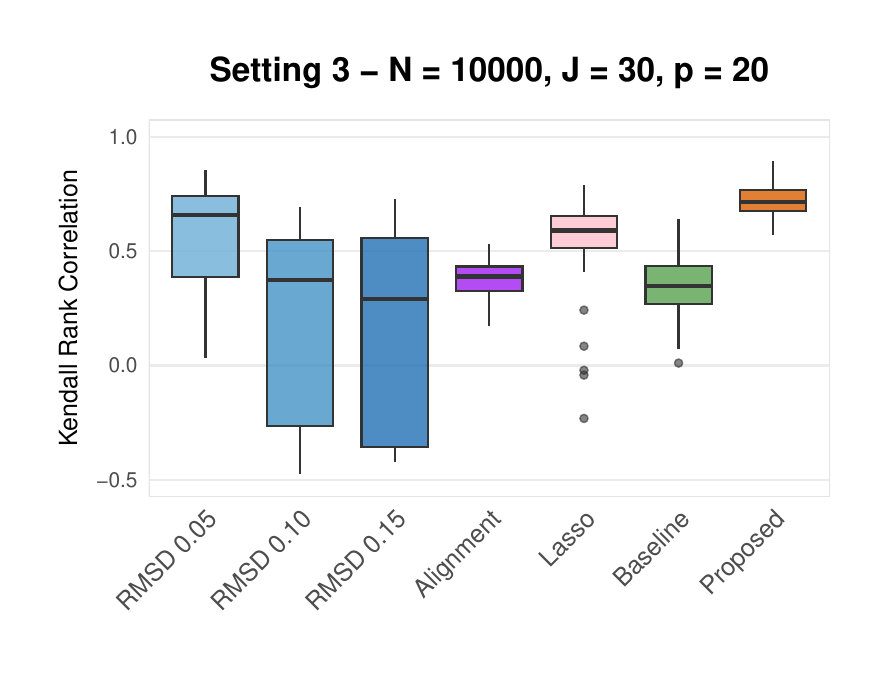}~
       \includegraphics[width=2.7in]{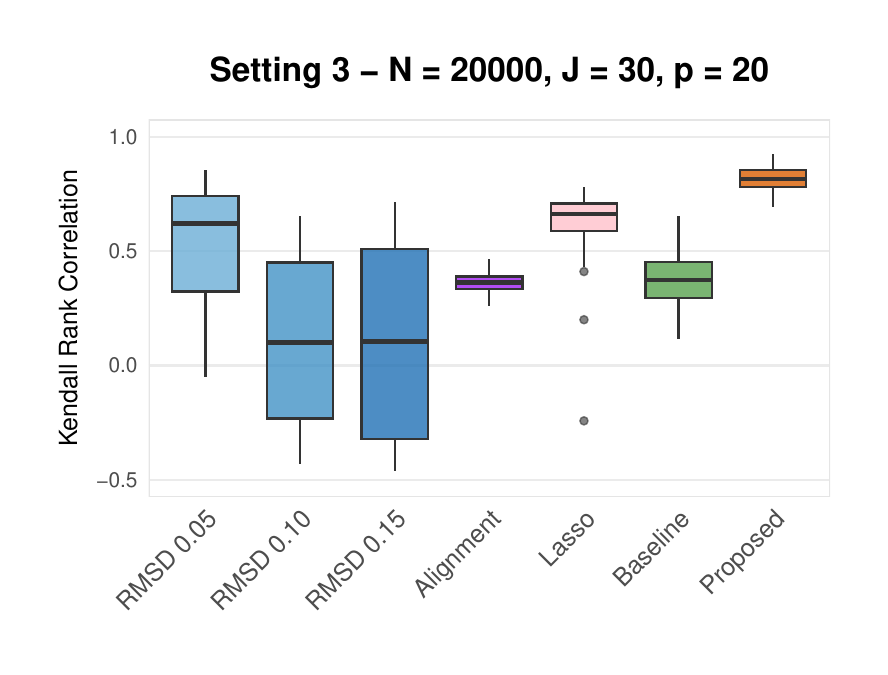} \\
       \includegraphics[width=2.7in]{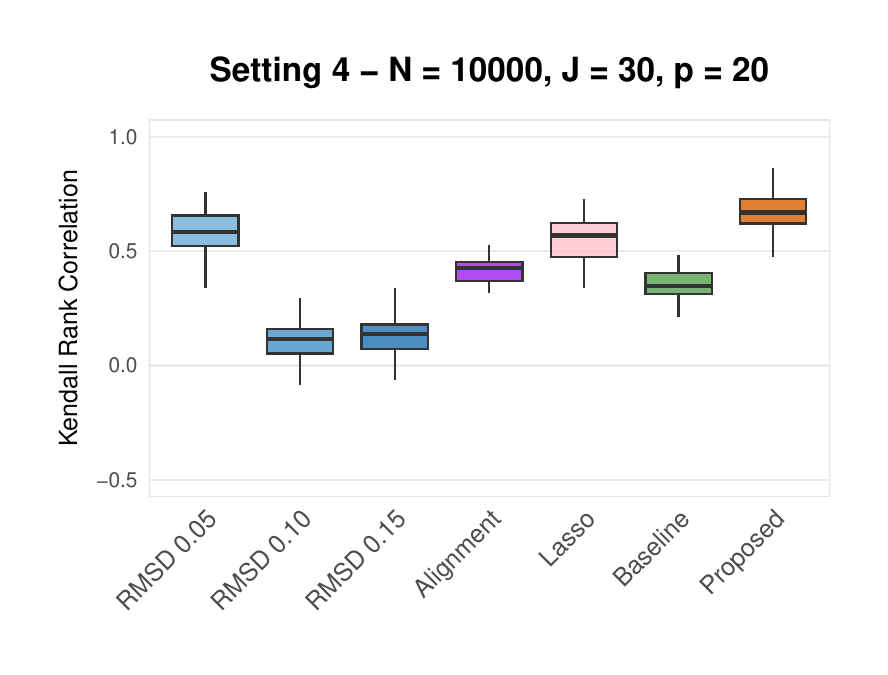}~
       \includegraphics[width=2.7in]{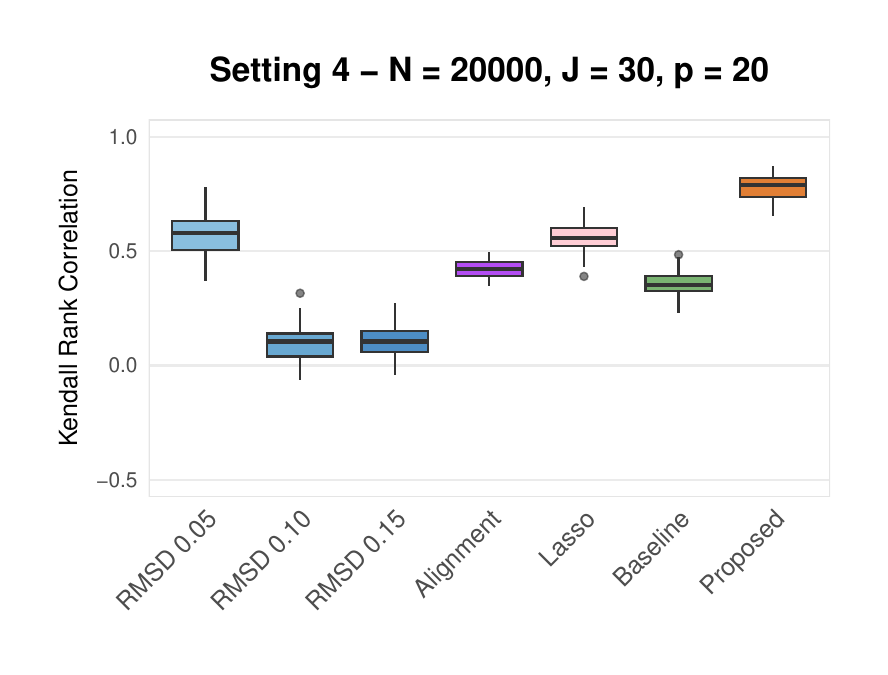}
        \caption{{Comparison of average Kendall rank correlation of the proposed method, baseline methods, alignment method, Lasso-regularization method, and the RMSD-based methods with different thresholds.}}
        \label{fig:kendall}
        \end{figure}

Figure~\ref{fig:kendall} presents the box plots for the Rank-Kendall statistics for different methods under different settings, where each boxplot is based on 100 independent simulations. These box plots show that the proposed method tends to have a higher Rank-Kendall than the competing methods under all the settings, meaning that the proposed method gives more accurate ranking results.  
Comparing the left panels ($N=10000$) to the right panels ($N=20000$), we see that the ranking quality of the proposed method improves as the sample size increases, which is not the case with the competing methods. Moreover, in some settings such as Setting 2 and Setting 4 where the nonzero entries are randomly dispersed in $(\gamma_{jk}^*)_{J\times p}$, the ranking performance of RSMD method, particularly with thresholds 0.10 and 0.15, fails to outperform even the baseline method. The proposed method also consistently outperforms the iterative RMSD method across all settings. In addition, the iterative RMSD method does not systematically outperform the non-iterative one; see Section~\ref{sec:iterRMSD} for more details.

{In addition, we also investigate the performances of the different methods when the true DIF structure violates Assumption~\ref{cond:ML1} and found that our method remains reasonably robust against model misspecification and often outperforms the competing methods; see sensitivity analysis in Section~\ref{sec:violate DIF}.}




The MSEs of estimated model parameters are given in Table~\ref{tab:mse mu sigma gamma}.
These tables show that the proposed method has substantially smaller MSEs for $\hat{\mu}_k$s, $\hat{\gamma}_{jk}$s, and $\hat{d}_j$s than the competing methods for all the parameters across all the settings. For $\hat{a}_j$s and $\hat{\sigma}_k$s, the MSEs from both the proposed method and the RMSD methods are extremely small and comparable, and both are significantly lower than those from the baseline methods.


\begin{table}[!h]
\fontsize{8.5}{8.5}\selectfont 
\begin{center}
 \resizebox{\textwidth}{!}{\begin{tabular}{cc|cc|cc|cc|cc|cc}
 \hline
 && \multicolumn{2}{c|}{$\mu_k$} & \multicolumn{2}{c|}{$\sigma_k$} &  \multicolumn{2}{c|}{$\gamma_{jk}$} & \multicolumn{2}{c|}{$a_j$} & \multicolumn{2}{c}{$d_j$} \\
 \cline{3-12}
  & $N =$  &  $ 10^4$ & $ 2\times 10^4$  & $ 10^4$ & $ 2\times 10^4$  & $ 10^4$ & $ 2\times 10^4$ & $ 10^4$ & $ 2\times 10^4$ & $ 10^4$ & $ 2\times 10^4$ \\ 
  \hline
  \multirow{5}{*}{$p=10$, $S1$}  & Proposed Method   &  0.002     & 0.001     & 0.001  & 0.001  &   0.018     &  0.009   &  0.004    &    0.003     &   0.009     &  0.004        \\
\cline{2-12}
 & RMSD 0.05      &  0.117     &  0.117     & 0.002  &  0.002  &   0.529     &  0.524     &  0.018    &   0.016      &   0.174     & 0.171       \\
\cline{2-12}
 & RMSD 0.10    &   0.186    &  0.185    & 0.005  & 0.005  &   0.743     &0.748 &  0.036    &   0.036      &   0.178      & 0.183  \\
\cline{2-12}
& RMSD 0.15    &0.127 &  0.130    & 0.013   &  0.013  & 0.825       &  0.834    &  0.184    &    0.187     &   0.099     & 0.096   \\
\cline{2-12}
& Baseline Method  & 0.102 &  0.102    & 0.026  &  0.026  &  $\setminus$      &   $\setminus$   &  0.583    &   0.582      &   0.104     &  0.104  \\

     \hline
\multirow{5}{*}{$p=10$, $S2$}  & Proposed Method  &   0.002    &  0.001    & 0.001 & 0.001  &  0.017      & 0.009    &   0.005   &   0.004      &   0.008     & 0.004     \\
\cline{2-12}
 & RMSD 0.05     &  0.005     &  0.004    & 0.001 & 0.001  &      0.035   &  0.031    &   0.003   &  0.001       &  0.014      & 0.014       \\
\cline{2-12}
 & RMSD 0.10     &  0.082     &  0.085    &0.001  & 0.001  &      0.320   &0.332 &0.006&   0.004      &   0.056     & 0.059    \\
\cline{2-12}
& RMSD 0.15   &   0.098    &  0.100    & 0.003 & 0.003  &      0.557  &  0.560     &  0.016    &   0.015      &  0.047      &  0.046   \\
\cline{2-12}
& Baseline Method  &  0.112     &  0.112    & 0.008 & 0.008  &     $\setminus$   & $\setminus$    &  0.058    & 0.057        & 0.073       &  0.073    \\
\hline

 \multirow{5}{*}{$p=20$, $S3$}  & Proposed Method    &  0.002     &  0.001    & 0.002  & 0.002  &  0.038      & 0.019     &  0.006    &   0.005      &   0.020     & 0.010     \\
\cline{2-12}
 & RMSD 0.05     &  0.018     & 0.015     & 0.001  & 0.001  &   0.100     &  0.076    &  0.003    &   0.001      &   0.038     & 0.029    \\
\cline{2-12}
 & RMSD 0.10   &  0.106     &  0.098    & 0.001 &0.001&  0.411      & 0.390    &0.006 &   0.003      &   0.081     & 0.082       \\
\cline{2-12}
& RMSD 0.15   &  0.126     &  0.122    & 0.002 & 0.001   & 0.632       & 0.623 &  0.029    &   0.021      &   0.089     & 0.093    \\
\cline{2-12}
& Baseline Method  &  0.151     &  0.150    & 0.004 & 0.004  &    $\setminus$    &  $\setminus$    &0.026 &   0.024      &   0.061     &0.061 \\

\hline
\multirow{5}{*}{$p=20$, $S4$}  & Proposed Method   &  0.003     &   0.001   & 0.002 &  0.001 &   0.042     &  0.021   &  0.005    &     0.004    &   0.022     & 0.011     \\
\cline{2-12}
 & RMSD 0.05      &   0.011    &  0.012    & 0.001 & 0.001  &    0.068    &  0.064   &0.002&           0.001      &0.024 & 0.025     \\
\cline{2-12}
 & RMSD 0.10    &  0.115     & 0.125     & 0.001 & 0.001  &     0.446   &  0.467    &  0.003    &   0.002      &0.084&      0.087    \\
\cline{2-12}
& RMSD 0.15   &   0.093    &  0.095    &0.002  & 0.001  &       0.566 &  0.568    & 0.006     &   0.006      &    0.063    & 0.065    \\
\cline{2-12}
& Baseline Method  &  0.174     &  0.174    & 0.004  &  0.003 &      $\setminus$  &   $\setminus$    &  0.032    &  0.031       &  0.057      & 0.056    \\

\hline
\end{tabular}}
\caption{Average MSEs of the estimated parameters for $\mu_k$, $\sigma_k$, $\gamma_{jk}$, $a_j$ and $d_j$ respectively, for different methods, where MSEs are averaged over all $k=1,..., p$ and $j = 1,...,J$ across 100 simulations. }
\label{tab:mse mu sigma gamma}
\end{center}
\end{table}

The inferior performance of the RMSD method and the baseline method is likely due to their high bias. For the baseline method, the bias comes from ignoring the DIF effects of the items by setting all $\gamma_{jk}$s. For the RMSD method, the bias comes from the model selection step using the RMSD statistics, where some true DIF effects may be masked in the model selection. In Table~\ref{tab: false discovery}, we present the number of zeros selected by the RMSD method and the false positive and false negative error rates under all the settings.

\begin{table}[H]
\fontsize{8.5}{8.5}\selectfont 
\begin{center}
\begin{tabular}{cc|ccc|ccc}
 \hline
 && \multicolumn{3}{c|}{$N=10000$} & \multicolumn{3}{c}{$N=20000$} \\
   & &   Detections  &  FPR & FNR  &   Detections  &  FPR & FNR   \\ 
     \hline
\multirow{3}{*}{$p=10$, $S1$}  & RMSD 0.05   &  35.53     &   0.77  &  0.25   &36.14  &  0.76  &    0.25      \\
  \cline{2-8}
 & RMSD 0.10  &  64.82    &  0.51    & 0.35    & 64.65  &   0.52     &  0.35       \\
  \cline{2-8}
& RMSD 0.15  &  96.10     & 0.27    &  0.50   &95.74 &   0.27   &   0.50      \\

\hline
\multirow{3}{*}{$p=10$, $S2$}  & RMSD 0.05   &  7.66    &  0.93  &   0.02  &  7.34  &   0.93  &  0.02       \\
  \cline{2-8}
 & RMSD 0.10  &  40.23     & 0.64    &  0.13   & 39.52  &  0.65      &  0.13       \\
  \cline{2-8}
& RMSD 0.15   &  90.94      & 0.23   & 0.35    &  91.16   & 0.23    &  0.35        \\

\hline
\multirow{3}{*}{$p=20$, $S3$} & RMSD 0.05  &  35.11   &  0.93     & 0.04    &  35.91   &   0.92  &   0.03       \\
  \cline{2-8}
 & RMSD 0.10   &  159.57    &   0.67    &  0.17    &  157.59    &  0.67   &     0.16      \\
  \cline{2-8}
& RMSD 0.15   &  342.52     &   0.31   &  0.38    & 345.49     & 0.29      &   0.38      \\
\hline
\multirow{3}{*}{$p=20$, $S4$} & RMSD 0.05  &  11.67     &  0.98  & 0.03    &   11.66    & 0.99   &    0.03      \\
  \cline{2-8}
 & RMSD 0.10   &  103.61      &  0.82  &    0.16   & 100.84      &  0.83    &    0.16     \\
  \cline{2-8}
& RMSD 0.15   &  311.57    &   0.37    &    0.37  & 315.91     & 0.36     &  0.37        \\
\hline
\end{tabular}
\caption{Model selection accuracy of RMSD method: the detection of zero elements by the RMSD statistics, the false positive rates (FPR) and the false negative rates (FNR).} 
\label{tab: false discovery}
\end{center}
\end{table}

\subsection{Additional Details of Simulation Setting}
\label{sec-sim}

Below are detailed specifications for the parameter values in Section~\ref{sec-simulation settings}. 
The generation of DIF effect matrix $(\gamma_{jk}^*)_{J\times p}$ of $S3$--$S4$ at (ii) $p=20$ and $J=30$ are given in Table~\ref{tab: J30 dif setting 3}--\ref{tab: J30 dif setting 4}.

\begin{landscape}
\begin{table}[!ht]
\fontsize{8.5}{8.5}\selectfont 
\begin{center}
\begin{tabular}{cccccccccccccccccccc}
\hline
\multicolumn{20}{c}{$S3$} \\
\hline 0.00 & 0.00 & 0.00 & 0.00 & 0.00 & 0.00 & 0.00 & 0.00 & 0.00 & 0.00 & 2.17 & 0.00 & 1.20 & 2.92 & 2.30 & 2.13 & 1.92 & 0.00 & 2.45 & 2.64 \\
\hline 0.00 & 0.00 & 0.00 & 0.00 & 0.00 & 0.00 & 0.00 & 0.00 & 0.00 & 0.00 & 0.00 & 2.81 & 2.91 & 0.00 & 2.66 & 1.48 & 0.00 & 1.36 & 2.13 & 0.00 \\
\hline 0.00 & 0.00 & 0.00 & 0.00 & 0.00 & 0.00 & 0.00 & 0.00 & 0.00 & 0.00 & 1.59 & 2.90 & 1.83 & 0.00 & 1.96 & 2.76 & 1.78 & 2.81 & 1.78 & 1.35 \\
\hline 0.00 & 0.00 & 0.00 & 0.00 & 0.00 & 0.00 & 0.00 & 0.00 & 0.00 & 0.00 & 1.55 & 1.15 & 0.00 & 2.99 & 2.68 & 2.31 & 1.01 & 1.17 & 0.00 & 0.00 \\
\hline 0.00 & 0.00 & 0.00 & 0.00 & 0.00 & 0.00 & 0.00 & 0.00 & 0.00 & 0.00 & 2.63 & 2.51 & 2.94 & 0.00 & 0.00 & 1.96 & 2.88 & 0.00 & 2.79 & 2.79 \\
\hline 0.00 & 0.00 & 0.00 & 0.00 & 0.00 & 0.00 & 0.00 & 0.00 & 0.00 & 0.00 & 1.52 & 1.57 & 2.17 & 1.98 & 0.00 & 2.94 & 1.48 & 2.78 & 2.62 & 2.49 \\
\hline 2.56 & 0.00 & 0.00 & 0.00 & 0.00 & 0.00 & 0.00 & 0.00 & 0.00 & 0.00 & 2.89 & 1.67 & 0.00 & 0.00 & 0.00 & 0.00 & 0.00 & 0.00 & 0.00 & 0.00 \\
\hline 0.00 & 2.12 & 2.26 & 2.14 & 1.54 & 1.94 & 2.02 & 2.96 & 1.55 & 2.53 & 0.00 & 0.00 & 0.00 & 0.00 & 0.00 & 0.00 & 0.00 & 0.00 & 0.00 & 0.00 \\
\hline 0.00 & 1.14 & 1.82 & 2.83 & 1.71 & 2.15 & 1.88 & 1.40 & 1.86 & 1.89 & 0.00 & 0.00 & 0.00 & 0.00 & 0.00 & 0.00 & 0.00 & 0.00 & 0.00 & 0.00 \\
\hline 0.00 & 2.71 & 1.69 & 2.80 & 1.17 & 2.34 & 2.58 & 2.10 & 1.03 & 1.71 & 0.00 & 0.00 & 0.00 & 0.00 & 0.00 & 0.00 & 0.00 & 0.00 & 0.00 & 0.00 \\
\hline 0.00 & 2.83 & 2.65 & 1.12 & 2.53 & 1.42 & 2.82 & 1.69 & 2.13 & 1.07 & 0.00 & 0.00 & 0.00 & 0.00 & 0.00 & 0.00 & 0.00 & 0.00 & 0.00 & 0.00 \\
\hline 0.00 & 1.45 & 2.38 & 1.09 & 1.89 & 2.95 & 1.14 & 1.82 & 2.34 & 2.19 & 0.00 & 0.00 & 0.00 & 0.00 & 0.00 & 0.00 & 0.00 & 0.00 & 0.00 & 0.00 \\
\hline 0.00 & 2.26 & 1.64 & 2.98 & 1.07 & 2.62 & 1.29 & 1.38 & 1.63 & 2.92 & 0.00 & 0.00 & 0.00 & 0.00 & 0.00 & 0.00 & 0.00 & 0.00 & 0.00 & 0.00 \\
\hline 0.00 & 1.14 & 1.88 & 1.41 & 2.40 & 1.55 & 1.79 & 1.69 & 2.90 & 1.51 & 0.00 & 0.00 & 0.00 & 0.00 & 0.00 & 0.00 & 0.00 & 0.00 & 0.00 & 0.00 \\
\hline 0.00 & 2.03 & 1.52 & 2.86 & 1.51 & 1.80 & 2.71 & 1.37 & 1.34 & 2.69 & 0.00 & 0.00 & 0.00 & 0.00 & 0.00 & 0.00 & 0.00 & 0.00 & 0.00 & 0.00 \\
\hline 0.00 & 2.61 & 1.27 & 1.39 & 2.26 & 1.58 & 1.53 & 2.63 & 1.88 & 1.43 & 0.00 & 0.00 & 0.00 & 0.00 & 0.00 & 0.00 & 0.00 & 0.00 & 0.00 & 0.00 \\
\hline 0.00 & 2.94 & 2.82 & 1.25 & 1.53 & 1.99 & 1.19 & 2.76 & 2.14 & 1.56 & 0.00 & 0.00 & 0.00 & 0.00 & 0.00 & 0.00 & 0.00 & 0.00 & 0.00 & 0.00 \\
\hline 0.00 & 1.69 & 2.42 & 2.08 & 2.06 & 2.51 & 2.45 & 2.40 & 2.95 & 1.96 & 0.00 & 0.00 & 0.00 & 0.00 & 0.00 & 0.00 & 0.00 & 0.00 & 0.00 & 0.00 \\
\hline 2.80 & 0.00 & 0.00 & 0.00 & 0.00 & 0.00 & 0.00 & 0.00 & 0.00 & 0.00 & 1.83 & 1.57 & 0.00 & 0.00 & 0.00 & 0.00 & 0.00 & 0.00 & 0.00 & 0.00 \\
\hline 0.00 & 0.00 & 0.00 & 0.00 & 0.00 & 0.00 & 0.00 & 0.00 & 0.00 & 0.00 & 1.14 & 0.00 & 0.00 & 1.78 & 1.07 & 2.33 & 0.00 & 1.32 & 0.00 & 1.33 \\
\hline 0.00 & 0.00 & 0.00 & 0.00 & 0.00 & 0.00 & 0.00 & 0.00 & 0.00 & 0.00 & 1.14 & 0.00 & 2.51 & 1.19 & 1.08 & 2.29 & 1.43 & 2.24 & 2.27 & 1.01 \\
\hline 0.00 & 0.00 & 0.00 & 0.00 & 0.00 & 0.00 & 0.00 & 0.00 & 0.00 & 0.00 & 0.00 & 1.84 & 2.92 & 2.51 & 2.80 & 2.54 & 1.37 & 2.94 & 2.21 & 1.05 \\
\hline 0.00 & 0.00 & 0.00 & 0.00 & 0.00 & 0.00 & 0.00 & 0.00 & 0.00 & 0.00 & 1.30 & 2.43 & 1.96 & 2.59 & 2.88 & 1.58 & 0.00 & 1.11 & 1.55 & 2.11 \\
\hline 0.00 & 0.00 & 0.00 & 0.00 & 0.00 & 0.00 & 0.00 & 0.00 & 0.00 & 0.00 & 0.00 & 0.00 & 0.00 & 1.91 & 2.26 & 0.00 & 2.16 & 1.74 & 3.00 & 1.56 \\
\hline 0.00 & 0.00 & 0.00 & 0.00 & 0.00 & 0.00 & 0.00 & 0.00 & 0.00 & 0.00 & 0.00 & 0.00 & 2.98 & 1.72 & 2.56 & 0.00 & 2.91 & 2.43 & 1.02 & 2.84 \\
\hline 0.00 & 0.00 & 0.00 & 0.00 & 0.00 & 0.00 & 0.00 & 0.00 & 0.00 & 0.00 & 2.22 & 1.42 & 1.46 & 1.72 & 2.50 & 0.00 & 1.50 & 2.67 & 2.76 & 2.67 \\
\hline 0.00 & 0.00 & 0.00 & 0.00 & 0.00 & 0.00 & 0.00 & 0.00 & 0.00 & 0.00 & 1.40 & 2.79 & 1.56 & 1.12 & 1.27 & 1.94 & 1.47 & 1.54 & 0.00 & 0.00 \\
\hline 0.00 & 0.00 & 0.00 & 0.00 & 0.00 & 0.00 & 0.00 & 0.00 & 0.00 & 0.00 & 0.00 & 0.00 & 2.41 & 0.00 & 0.00 & 1.91 & 1.54 & 2.68 & 1.22 & 2.42 \\
\hline 0.00 & 0.00 & 0.00 & 0.00 & 0.00 & 0.00 & 0.00 & 0.00 & 0.00 & 0.00 & 1.26 & 1.50 & 2.67 & 1.97 & 1.08 & 2.94 & 1.89 & 1.19 & 0.00 & 2.16 \\
\hline 0.00 & 0.00 & 0.00 & 0.00 & 0.00 & 0.00 & 0.00 & 0.00 & 0.00 & 0.00 & 1.01 & 2.46 & 1.93 & 2.04 & 2.33 & 2.33 & 1.73 & 0.00 & 1.24 & 1.93 \\
\hline
\end{tabular}

\caption{DIF effect parameter at $(ii)~p = 20, J=30$ under Setting 3. Each nonzero are randomly generated from Unif$[1,3]$.}
\label{tab: J30 dif setting 3}
\end{center}
\end{table}
\end{landscape}

\begin{landscape}
\begin{table}[!ht]
\fontsize{9}{9}\selectfont 
\begin{center}
\begin{tabular}{cccccccccccccccccccc}
\hline
\multicolumn{20}{c}{$S4$} \\
\hline 0.00 & 0.00 & 1.20 & 0.00 & 2.17 & 2.94 & 2.91 & 0.00 & 0.00 & 2.92 & 2.51 & 1.83 & 0.00 & 0.00 & 0.00 & 0.00 & 0.00 & 0.00 & 1.57 & 0.00 \\
\hline 0.00 & 1.01 & 2.31 & 0.00 & 2.24 & 0.00 & 0.00 & 0.00 & 0.00 & 0.00 & 2.76 & 2.88 & 1.96 & 0.00 & 0.00 & 2.94 & 1.92 & 1.78 & 0.00 & 0.00 \\
\hline 1.35 & 0.00 & 0.00 & 0.00 & 0.00 & 2.12 & 0.00 & 2.79 & 0.00 & 0.00 & 2.62 & 2.49 & 0.00 & 1.84 & 0.00 & 2.64 & 0.00 & 0.00 & 1.35 & 1.14 \\
\hline 0.00 & 2.82 & 0.00 & 1.64 & 1.82 & 0.00 & 0.00 & 1.52 & 1.69 & 2.38 & 0.00 & 0.00 & 0.00 & 1.88 & 0.00 & 0.00 & 1.27 & 2.65 & 0.00 & 0.00 \\
\hline 1.07 & 1.17 & 1.89 & 1.71 & 0.00 & 0.00 & 2.40 & 0.00 & 0.00 & 0.00 & 0.00 & 0.00 & 2.08 & 0.00 & 0.00 & 0.00 & 1.54 & 2.53 & 0.00 & 1.51 \\
\hline 2.02 & 1.79 & 0.00 & 1.14 & 0.00 & 0.00 & 0.00 & 0.00 & 0.00 & 0.00 & 0.00 & 1.88 & 0.00 & 0.00 & 2.51 & 1.99 & 0.00 & 0.00 & 1.29 & 2.58 \\
\hline 2.40 & 1.55 & 2.13 & 0.00 & 1.63 & 0.00 & 2.63 & 0.00 & 0.00 & 0.00 & 0.00 & 0.00 & 1.03 & 2.76 & 0.00 & 0.00 & 0.00 & 1.86 & 0.00 & 2.34 \\
\hline 0.00 & 0.00 & 1.43 & 0.00 & 0.00 & 2.80 & 0.00 & 0.00 & 1.56 & 2.56 & 0.00 & 0.00 & 1.96 & 1.83 & 0.00 & 0.00 & 0.00 & 2.89 & 2.69 & 1.67 \\
\hline 0.00 & 1.84 & 1.50 & 0.00 & 2.46 & 0.00 & 0.00 & 0.00 & 0.00 & 1.20 & 0.00 & 0.00 & 2.13 & 0.00 & 0.00 & 0.00 & 0.00 & 2.79 & 0.00 & 1.42 \\
\hline 0.00 & 0.00 & 1.12 & 2.04 & 2.80 & 0.00 & 1.92 & 0.00 & 0.00 & 1.07 & 0.00 & 0.00 & 0.00 & 0.00 & 1.72 & 1.08 & 1.97 & 0.00 & 1.72 & 0.00 \\
\hline 1.94 & 2.94 & 0.00 & 0.00 & 0.00 & 1.91 & 0.00 & 0.00 & 0.00 & 0.00 & 0.00 & 2.61 & 0.00 & 0.00 & 0.00 & 1.43 & 2.24 & 2.68 & 0.00 & 2.10 \\
\hline 0.00 & 1.19 & 2.68 & 0.00 & 0.00 & 0.00 & 0.00 & 0.00 & 0.00 & 2.94 & 1.35 & 0.00 & 2.43 & 2.67 & 1.11 & 0.00 & 1.74 & 0.00 & 1.54 & 0.00 \\
\hline 2.75 & 0.00 & 0.00 & 2.84 & 0.00 & 0.00 & 0.00 & 2.42 & 1.09 & 0.00 & 0.00 & 2.67 & 0.00 & 1.93 & 0.00 & 2.89 & 0.00 & 2.16 & 0.00 & 2.84 \\
\hline 1.91 & 0.00 & 0.00 & 0.00 & 0.00 & 0.00 & 1.59 & 2.98 & 0.00 & 0.00 & 1.88 & 2.30 & 0.00 & 1.23 & 0.00 & 0.00 & 1.02 & 0.00 & 1.61 & 1.65 \\
\hline 1.78 & 2.70 & 0.00 & 0.00 & 0.00 & 1.04 & 0.00 & 0.00 & 0.00 & 1.48 & 1.39 & 2.42 & 2.72 & 0.00 & 0.00 & 2.50 & 0.00 & 0.00 & 2.44 & 0.00 \\
\hline 1.82 & 0.00 & 0.00 & 2.04 & 0.00 & 0.00 & 1.46 & 0.00 & 0.00 & 0.00 & 2.93 & 0.00 & 0.00 & 2.86 & 1.78 & 1.44 & 0.00 & 2.01 & 0.00 & 1.46 \\
\hline 1.68 & 1.42 & 1.26 & 0.00 & 0.00 & 0.00 & 1.31 & 0.00 & 2.09 & 0.00 & 1.95 & 0.00 & 2.47 & 0.00 & 0.00 & 1.66 & 2.25 & 0.00 & 0.00 & 0.00 \\
\hline 2.30 & 0.00 & 0.00 & 1.95 & 0.00 & 1.33 & 2.20 & 0.00 & 1.86 & 0.00 & 0.00 & 0.00 & 1.43 & 0.00 & 2.34 & 0.00 & 0.00 & 0.00 & 1.70 & 0.00 \\
\hline 0.00 & 1.76 & 1.12 & 0.00 & 1.05 & 0.00 & 2.57 & 0.00 & 0.00 & 1.44 & 0.00 & 0.00 & 0.00 & 0.00 & 2.70 & 1.92 & 2.77 & 0.00 & 0.00 & 1.56 \\
\hline 0.00 & 0.00 & 1.30 & 0.00 & 1.63 & 0.00 & 0.00 & 0.00 & 0.00 & 2.89 & 1.80 & 2.96 & 0.00 & 2.70 & 2.96 & 0.00 & 0.00 & 0.00 & 1.99 & 2.77 \\
\hline 2.85 & 2.18 & 0.00 & 0.00 & 1.89 & 0.00 & 0.00 & 0.00 & 2.74 & 2.74 & 0.00 & 0.00 & 1.15 & 0.00 & 0.00 & 1.22 & 0.00 & 2.21 & 2.00 & 0.00 \\
\hline 0.00 & 2.39 & 0.00 & 0.00 & 2.67 & 1.37 & 0.00 & 0.00 & 0.00 & 2.63 & 0.00 & 1.72 & 0.00 & 0.00 & 2.47 & 0.00 & 0.00 & 2.05 & 0.00 & 0.00 \\
\hline 1.00 & 0.00 & 0.00 & 0.00 & 0.00 & 0.00 & 0.00 & 0.00 & 2.03 & 1.97 & 2.12 & 0.00 & 0.00 & 2.05 & 2.02 & 0.00 & 2.76 & 2.36 & 1.41 & 0.00 \\
\hline 0.00 & 0.00 & 0.00 & 0.00 & 2.70 & 1.27 & 0.00 & 0.00 & 0.00 & 0.00 & 1.05 & 0.00 & 0.00 & 2.79 & 2.10 & 1.05 & 0.00 & 0.00 & 0.00 & 2.85 \\
\hline 0.00 & 0.00 & 1.86 & 2.16 & 0.00 & 0.00 & 0.00 & 2.58 & 0.00 & 2.19 & 0.00 & 1.27 & 2.37 & 0.00 & 1.71 & 0.00 & 0.00 & 0.00 & 0.00 & 0.00 \\
\hline 0.00 & 0.00 & 0.00 & 1.12 & 0.00 & 0.00 & 0.00 & 2.78 & 0.00 & 0.00 & 1.48 & 1.83 & 1.40 & 1.13 & 0.00 & 1.82 & 0.00 & 1.37 & 2.58 & 0.00 \\
\hline 1.12 & 0.00 & 2.60 & 1.33 & 0.00 & 0.00 & 2.48 & 1.06 & 0.00 & 0.00 & 1.08 & 2.24 & 0.00 & 0.00 & 2.99 & 0.00 & 2.50 & 0.00 & 0.00 & 0.00 \\
\hline 0.00 & 1.31 & 1.00 & 2.47 & 0.00 & 2.85 & 0.00 & 0.00 & 0.00 & 2.48 & 1.31 & 0.00 & 1.00 & 2.39 & 0.00 & 0.00 & 2.87 & 0.00 & 0.00 & 0.00 \\
\hline 0.00 & 0.00 & 0.00 & 0.00 & 1.76 & 0.00 & 0.00 & 2.28 & 1.58 & 0.00 & 0.00 & 0.00 & 2.37 & 2.51 & 1.33 & 0.00 & 1.02 & 2.63 & 2.30 & 0.00 \\
\hline 0.00 & 0.00 & 0.00 & 2.70 & 0.00 & 0.00 & 1.84 & 2.23 & 1.51 & 0.00 & 0.00 & 2.51 & 0.00 & 0.00 & 0.00 & 1.81 & 1.59 & 0.00 & 0.00 & 1.34 \\
\hline
\end{tabular}

\caption{DIF effect parameter at $(ii)~p = 20, J=30$ under Setting 4. Each nonzero entry are randomly generated from Unif$[1,3]$.}
\label{tab: J30 dif setting 4}
\end{center}
\end{table}
\end{landscape}

{

\subsection{Discussions on Comparative Methods and Related Work}
\label{sec:alignment lasso}
 In this section, provide a comprehensive review of the two comparative methods used in simulation studies, including Lasso-MNFA methods and alignment method. 

\subsubsection{Lasso-MNFA Method}
Lasso-MNFA regularized methods for moderated nonlinear factor (Lasso-MNFA) analysis~\citep{huang2018penalized, belzak2020improving, bauer2020simplifying, schauberger2020regularization} has been widely studied for model selection and parameter estimation. Specifically, these approaches use $L_1$ regularizations to shrink or select DIF effects, which can be effective in detecting non-invariant items. However, they are very computationally demanding due to the need for parameter tuning, and they also impose more restrictive assumptions, particularly requiring the specification of a reference group assumed to be DIF-free. 
Moreover, the performance of Lasso-MNFA methods is highly sensitive to the specification of reference group and misspecification of it often leads to undesirable results. 
This issue is particularly severe in analyzing data from large-scale assessments, where specifying a reference group is typically difficult. Due to the need to specify a reference group, the Lasso-MNFA method does not yield results similar to those of the proposed method. Technically, it may be possible to extend the existing Lasso-MNFA method so that no reference group is needed. However, such an extension is not available yet, and the existing software packages for Lasso-MNFA cannot perform analysis without a reference group. 
In terms of numerical implementation in our simulation studies, as the Lasso-MNFA method requires a reference group, we randomly pick a group and set it as the reference for each dataset,  and then apply the function in the 
\texttt{regDIF} package.

Our method can be seen as a limiting case of the Lasso-MNFA approach when its tuning parameter approaches zero. Besides, we clarify that, to the best of our knowledge, the consistency of the Lasso-MNFA method has not yet been established in the literature. Intuitively, such a result would require at least the same sparsity conditions as those assumed in our proposed method and would involve similar arguments in the proof. Furthermore, the existing Lasso theory developed under standard regression settings is not directly applicable to the current problem because our model involves identifiability issues that do not arise in conventional regression frameworks.
Finally, we acknowledge that under certain settings, the Lasso approach may outperform our proposed method, as the tuning parameter in its penalty term can be adjusted to optimize empirical performance. However, identifying the optimal tuning parameter is both computationally and theoretically nontrivial, and we leave this as an interesting direction for future research.


\subsubsection{Alignment Method}
Another comparative method used in simulation studies is the alignment method, a linking approach for multi-group analysis, which has been extensively investigated~\citep{muthen2014irt, Alexander2020Lp, Alexander2023implementation}. Its goal is to maximize the extent of invariance across item parameters by minimizing a linking function used to determine the group parameters including the group-level means and variances.
Specifically, when item parameters differ across groups, the alignment method performs pairwise comparisons of these parameters and minimize their differences using robust loss functions. This approach has proven useful in practice, as demonstrated by \citet{muthen2014irt}, and has been further developed with $L_p$-based formulations and implementation strategies in subsequent works~\citep{Alexander2020Lp, Alexander2023implementation}.



At a high level, this approach is similar in spirit to the proposed method. However, it uses a different loss function based on pairwise differences in group-specific item parameters (as functions of group means).  By taking the pairwise differences, the parameters $h_j$ in the proposed loss function are cancelled out, so that the loss function only depends on group-specific parameters (i.e., $c_k$s under our notation). Although no theoretical results have been established, we believe that for certain loss functions, for example, the one based on the $L_1$ norm, consistency results can be established for the alignment method, under certain sparsity conditions about the true DIF structure. That said, however, we shall note that the alignment method has at least two disadvantages 
when compared with the proposed method. First, the alignment method does not directly estimate item-specific DIF parameters. It is thus more suitable for linking than for DIF detection, whereas the proposed method can be used for both purposes. Second, the alignment method tends to require a stronger sparsity condition on the true DIF structure for its result to be consistent. To see this, we consider the $L_1$-based alignment method under the same uniform DIF setting as the current paper. For the alignment method to be consistent, it can be shown that the following assumption needs to hold: 
\begin{assumption}
\label{cond:pairwise sparsity}
    The true DIF effects satisfy that
\begin{align}
\sum_{j=1}^J \sum_{k, k^{\prime}} |\gamma_{jk}^*- \gamma_{jk^{\prime}}^*| \leq & \sum_{j=1}^J \sum_{k, k^{\prime}}|(\gamma_{jk}^* - a_j^* c_k - h_j) -(\gamma_{jk^{\prime}}^* -a_j^* c_{k^{\prime}} - h_j )  | \nonumber  \\
= &   \sum_{j=1}^J \sum_{k, k^{\prime}} |(\gamma_{jk}^* - \gamma_{jk^{\prime}}^*) - a_j^* (c_k - c_{k^\prime})|.
\label{eq:quantile pairwise}
\end{align} 
for all $c_1, ..., c_p$ satisfying $\sum_{k=1}^p c_k =0$, where the equality holds when $c_1 =\cdots = c_p=0$.
\end{assumption}



Roughly speaking, the true $\gamma_{jk}^*$s need to be more sparse for this condition to hold than what is needed for Assumption 1 in the main text to hold. This is because the proportion of zero values for the pairwise differences,  $\gamma_{jk}^*- \gamma_{jk^{\prime}}^*$, $k\neq k'$, $j = 1, ..., J$, is typically smaller than the proportion of zero values in $\gamma_{jk}^*$, $j=1, ..., J, k=1, ..., p$, when the nonzero elements of $\gamma_{jk}^*$ are distinct. 
Thus any sparsity requirement placed on all pairwise differences implicitly enforces a much higher degree of sparsity on the individual entries themselves. In other words, the alignment method requires sparsity at the level of all $C_p^2$ pairwise comparisons, whereas Assumption 1 only requires sparsity of the $\gamma_{j k}^*$ themselves. Consequently, the pairwise sparsity condition underlying alignment is strictly stronger than our proposed assumption.

 To illustrate this, we provide an example with DIF effect structure $\gamma_{jk}^*$'s violating Assumption~\ref{cond:pairwise sparsity} but satisfying Assumption 1.
Consider the matrix $(\gamma_{jk}^*)_{J \times p}$ with $p=9$ columns and $J=C_q^6=84$ rows. Each row has exactly six zeros and three nonzeros in different locations. For each row $j$, let
let $\cS_j \subset\{1, \ldots, 9\},|\cS_j|=6$ denote the zero coordinates and let $\mathcal{T}_j=\mathcal{S}_j^c$ be the three nonzero coordinates. Define $\bGamma^*=(\gamma_{j k}^*)$ by
$$
\gamma_{j k}^*= \begin{cases}2, & k=9,9 \in \cT_j \\ 1, & k \in \cT_j \backslash\{9\}, 9 \in T_j \\ -1, & k \in T_j, 9 \in \cS_j \\ 0, & k \in \cS_j\end{cases}
$$

Then $\gamma_{j 9}^*$ is the maximum entry in each row, that is, $\gamma_{j 9}^*=\max _{1 \leq k \leq 9} \gamma_{j k}^*$ for all $j$. We use $*$ to denote the nonzero entries to show the schematic form of $\bGamma^*$:
\begin{equation}
\bGamma^*=\left(\begin{array}{rrrrrrrrr}
0 & 0 & 0  & 0 & 0 & 0 & * & * & * \\
0 & 0 & 0  & 0 & 0 & * & 0 & * & * \\
0 & 0 & 0  & 0 & 0 & * & * & 0 & * \\
  & & & & \vdots & & & & \\
\end{array}\right) \nonumber
\end{equation}

Furthermore, we have the following observations:
\begin{itemize}
    \item Among $J=84$ rows, there are $C_8^5 = 56$ rows in which item 9 is among the zero entries, that is, $\gamma_{j9}^* =0$.
    
    \item Among the $8 \times 84=672$ pairwise differences $\gamma_{j k}^*-\gamma_{j 9}^*$ with $k \neq 9$, only $5 \times C_8^5=280$ are zero.
\end{itemize}


Consider $a_j^*=1$ and set $\hat{c}_1=\cdots=\hat{c}_8=0$. Optimizing the pairwise objective for $c_9$ gives
$$
\min _{c_9} \sum_{j=1}^J \sum_{k=1}^8\left|\left(\gamma_{j k}^*-\gamma_{j 9}^*\right)+c_9\right|+\text { constant. }
$$

Since $\gamma_{j 9}^*$ is the largest value in each row, all nonzero differences $\gamma_{j k}^*-\gamma_{j 9}^*$ are negative. Because negative pair differences $(\gamma_{jk}^* - \gamma_{j9}^*)<0$ dominate among the 672 differences, the median  is strictly negative, forcing
$
\hat{c}_9 \neq 0.
$
Hence the minimum is not attained at $c_9=0$, so the pairwise sparsity condition is violated and the model is not identifiable under alignment method.

In the above, we show that such $\bGamma^*$ fails to satisfy Assumption~\ref{cond:pairwise sparsity}. In contrast, Proposition 1 is easily verified because every row and column of $\bGamma^*$ has median zero. Moreover, we numerically confirm condition~\ref{eq:stronger sparsity} of Proposition 1 by exhaustively checking all the condition across $w\in \{1, ..., 8\}$ across all $9!$ permutations. 
Therefore, the proposed Assumption 1 is satisfied.

\subsection{Model Selection Accuracy Comparison}
\label{sec:DIF detection}


In this section, we evaluate model selection accuracy of the proposed method and calculate the corresponding false negative rate (FNR) and false positive rate (FPR). We then compare the model selection accuracy of proposed method with the RMSD method using: (i) the true positive rate (TPR), computed as TPR $=1-\mathrm{FNR}$ and also referred to as the empirical DIF detection rate; 
 and (ii) the true negative rate (TNR), computed as $\mathrm{TNR}=1-\mathrm{FPR}$. 
For the proposed method, we apply a hard thresholding rule by setting $\hat{\gamma}_{jk}=0$ when $|\hat{\gamma}_{jk} | < \epsilon$ with $\epsilon \in \{0.2, 0.5, 0.8\}$. The results are presented in Table~\ref{tab: TNR} below. 


Our proposed method outperforms RMSD methods across diverse settings and various threshold choices, showing great detection ability of both nonzero and zero entries of $(\gamma_{jk}^*)_{J \times p}$. 
Nonetheless, we emphasize that in large-scale assessments, DIF items largely exist. In this work, our focus is not on detecting DIF items, but rather on addressing other important tasks while accounting for measurement non-invariance or DIF. Specifically, we aim to recover the mean parameters and producing rankings of country performance.
Motivated by this, we propose new approach that does not require prior knowledge of anchor items or the specification of reference groups, providing a more computationally efficient, theoretically sound and broadly applicable method for large-scale educational assessments.

\begin{table}[!ht]
\fontsize{8.5}{8.5}\selectfont 
\begin{center}
\begin{tabular}{c|c|cc|cc|cc|cc|cc|cc}
&& \multicolumn{4}{c|}{$\epsilon = 0.2$} & \multicolumn{4}{c|}{$\epsilon = 0.5$} & \multicolumn{4}{c|}{$\epsilon = 0.8$} \\
 \hline
 & $N = $& \multicolumn{2}{c|}{$10^4$} & \multicolumn{2}{c|}{$2\times 10^4$} &  \multicolumn{2}{c|}{$10^4$} & \multicolumn{2}{c|}{$2\times 10^4$}&  \multicolumn{2}{c|}{$10^4$} & \multicolumn{2}{c|}{$2\times 10^4$} \\
 \hline
   & &   TPR   & TNR  &     TPR   & TNR & TPR   & TNR  &     TPR   & TNR & TPR   & TNR  &     TPR   & TNR     \\ 
     \hline
\multirow{4}{*}{$S1$}  & Proposed &  1.000    &  0.900  &  1.000    & 0.974   &   1.000 & 1.000  &   1.000 & 1.000  &   0.996 & 1.000  &    0.999 & 1.000     \\
  \cline{2-14}
 & RMSD 0.05 &    0.754 & 0.267  &   0.752   & 0.260   &   0.746 & 0.712 &    0.743 &  0.711 &    0.724 & 0.969  &   0.727 & 0.972      \\
  \cline{2-14}
 & RMSD 0.10  &   0.652 & 0.508 &   0.649   & 0.503   &     0.652 & 0.621  &   0.649 & 0.610  &   0.648 & 0.869 &    0.647 & 0.869    \\
  \cline{2-14}
& RMSD 0.15 &    0.502 & 0.749 &   0.503   & 0.746    &   0.502 & 0.793 &    0.503 &  0.791  &   0.502 & 0.858  &   0.503 & 0.852      \\
\hline
\multirow{4}{*}{$S2$}  & Proposed   &     1.000 & 0.919  &   1.000 & 0.977 &     1.000 & 1.000 &     1.000 & 1.000  &  0.995 & 1.000 &    0.998 & 1.000      \\
  \cline{2-14}
 & RMSD 0.05  & 0.984 & 0.883   &    0.984 & 0.883 &     0.983 & 0.981  &     0.984 & 0.986 &   0.982 & 0.999 &  0.984 & 0.999   \\
  \cline{2-14}
 & RMSD 0.10 &    0.873 &  0.509   &    0.873 & 0.509  &   0.871 & 0.737&     0.872 & 0.716  & 0.837 & 0.906 &  0.836 & 0.904  \\
  \cline{2-14}
& RMSD 0.15 &    0.647 & 0.791  &   0.647 & 0.791  &   0.648 & 0.842 &     0.647 & 0.843  & 0.639 & 0.895 &  0.637 & 0.892  \\
\hline
\multirow{4}{*}{$S3$} & Proposed  &   1.000 & 0.754  &   1.000 & 0.896 &     1.000 & 0.995  &   1.000 & 1.000  &    0.979 & 1.000  &   0.993 & 1.000        \\
  \cline{2-14}
 & RMSD 0.05  &    0.961 & 0.661  &     0.967 & 0.722  &    0.958 & 0.946  &    0.965 & 0.962  &   0.938 & 0.996  &   0.948 & 0.998       \\
  \cline{2-14}
 & RMSD 0.10 &     0.832 & 0.423   &      0.838 & 0.425 &    0.826 & 0.662  &   0.832 & 0.653 &    0.786 & 0.855 &  0.792 & 0.861   \\
  \cline{2-14}
& RMSD 0.15 &    0.618 & 0.701   &   0.621 & 0.710  &    0.617   & 0.747 &    0.621 & 0.752  &    0.608 & 0.837  &   0.608 & 0.836     \\
\hline
\multirow{4}{*}{$S4$} & Proposed   &   1.000 & 0.727  &     1.000 & 0.874  &     0.999 & 0.994  &    1.000 & 1.000  &    0.975 & 1.000  &     0.990 & 1.000     \\
  \cline{2-14}
 & RMSD 0.05   &   0.975 & 0.709   &   0.974 & 0.728  &    0.974 & 0.957  &    0.974 & 0.959 &    0.963 & 0.997 &    0.964 & 0.998     \\
  \cline{2-14}
 & RMSD 0.10 &    0.838 & 0.307  &    0.835 & 0.291  &   0.825 & 0.547  &   0.822 & 0.525 &    0.786 & 0.797 &    0.783 & 0.775     \\
  \cline{2-14}
& RMSD 0.15 &    0.631 & 0.641 &    0.628 & 0.652  &   0.631 & 0.707 &    0.627 & 0.709  &     0.627 &  0.865  &     0.623 & 0.861     \\
\hline
\end{tabular}
\caption{True positive rates (TPRs), computed as TPR = 1 $-$ FNR, and true negative rates (TNR), computed as TNR = 1 $-$ FPR, for the proposed method and the RMSD methods.}
\label{tab: TNR}
\end{center}
\end{table}

\subsection{Simulation Results using Iterative RMSD Methods}
\label{sec:iterRMSD}
In this section, we implement the RMSD methods in both non-iterative and iterative manners. 
The comparison results are given in Figures~\ref{fig:kendall iter S1-2} and \ref{fig:kendall iter S3-4}.
{Based on the comparison results in Figures~\ref{fig:kendall iter S1-2} and \ref{fig:kendall iter S3-4}, we observe that the proposed method consistently outperforms RMSD methods across all settings, whether an iterative strategy is applied or not. In addition, the iterative strategy does not always improve the performance of the RMSD method.} 
More specifically, under setting $S1$, the iterative strategy improves the performance of RMSD with threshold 0.05 (hereafter RMSD 0.05). However, in settings $S2$-$S4$, iterative RMSD 0.05 performs no better than the non-iterative version. For RMSD with thresholds 0.10 and 0.15, the iterative version performs substantially worse than the non-iterative one under $S1$, and does not show a significant advantage over the non-iterative method under $S2$–$S4$.
We also observe that the variance of Kendall rank correlation from the iterative method is higher than that from the non-iterative method in $S1$, whereas in $S3$, the variance from the iterative method is lower. These findings suggest that the iterative strategy does not systematically outperform the non-iterative one. 
Moreover, for both iterative and non-iterative RMSD methods, the performance is sensitive to the choice of the RMSD threshold. 

The results in Figures~\ref{fig:kendall iter S1-2}--\ref{fig:kendall iter S3-4} are based on a maximum of three iterative updates (the algorithm stops if all RMSD values are less than a pre-specified threshold or the maximum number of iterations has been reached). To check if the results are affected by the number of iterations, we further conducted a sensitivity analysis by varying the number of iterations across the values of 3, 5, and 7. The results, presented in Figure~\ref{fig:iter SA}, indicate that increasing the maximum number of iterations does not significantly improve the performance of the iterative RMSD methods.

\begin{figure}[!htbp]
\centering    
        \includegraphics[width=3.3in]{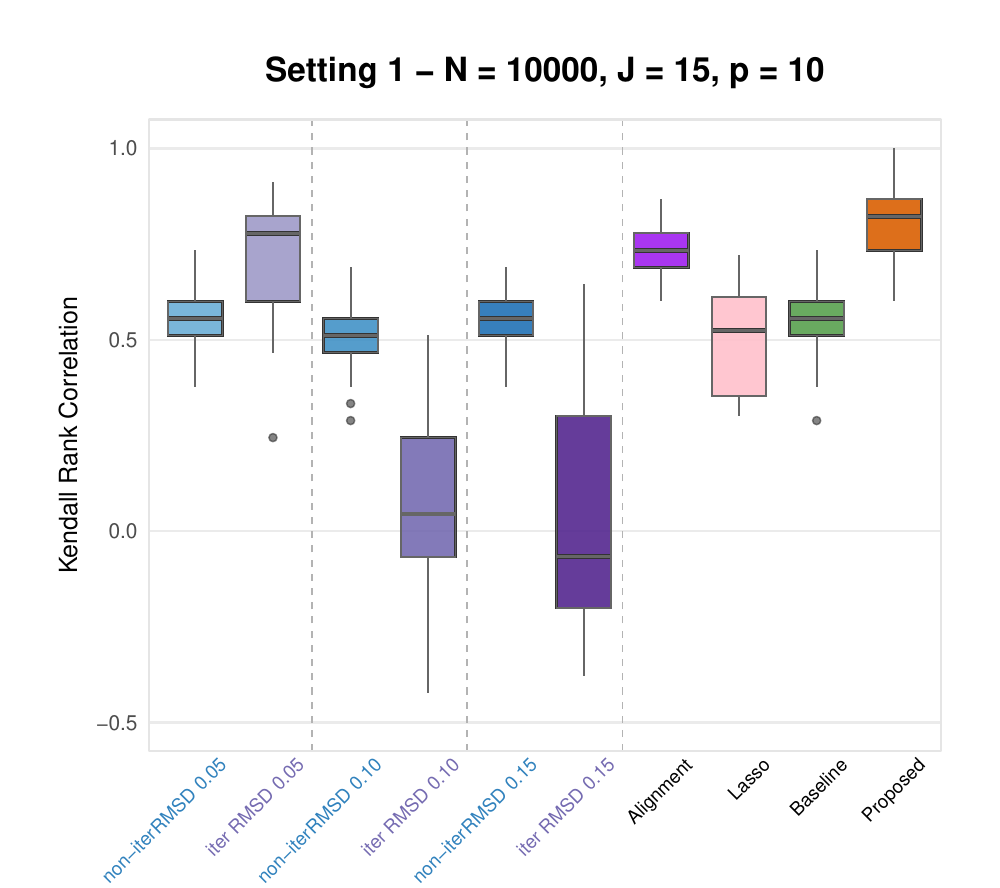}~
       \includegraphics[width=3.3in]{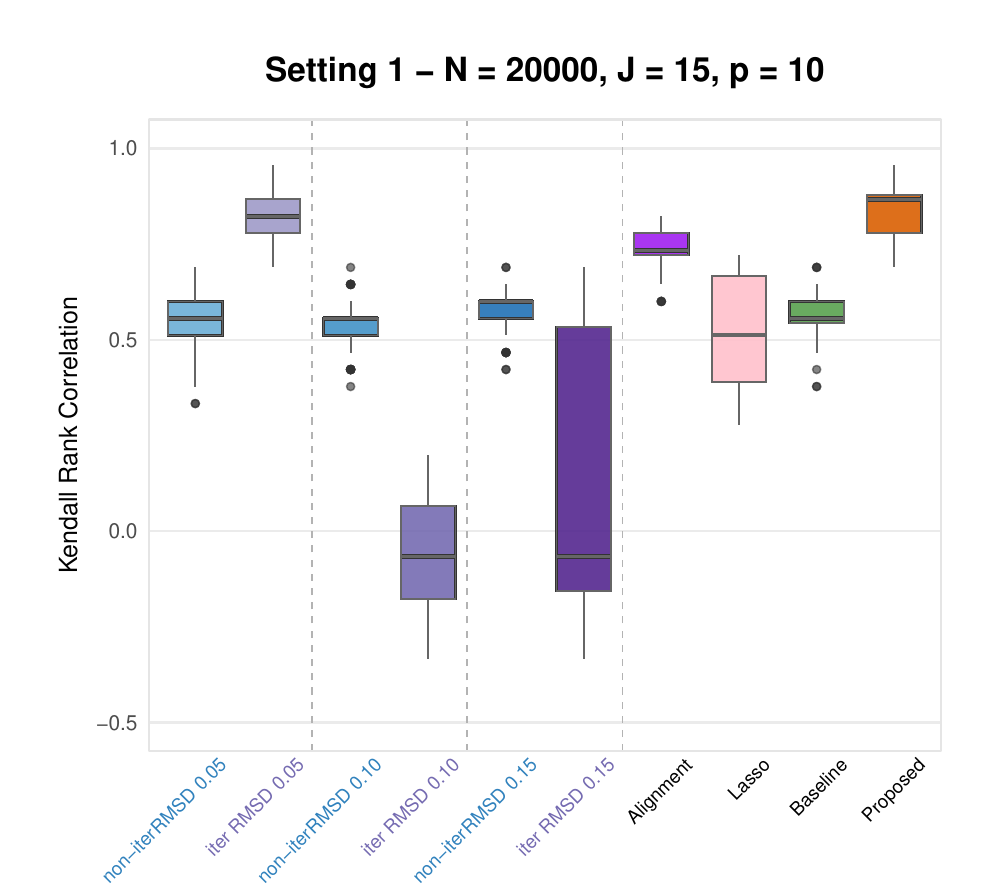}\\
       \includegraphics[width=3.3in]{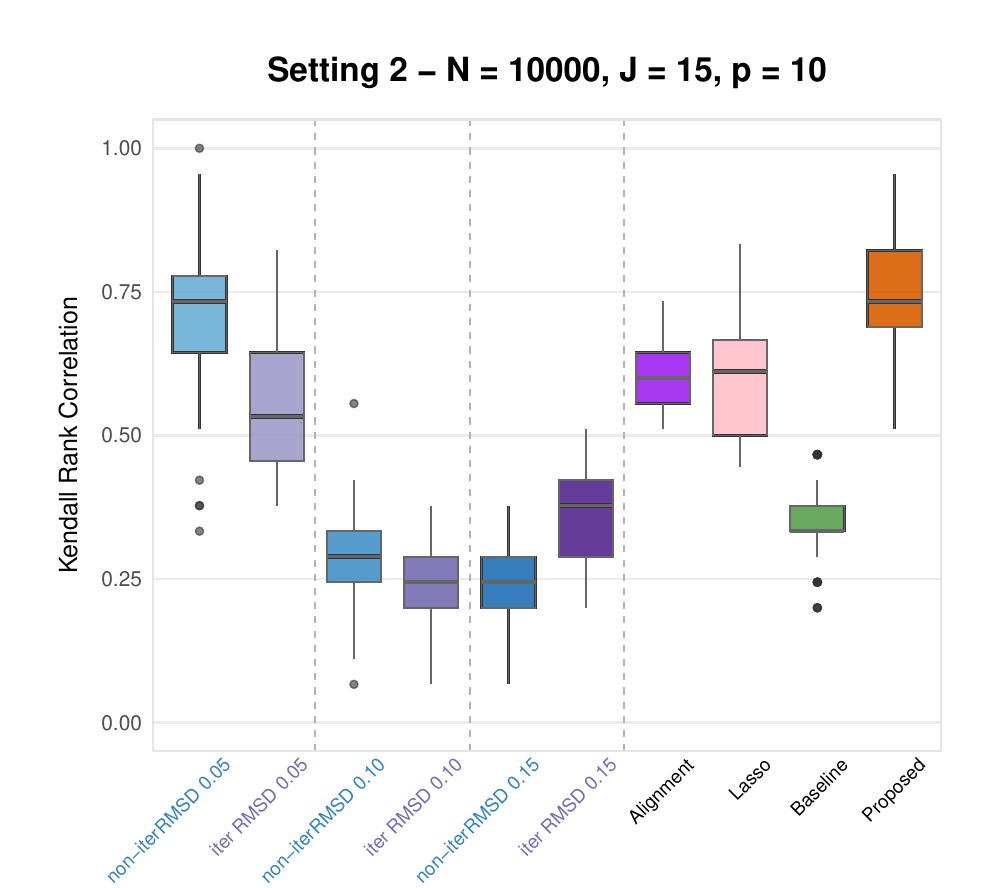}~
       \includegraphics[width=3.3in]{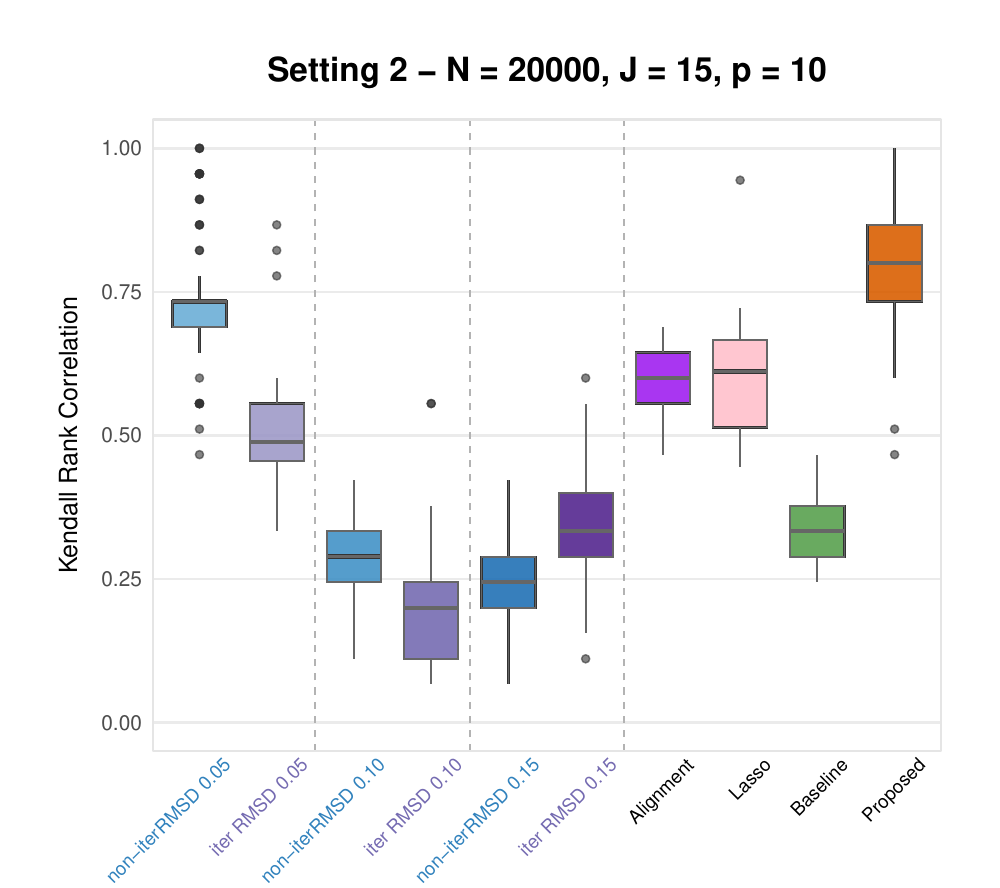} 
        \caption{Comparison of average Kendall rank correlation of the proposed method, baseline methods, alignment method, Lasso-type method, and the non-iterative and iterative RMSD-based methods with different thresholds under $S1$ and $S2$.}
        \label{fig:kendall iter S1-2}
        \end{figure}

        \begin{figure}[!htbp]
\centering    
       \includegraphics[width=3.3in]{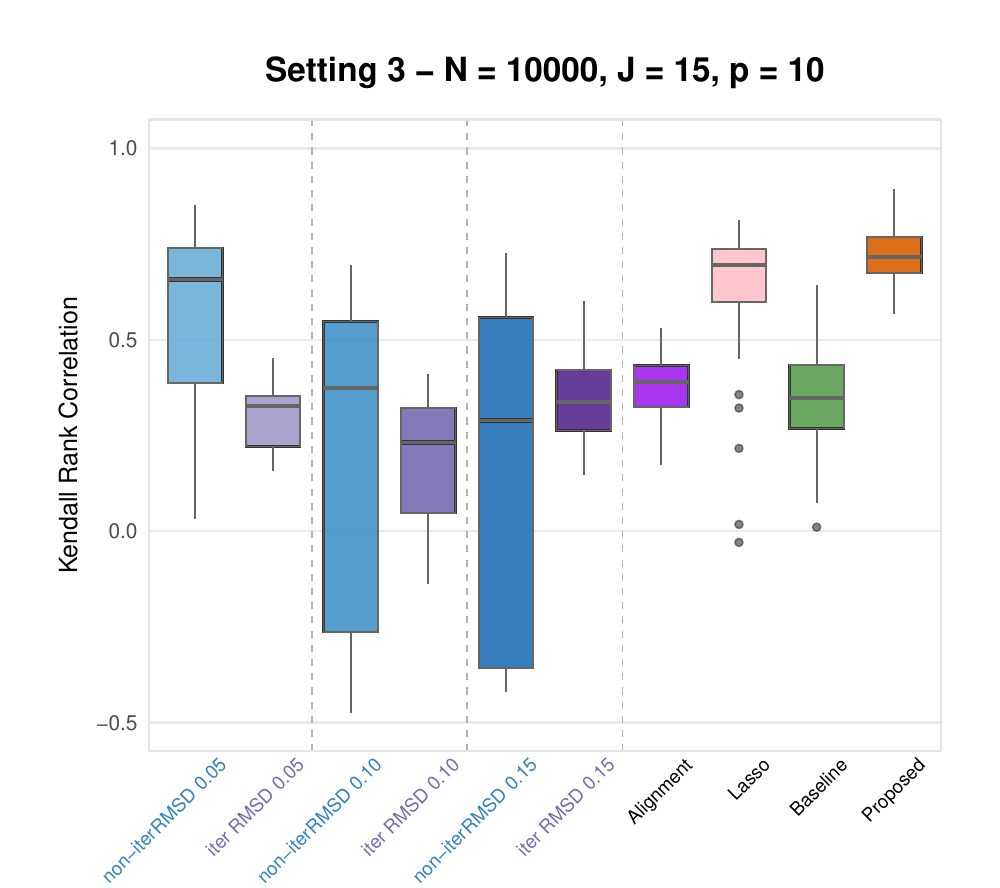}~
       \includegraphics[width=3.3in]{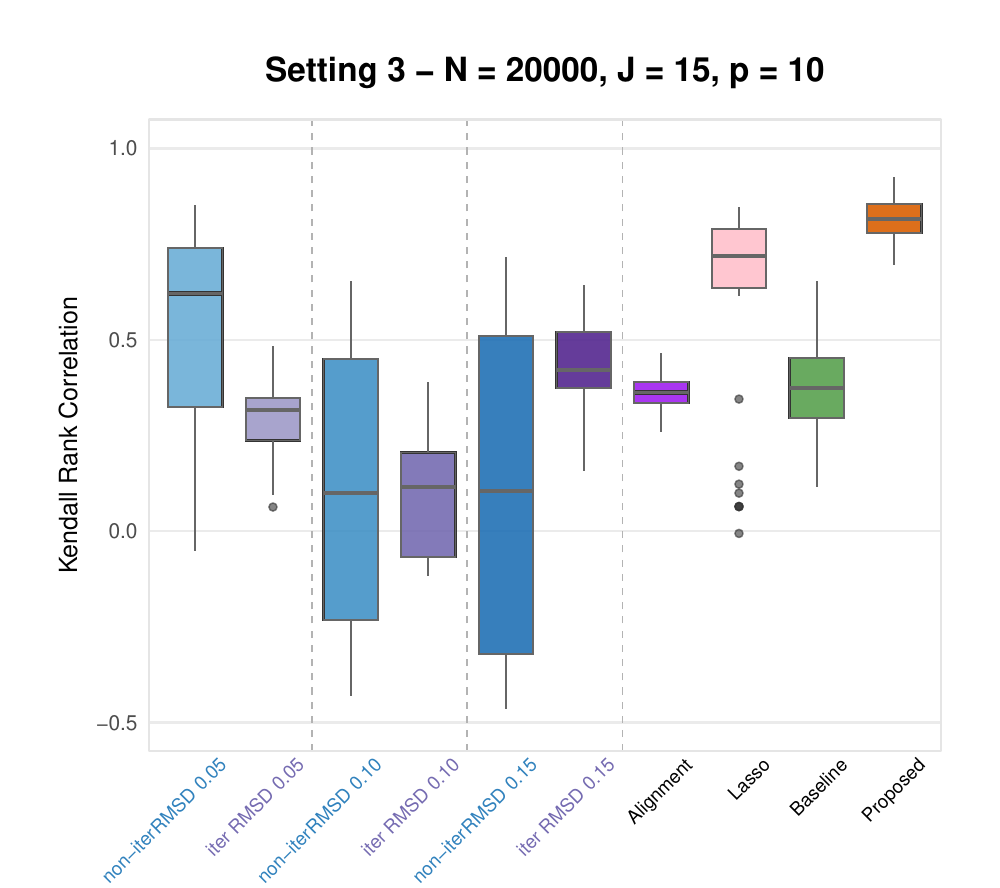} \\
       \includegraphics[width=3.3in]{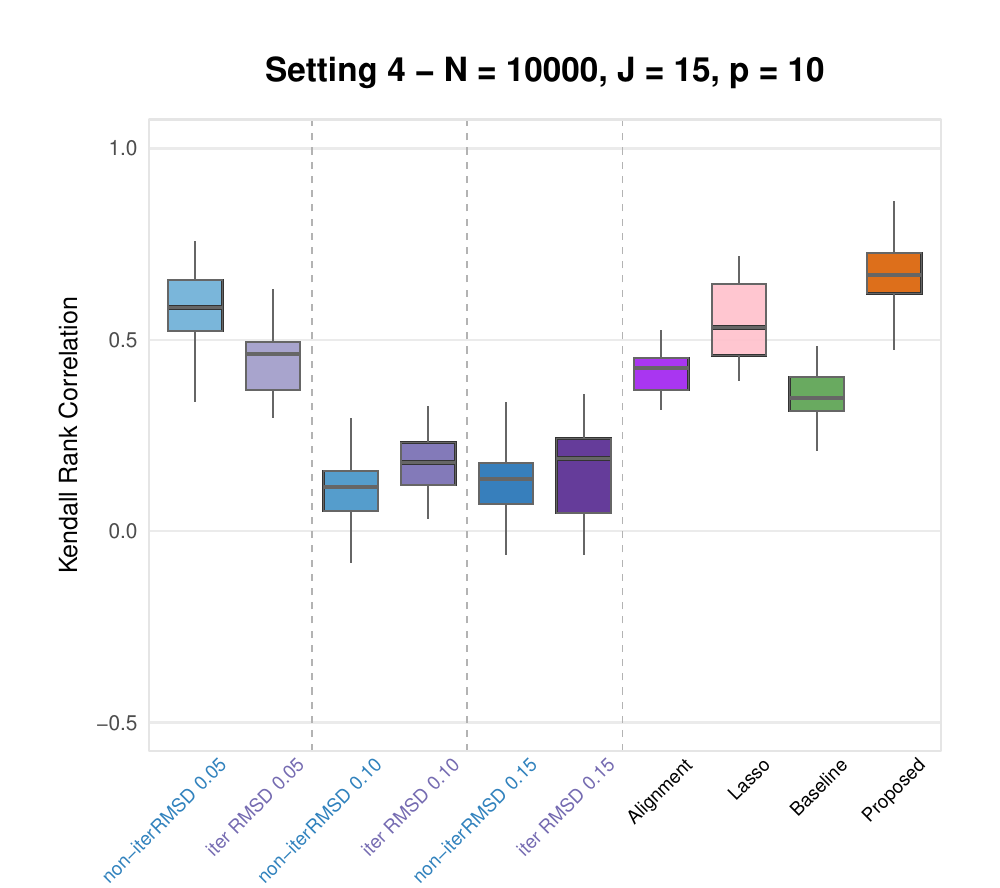}~
       \includegraphics[width=3.3in]{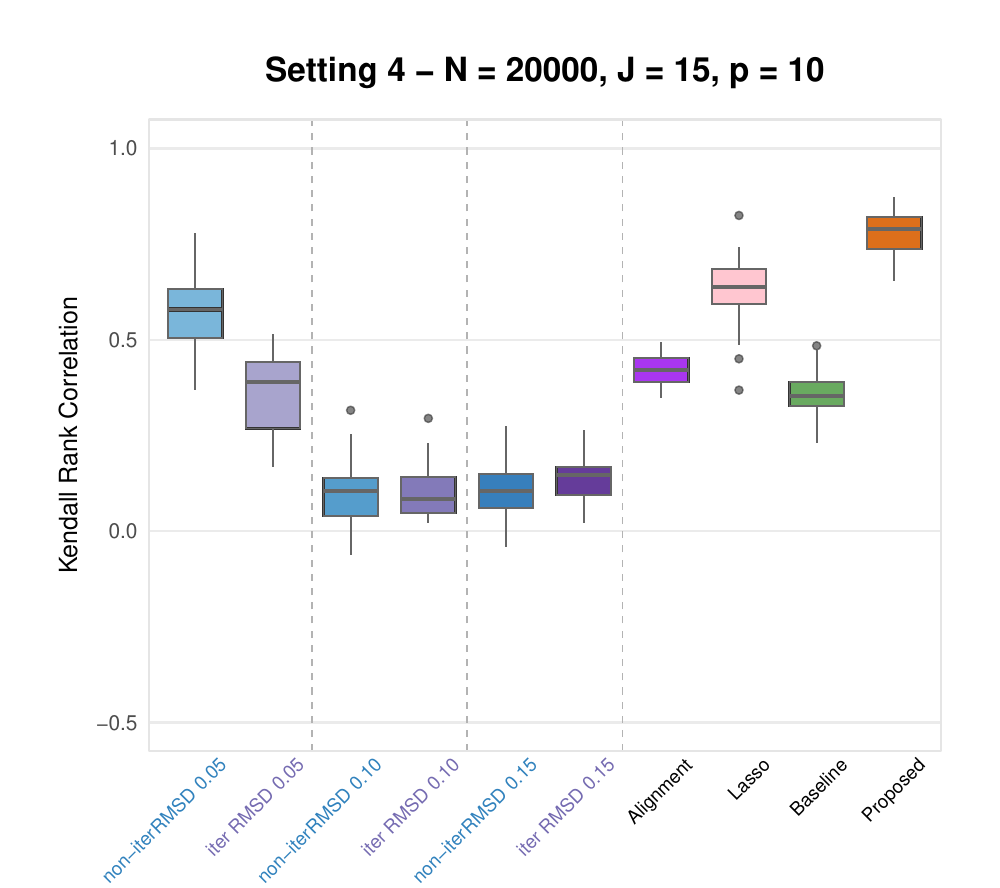} 
        \caption{Comparison of average Kendall rank correlation of the proposed method, baseline methods, alignment method, Lasso-type method, and the non-iterative and iterative RMSD-based methods with different thresholds under $S3$ and $S4$.}
        \label{fig:kendall iter S3-4}
        \end{figure}

\begin{figure}
    \centering
    \includegraphics[width=0.45\linewidth]{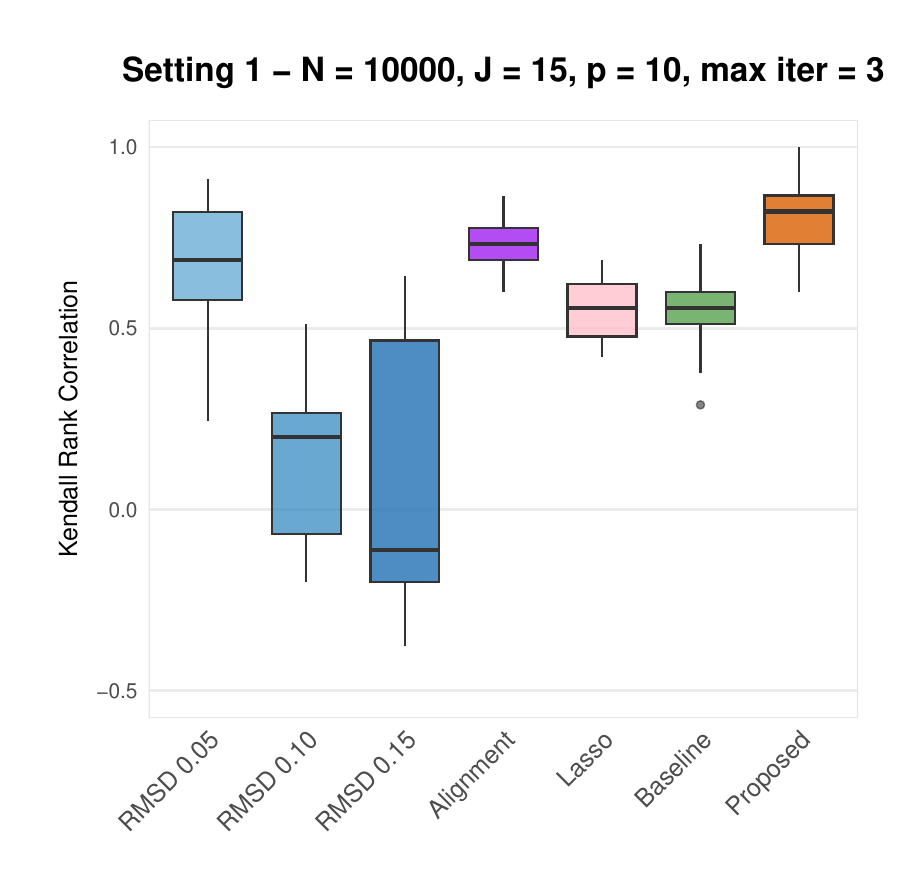}\\
    \includegraphics[width=0.45\linewidth]{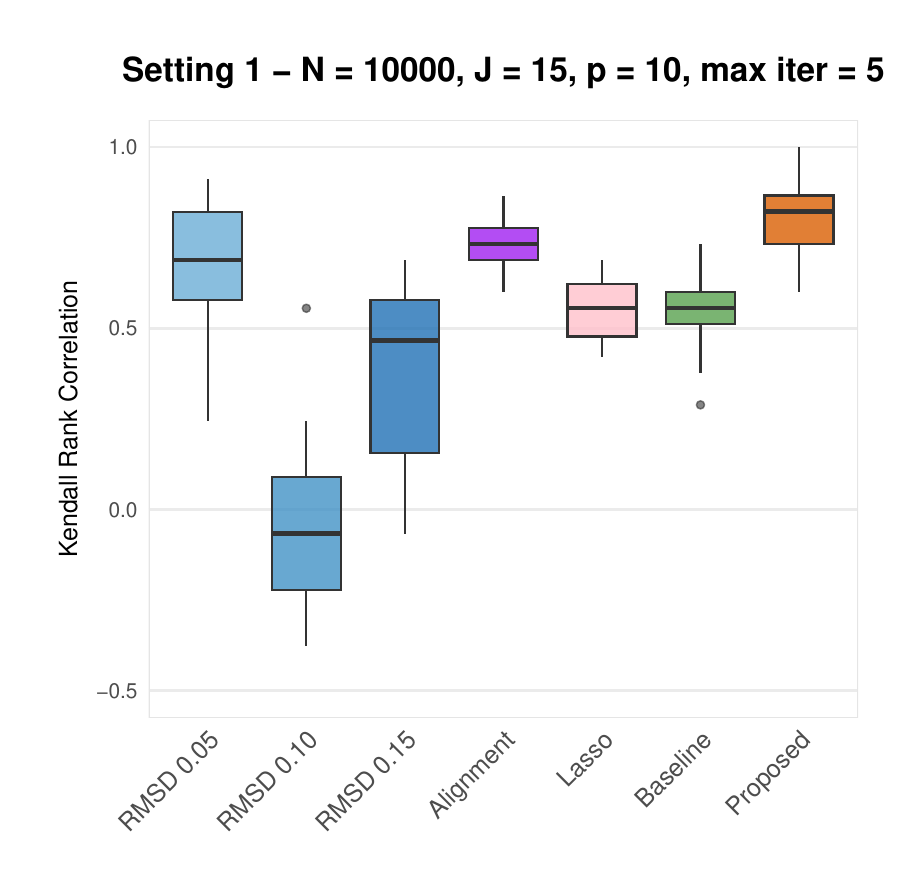}\\
    \includegraphics[width=0.45\linewidth]{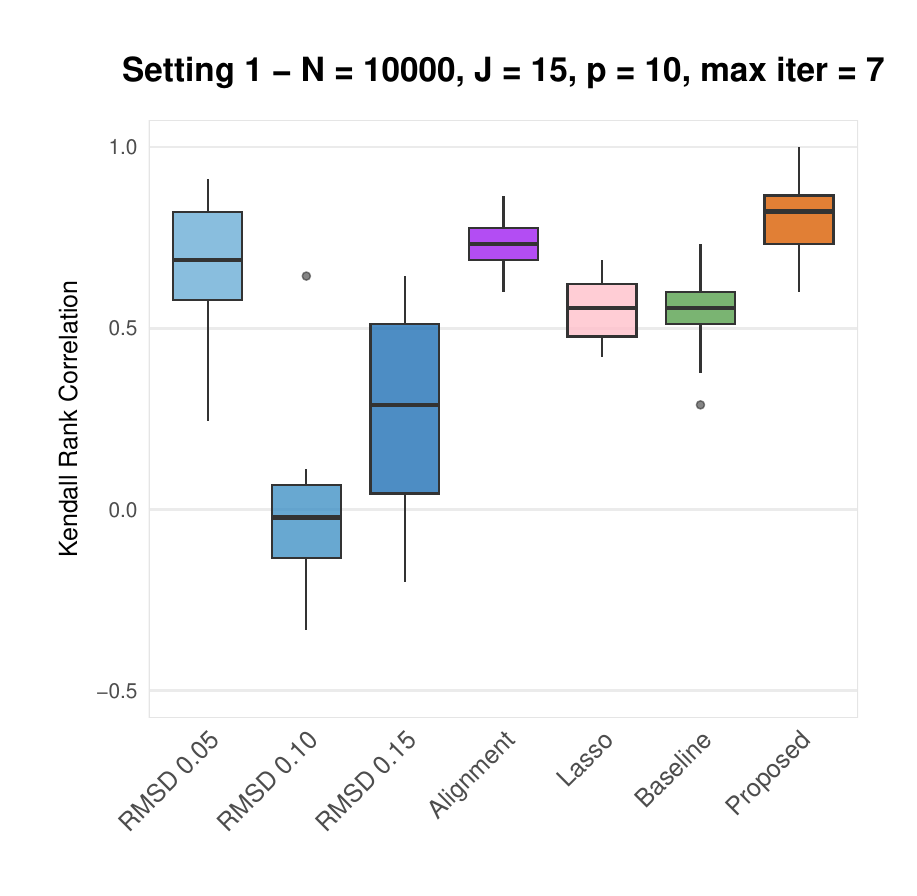}
    \caption{Comparison of average Kendall rank correlation of the proposed method, baseline methods, alignment method, Lasso-type method, and {\bf iterative} RMSD-based methods with maximum iterations = $\{ 3, 5, 7\}$ under $S1$.}
    \label{fig:iter SA}
\end{figure}

\subsection{Simulation Results with DIF Structure Violating Sparsity Assumption}
\label{sec:violate DIF}

In this section, we conduct sensitivity analysis to investigate the method's performance when the sparsity assumption (Assumption 1 in the main text) holds only approximately. 
  Specifically, we considered similar settings as scenario $S1$ in the main text manuscript. In these settings, the nonzero DIF parameters (i.e., $\gamma_{jk}^*$) are the same as in $S1$ (generated from Unif$[1,3]$), but the zero DIF parameters are replaced with nonzero values, generated from Unif$[-m, m]$, where $m \in \{0.1, 0.2, 0.3\}$ controls the degree of violation to the original setting. Under these new settings, there is no $\gamma_{jk}^* = 0$, and thus, the sparsity assumption is violated for the proposed method. 
  The results, presented in Figure~\ref{fig:comparison of kendall rank}, show that under mild violation case ($m= 0.1, 0.2$), the proposed method outperforms all other comparative methods; the alignment method 
  performs the worst, and the RMSD methods with various thresholds and the Lasso-MNFA method perform similarly and are better than the alignment method. Under more severe violations ($m = 0.3$), our method performs comparably to and sometimes slightly worse than RMSD methods with various thresholds, the Lasso-MNFA method, and the benchmark method, while still outperforming the alignment method. 
Overall, we see from the sensitivity analyses that the proposed method is reasonably robust against model misspecification and often outperforms the competing methods. These results also suggest that there may not exist a method that always outperforms the others.

\begin{figure}[htbp]
    \centering
   \includegraphics[width=3in]{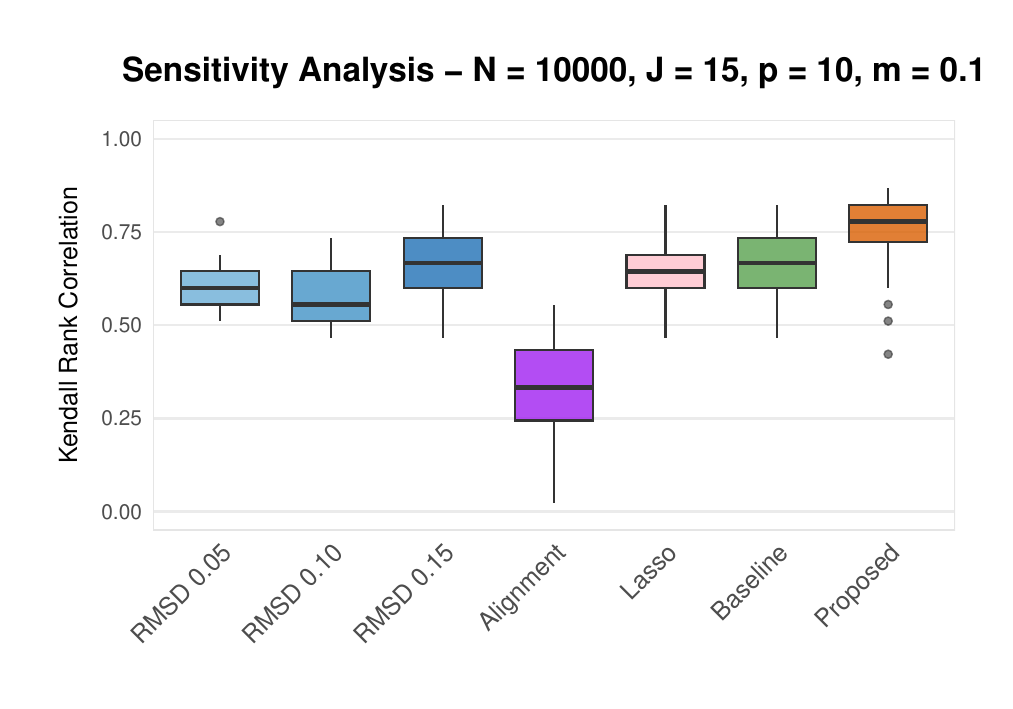}~\includegraphics[width=3in]{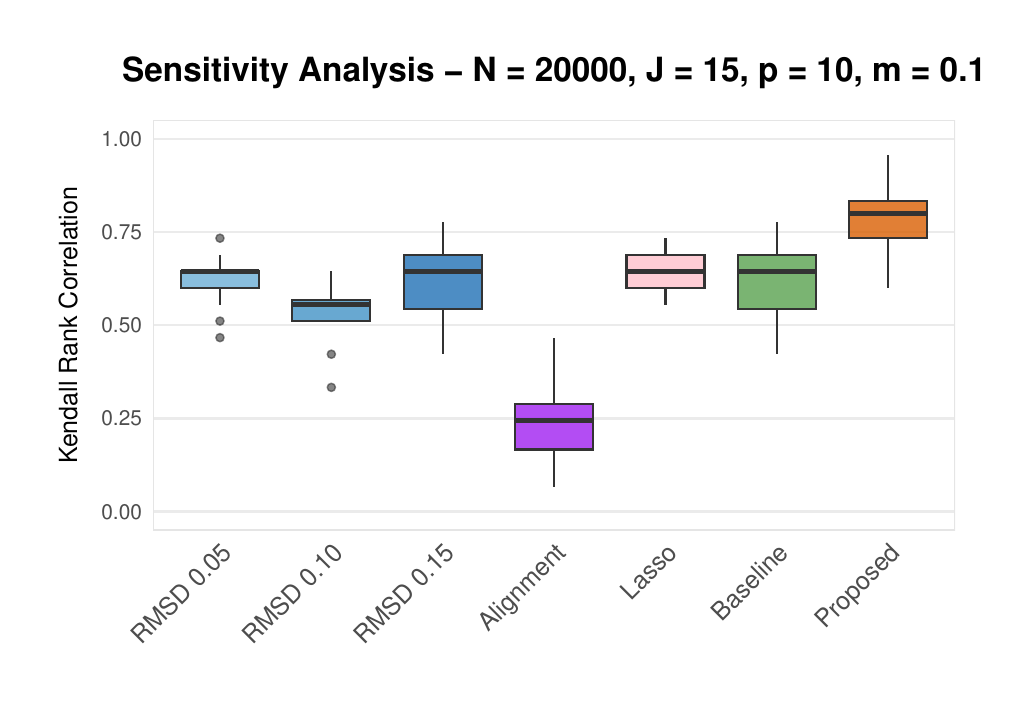}\\
       \includegraphics[width=3in]{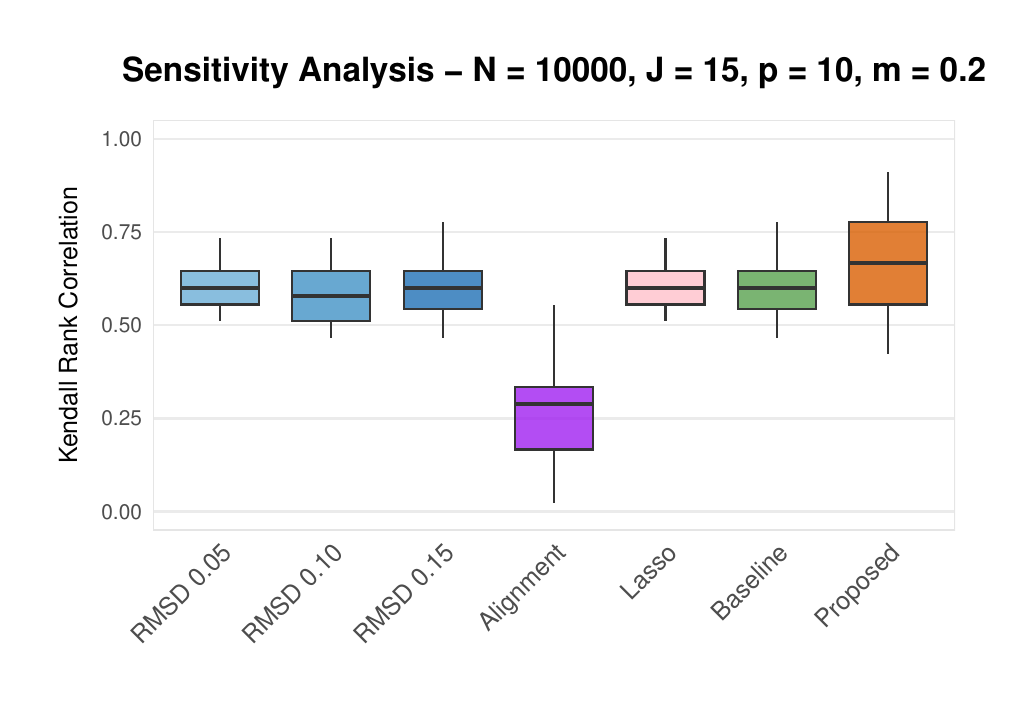}~\includegraphics[width=3in]{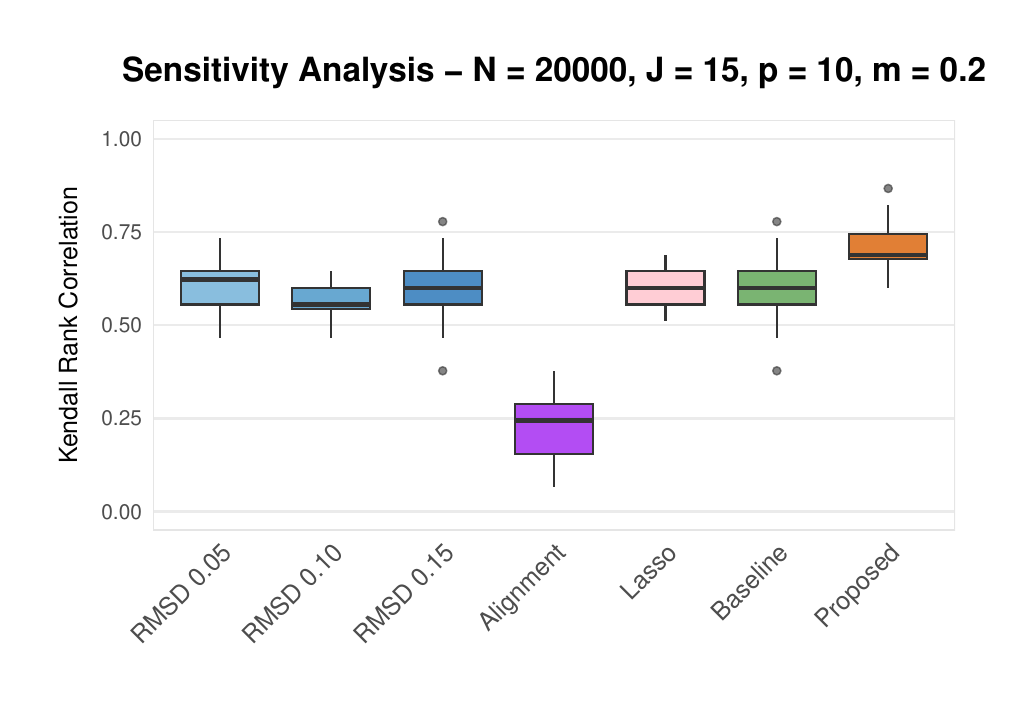}\\
      \includegraphics[width=3in]{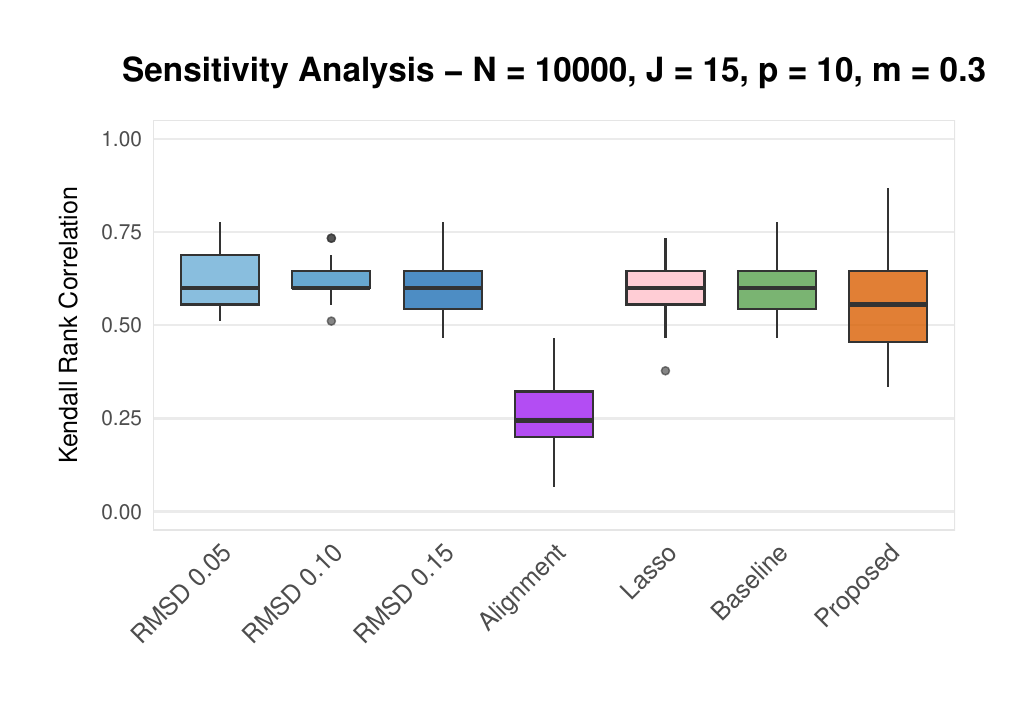}~\includegraphics[width=3in]{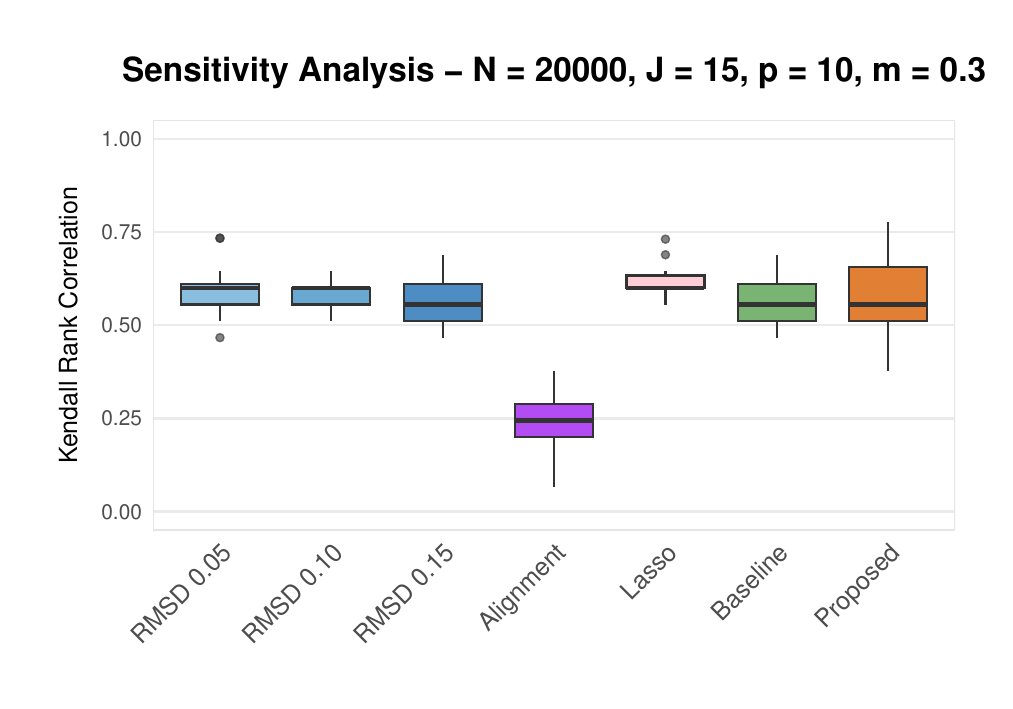}\\
    \caption{Comparison of average Kendall rank correlation of the proposed method, baseline methods, RMSD-based methods with different thresholds, alignment methods, and Lasso-type methods under DIF structure violating sparsity assumptions.} 
    \label{fig:comparison of kendall rank}
\end{figure}

\subsection{Simulation Results under Enlarged Range of $d_j^*$}
\label{sec:enlarge d}

To illustrate that our method is robust to the range of easiness parameters, we conduct a sensitivity analysis with $d_j^*$ defined within $ [-3, 3]$ while the other settings for $S1$-$S4$ remain the same. The detailed settings of easiness parameter are given in Table~\ref{tab:d setting large}. We compare the performances of proposed methods with the baseline method and the RMSD methods and present the comparison results, in terms of Kendall rank correlation, in Figure~\ref{fig:large d}. The results imply that our method is robust to the range of easiness parameters, and our proposed method still outperform RMSD methods and baseline methods under an enlarged range $d_j^* \in [-3,3]$ for nearly all the settings, with the exception of $S3$ at $N=10000$. Nonetheless, in this case, the proposed method performs comparably to RMSD methods with thresholds 0.10 and 0.15, while still outperforming RMSD with threshold 0.05 and the baseline method. 
Furthermore, when comparing our proposed method under original range of $d_j^* \in [-1, 1]$ with that under enlarged range $d_j^* \in [-3, 3]$, we find out that increasing the range of $d_j^*$  leads to a slight reduction in Kendall rank correlation across all settings $S1$-$S4$.
Intuitively, when the range of the difficulty parameter becomes too wide (i.e., items that are extremely easy or extremely difficult), the information available for accurately estimating the students' latent abilities may not be sufficient, leading to this slight decrease in country performance.

\begin{figure}[H]
    \centering
    \includegraphics[width=2.4in]{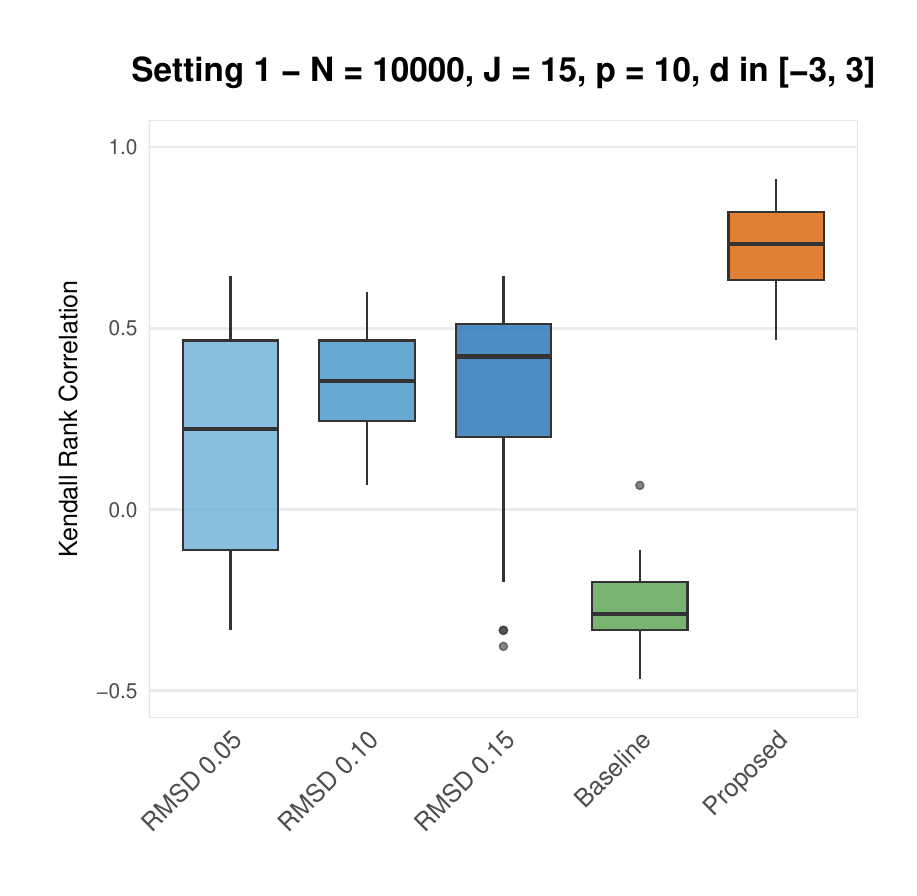}~\includegraphics[width=2.4in]{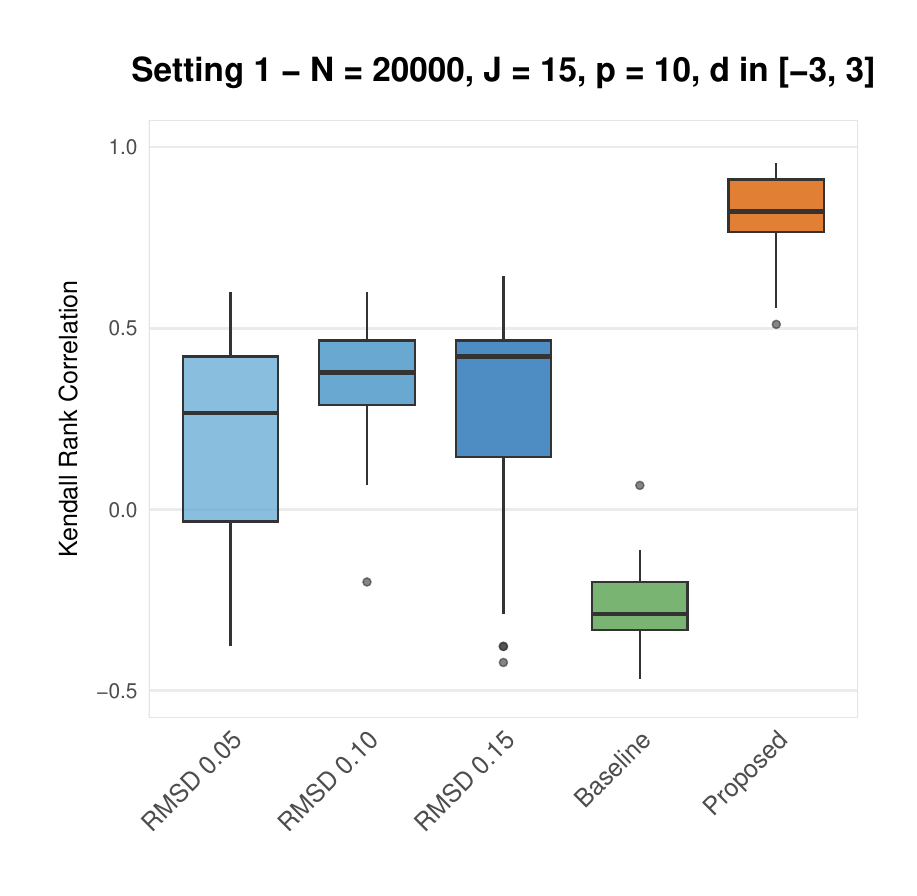}\\
     \includegraphics[width=2.4in]{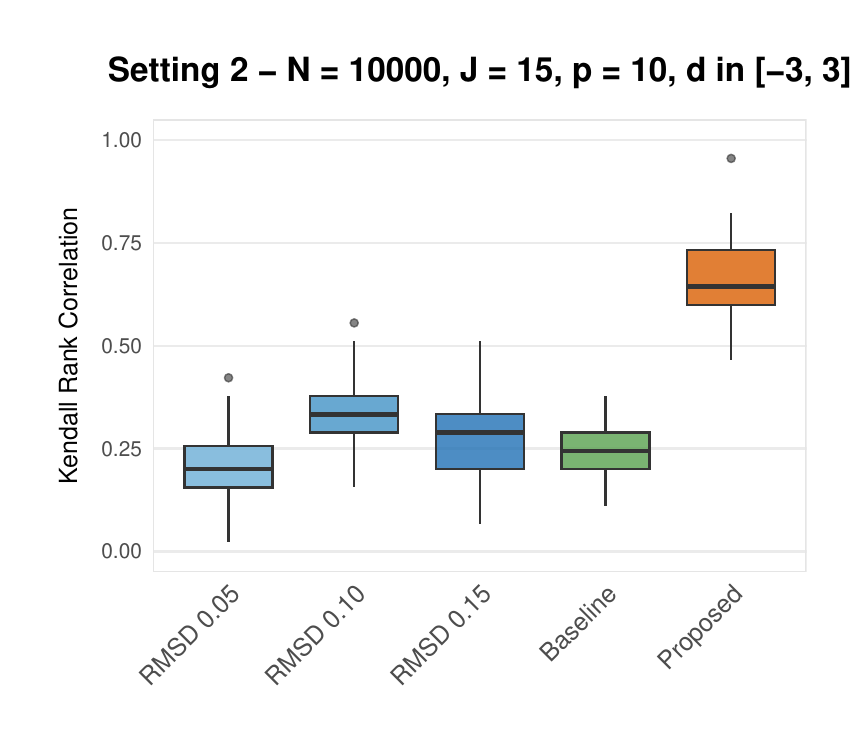}~\includegraphics[width=2.4in]{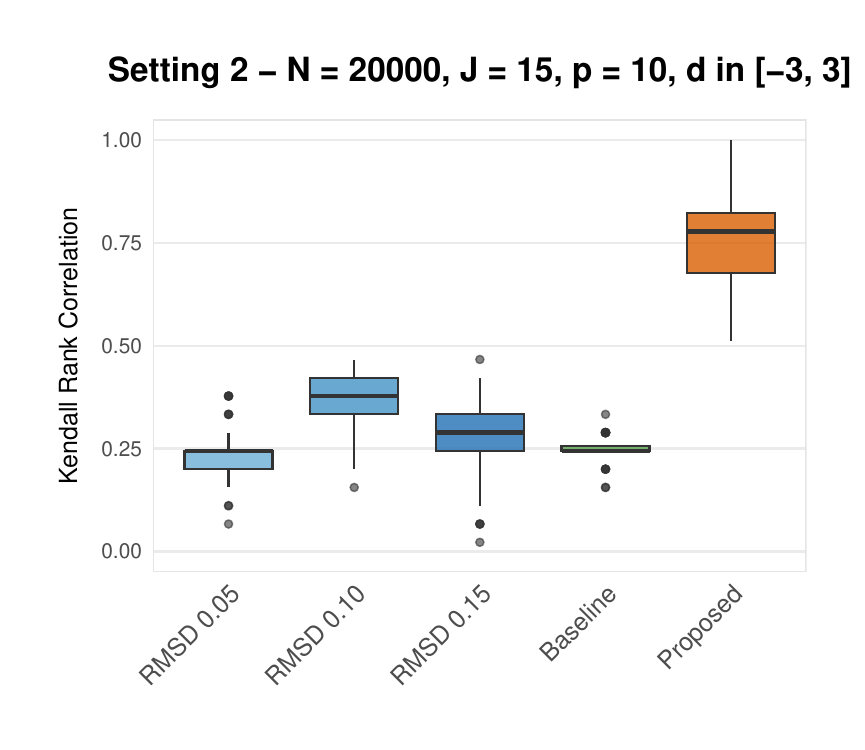}\\
      \includegraphics[width=2.4in]{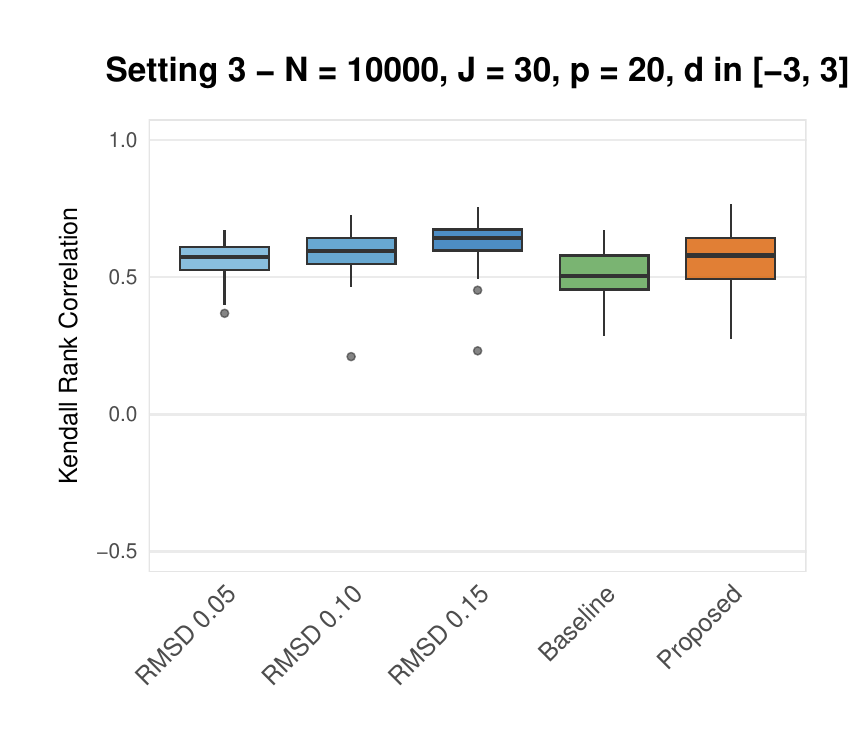}~\includegraphics[width=2.4in]{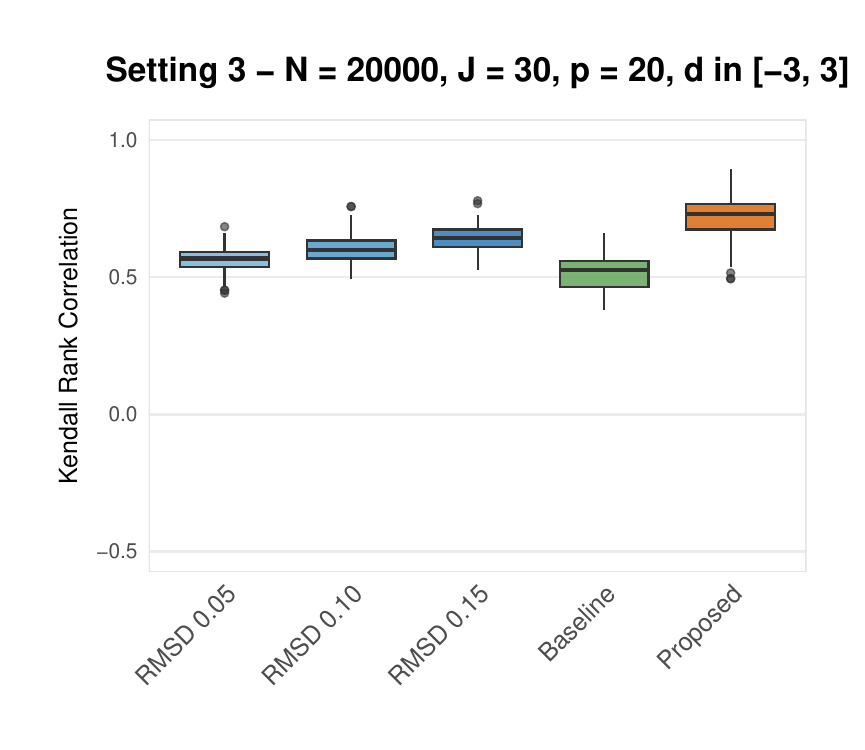}\\
       \includegraphics[width=2.4in]{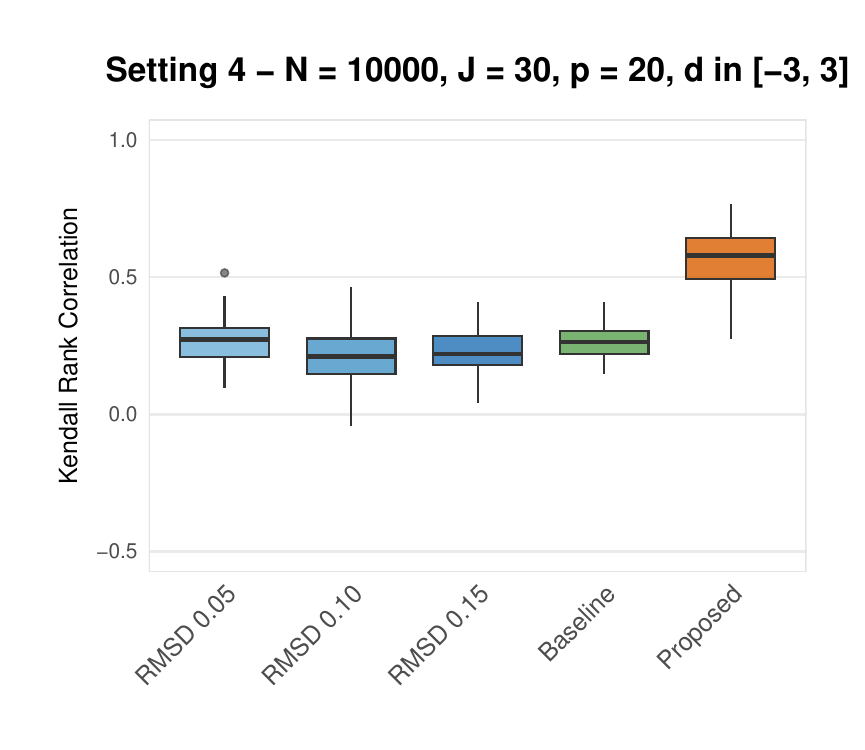}~\includegraphics[width=2.4in]{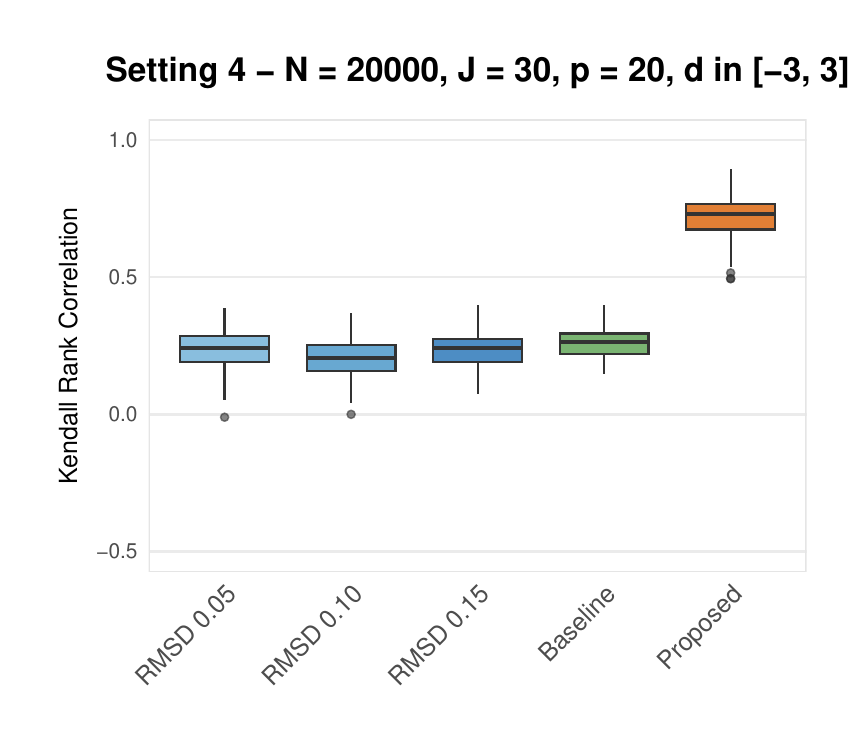}
    \caption{Comparison of average Kendall rank correlation of the proposed method, baseline methods, and the RMSD-based methods with different thresholds under $d_j^* \in [-3, 3]$.}
    \label{fig:large d}
\end{figure}

}

\section{Additional Results for Real Data}
\label{sec-real}

\subsection{Additional results for Mathematics Data}
\label{sec-real-additional-math}

\begin{figure}[H]
    \centering
    \includegraphics[scale = 0.4]{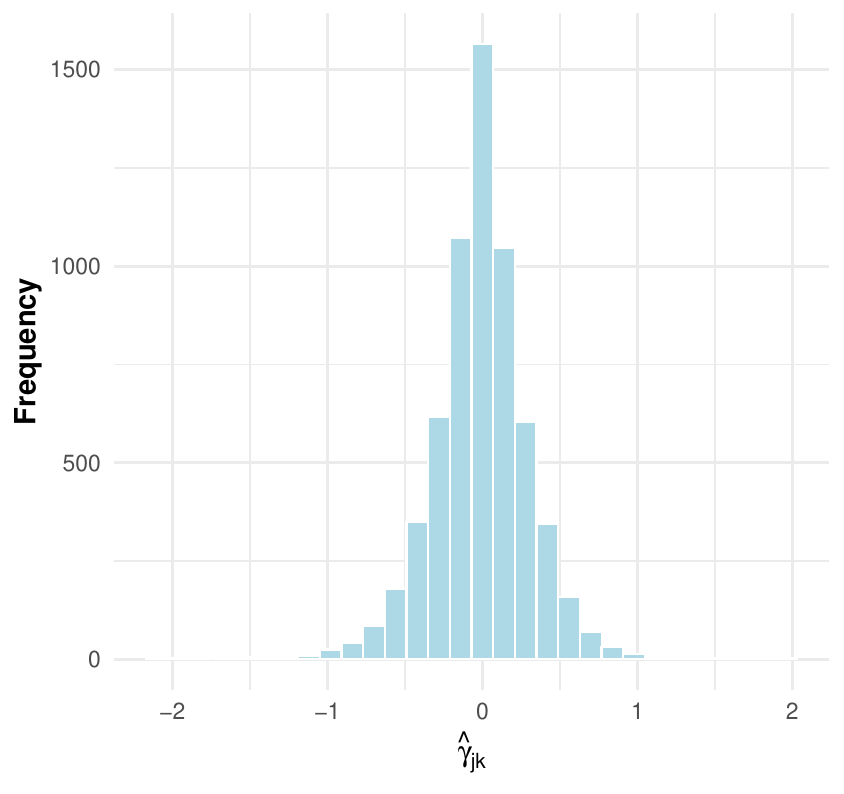}
    \caption{Histogram of estimated $\hat{\gamma}_{jk}$ for the mathematics domain.}
    \label{fig: hist-gamma-math}
\end{figure}

We present the ranking of country-wise mean proficiencies by RMSD methods with thresholds 0.1 and 0.15, and by the baseline method in Table~\ref{tab: math 0.1 0.15 baseline}. Additionally, the estimated mean $\hat{\mu}_k$s and variance parameters $\hat{\sigma}_k$s from all methods are also given in Tables~\ref{tab: math with variance}--\ref{tab: math 0.1 0.15 baseline}. Table~\ref{tab:math a d} provides the estimated $\hat{a}_j$s and $\hat{d}_j$s from proposed method using the math data. In addition, Tables~\ref{tab:math gamma 1}--\ref{tab:math gamma 2} presents the estimated $\hat{\gamma}_{jk}$s based on the math data. We examine the estimated DIF parameters  $\hat \gamma_{jk}$ and present the histogram of the $\hat \gamma_{jk}$s in Figure~\ref{fig: hist-gamma-math}.

{ In addition, we examined the country differences in terms of estimated means and found several patterns that serve as valuable complements to the country rankings. For example, our proposed method estimates the difference in means $\hat{\mu}_k$ for Hungary and Denmark as $0.061 - 0.032 = 0.029$, whereas the RMSD method gives $-0.032- (-0.034)=0.002$. Thus, while both methods suggest that Hungary performs slightly better than Denmark, our method indicates a more noticeable difference. A similar pattern appears for Ireland and Sweden: our method estimates a difference of $0.116-0.091=0.025$, but RMSD yields $0.0228-0.0226=0.002$. In other words, our method suggests a larger performance gap in favor of Ireland, while RMSD implies the two countries are at essentially the same level.}

\begin{table}[!h]
\begin{center}
\fontsize{8}{8}\selectfont
\renewcommand{\arraystretch}{0.9}
\begin{tabular}{c|c|c|c||c|c|c|c}
 \hline
Rank& Proposed&  $\hat{\mu}_k$ & $\hat{\sigma}_k$ &  Rank &RMSD& $\hat{\mu}_k$ & $\hat{\sigma}_k$ \\
 \hline
      1 & Japan  &   0.644        &0.968    & 1  & Japan   &  0.692 & 1.150       \\ 
  2 &  Korea &    0.478      & 0.997   & 2 & Korea  &  0.481& 1.197  \\ 
  3 &  Estonia   &   0.356      & 0.864  & 3 &  Estonia  &   0.334&  0.998      \\ 
  4 & Switzerland  &  0.345    &  0.931    & 4  & Switzerland   &  0.320&  1.142      \\ 
  5 &    Czech Republic&  0.283     &  0.992   & 5  & Czech Republic   &0.278 & 1.188 \\ 
  6 &   Belgium &  0.197     & 0.946    &  6 &  Poland  &  0.151&  1.027       \\ 
  7 &  Poland   &   0.188   &  0.866  & 7  & Belgium   &  0.144 &  1.115      \\
\hline
\end{tabular}
\caption{Comparison of rankings of country-wise average latent skill levels between proposed method and RMSD method for math domain.}
\label{tab: math with variance}
\end{center}
\end{table}

\begin{landscape}
    
\begin{table}[!htbp]
\begin{center}
\fontsize{10}{10}\selectfont
\begin{tabular}{c|c|c|c|c|c|c|c|c|c|c|c}
 \hline
Rank& RMSD 0.1 &  $\hat{\mu}_k$ & $\hat{\sigma}_k$ &  Rank &RMSD 0.15 & $\hat{\mu}_k$ & $\hat{\sigma}_k$ &  Rank &Baseline   & $\hat{\mu}_k$ & $\hat{\sigma}_k$ \\
 \hline
   1 & Japan & 0.825 & 1.173 & 1 & Japan & 0.705 & 1.115 & 1 & Japan & 0.616 & 0.881 \\ 
  2 & Korea & 0.608 & 1.198 & 2 & Korea & 0.51 & 1.141 & 2 & Korea & 0.452 & 0.888 \\ 
  3 & Estonia & 0.455 & 1.017 & 3 & Estonia & 0.337 & 0.947 & 3 & Estonia & 0.319 & 0.721 \\ 
  4 & Switzerland & 0.418 & 1.143 & 4 & Switzerland & 0.314 & 1.083 & 4 & Switzerland & 0.312 & 0.854 \\ 
  5 & Czech Republic & 0.365 & 1.195 & 5 & Czech Republic & 0.247 & 1.108 & 5 & Czech Republic & 0.252 & 0.87 \\ 
  6 & Poland & 0.267 & 1.026 & 6 & Poland & 0.152 & 0.952 & 6 & Poland & 0.184 & 0.75 \\ 
  7 & Belgium & 0.246 & 1.132 & 7 & Belgium & 0.139 & 1.055 & 7 & Belgium & 0.171 & 0.825 \\
\hline
\hline
\end{tabular}
\caption{Estimated mean and variance parameters and Rankings of country-wise average latent skill levels obtained from RMSD methods and the Baseline method for math domain.} 
\label{tab: math 0.1 0.15 baseline}
\end{center}
\end{table}

\end{landscape}

\begin{table}[!h]
\centering
\fontsize{10}{10}\selectfont
\begin{tabular}{cc|cc|cc}
  \hline
  $j$ & $\hat{a}_j$ & $j$ & $\hat{a}_j$ & $j$ & $\hat{a}_j$ \\ 
  \hline
1   & 0.47 & 58 & 1.71 & 115 & 1.23 \\ 
  2   & 0.78 & 59 & 1.55 & 116 & 1.14 \\ 
  3   & 1.24 & 60 & 2.20 & 117 & 1.59 \\ 
  4   & 1.07 & 61 & 1.39 & 118 & 0.92 \\ 
  5   & 1.57 & 62 & 1.37 & 119 & 1.66 \\ 
  6   & 0.98 & 63 & 1.70 & 120 & 0.74 \\ 
  7   & 1.93 & 64 & 2.37 & 121 & 0.54 \\
\hline
\end{tabular}~\quad\quad\quad\begin{tabular}{rr|rr|rr}
  \hline
 $j$ & $\hat{d}_j$ & $j$ & $\hat{d}_j$ & $j$ & $\hat{d}_j$ \\ 
  \hline
1  & 1.88 & 58 & -4.83 & 115 & 1.13 \\ 
  2  & 0.52 & 59 & -0.42 & 116 & 0.42 \\ 
  3  & 1.07 & 60 & -0.11 & 117 & 0.83 \\ 
  4   & 0.17 & 61 & 2.76 & 118 & 0.72 \\ 
  5   & -0.27 & 62 & 1.12 & 119 & 0.74 \\ 
  6   & -0.37 & 63 & -2.00 & 120 & -0.07 \\ 
  7   & -1.78 & 64 & 0.01 & 121 & -0.49 \\
\hline
\end{tabular}
\caption{Estimated $\hat{a}_j$ and $\hat{d}_j$ for Model~\eqref{eq:irf-dif} from proposed method based on the math data.}
\label{tab:math a d}
\end{table}

\begin{landscape}
{\small
\begin{longtable}{c|cccccccccccccccccc}
\caption{Estimated $\hat{\gamma}_{jk}$ parameters for $k=1,...,18$ for Model \eqref{eq:irf-dif} on the math data.} \label{tab:math gamma 1} \\
 & 1 & 2 & 3 & 4 & 5 & 6 & 7 & 8 & 9 & 10 & 11 & 12 & 13 & 14 & 15 & 16 & 17 & 18 \\ 
  \hline
1 & 0.04 & 0.03 & 0.12 & -0.05 & -0.11 & -0.66 & -0.57 & 0.29 & 0.40 & 0.29 & 0.20 & -0.17 & -0.05 & -0.44 & -0.00 & 0.50 & -0.36 & -0.76 \\ 
  2 & -0.23 & 0.15 & -0.22 & -0.20 & -0.49 & -0.66 & -0.42 & -0.19 & -0.00 & 0.42 & 0.24 & -0.00 & 0.15 & 0.08 & 0.18 & 0.21 & -0.35 & -0.36 \\ 
  3 & 0.07 & -0.20 & 0.07 & 0.14 & 0.32 & -0.16 & 0.00 & 0.18 & 0.10 & 0.25 & -0.18 & -0.11 & -0.01 & 0.30 & 0.20 & 0.63 & 0.37 & -0.29 \\ 
  4 & 0.22 & -0.29 & -0.09 & 0.23 & 0.00 & -0.42 & -0.35 & 0.15 & 0.23 & 0.28 & 0.29 & 0.12 & -0.17 & -0.10 & 0.07 & 0.23 & 0.62 & -0.64 \\ 
  5 & 0.26 & 0.04 & -0.08 & 0.14 & -0.23 & -0.40 & -1.26 & -0.31 & 0.30 & -0.11 & -0.09 & 0.19 & -0.17 & -0.13 & 0.22 & -0.19 & 0.17 & 0.50 \\ 
  6 & 0.40 & 0.28 & 0.41 & 0.07 & 0.11 & -0.04 & -0.05 & -0.46 & -0.15 & -0.08 & -0.07 & -0.49 & -0.01 & 0.50 & -0.19 & -0.27 & 0.25 & 0.04 \\
\end{longtable}
}
\end{landscape}

\begin{landscape}
    
    {\small
\begin{longtable}{c|ccccccccccccccccccc}
\caption{Estimated $\hat{\gamma}_{jk}$ parameters for $k=19,...,37$ for Model \eqref{eq:irf-dif} on the math data.}\label{tab:math gamma 2} \\
  \hline
 & 19 & 20 & 21 & 22 & 23 & 24 & 25 & 26 & 27 & 28 & 29 & 30 & 31 & 32 & 33 & 34 & 35 & 36 & 37 \\ 
  \hline
1 & 0.12 & 0.00 & 0.35 & 0.09 & -0.08 & -0.59 & 0.01 & 0.28 & 0.21 & -0.01 & -0.10 & -0.26 & -0.03 & -0.04 & 0.17 & 0.18 & -0.58 & -0.35 & 0.02 \\ 
  2 & 0.11 & 0.99 & 0.06 & 0.29 & 0.00 & -0.37 & -0.41 & -0.11 & 0.12 & 0.44 & 0.13 & 0.11 & 0.19 & -0.02 & 0.11 & 0.10 & -0.18 & -0.29 & -0.26 \\ 
  3 & 0.24 & -0.64 & -0.07 & 0.22 & -0.04 & -0.14 & -0.04 & -0.04 & -0.18 & -0.14 & -0.03 & -0.10 & 0.44 & 0.10 & -0.06 & 0.03 & -0.31 & 0.00 & 0.30 \\ 
  4 & -0.24 & -0.57 & -0.42 & 0.32 & -0.01 & -0.05 & -0.72 & 0.03 & 0.23 & -0.05 & -0.09 & -0.04 & 0.20 & 0.08 & 0.28 & -0.14 & -0.31 & 0.12 & 0.13 \\ 
  5 & 0.57 & -0.29 & -0.38 & 0.22 & 0.04 & -0.36 & -0.11 & 0.16 & 0.26 & 0.00 & 0.21 & 0.18 & -0.22 & 0.49 & -0.21 & 0.15 & -0.69 & 0.25 & -0.03 \\ 
  6 & 0.07 & 0.10 & 0.24 & -0.14 & -0.01 & -0.44 & 0.29 & 0.29 & 0.09 & -0.13 & -0.02 & -0.46 & 0.14 & 0.23 & -0.00 & -0.03 & -0.06 & 0.07 & 0.15 \\
\end{longtable}
}
\end{landscape}

\subsection{Additional results for Science Data}
\label{sec-real-additional-science}

\begin{figure}
    \centering
    \includegraphics[width=0.4\linewidth]{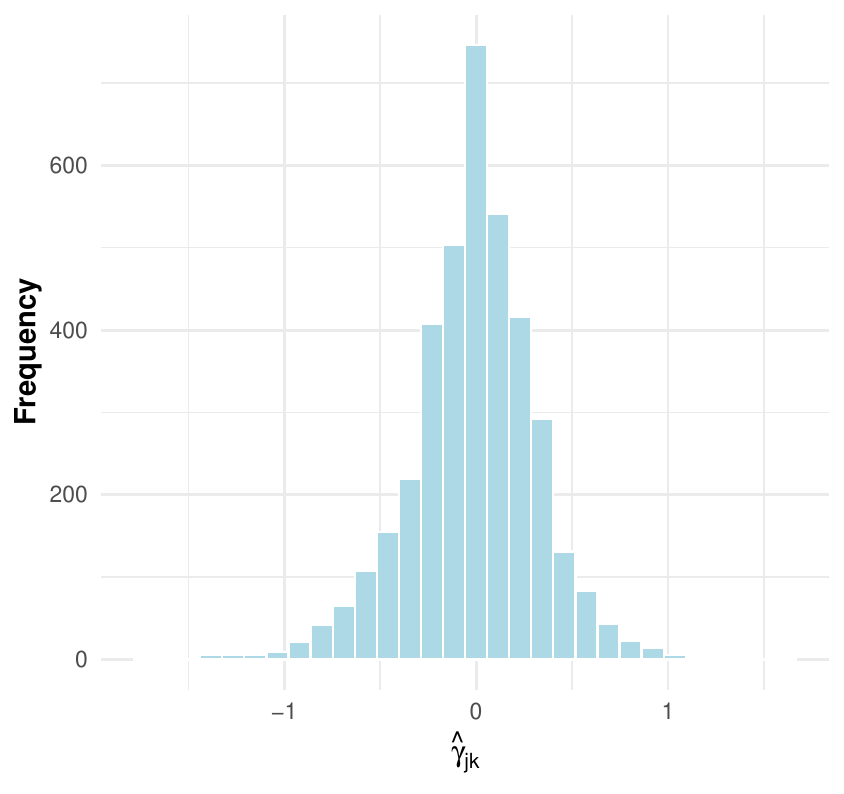}
    \caption{Histogram of estimated $\hat{\gamma}_{jk}$s based on the science data.}
    \label{fig:hist-science}
\end{figure}

The estimated mean parameters $\hat{\mu}_k$ and variance parameters $\hat{\sigma}_k$ from all methods are also provided in Tables~\ref{tab: science with variance}--\ref{tab: rmsd 0.1 0.15 science}.
We also report the country rankings based on the RMSD methods with thresholds of 0.10 and 0.15, as well as the baseline method, in Table~\ref{tab: rmsd 0.1 0.15 science}.  Table~\ref{tab:science a d} presents the estimated $\hat{a}_j$ and $\hat{d}_j$ parameters from the proposed method for the science data. In addition, Tables~\ref{tab:science gamma 1}--\ref{tab:science gamma 2} report the estimated $\hat{\gamma}_{jk}$ values for the science data. We further examine the estimated DIF parameters $\hat{\gamma}_{jk}$ and observe that many of them are close to zero, as illustrated by the histogram in Figure~\ref{fig:hist-science}.

Besides the country rankings, we explored the differences in estimated means across different countries and found several interesting patterns. For instance, our method estimates the difference in means $\hat{\mu}_k$ between Sweden and the United States to be $0.081 - 0.059 = 0.022$, whereas the RMSD method gives a difference of $0.018 - 0.003 = 0.015$. Hence, while both methods indicate that Sweden outperforms the United States, our method indicates a larger gap in performance. Another case is the comparison between Ireland and Austria: our method estimates the difference of means to be $0.185 - 0.106 = 0.079$, which is noticeably larger than the differences of means from RMSD method: $0.137 - 0.094 = 0.043$.

\begin{table}[!h]
\begin{center}
\fontsize{8}{8}\selectfont
\renewcommand{\arraystretch}{0.9}
\begin{tabular}{c|c|c|c|c|c|c|c}
 \hline
Rank& Proposed&  $\hat{\mu}_k$ & $\hat{\sigma}_k$ &  Rank &RMSD& $\hat{\mu}_k$ & $\hat{\sigma}_k$ \\
 \hline
    1 & Japan & 0.553 & 0.985 & 1 & Japan & 0.624 & 1.174 \\ 
  2 & Korea & 0.465 & 1.03 & 2 & Korea & 0.573 & 1.237 \\ 
  3 & Estonia & 0.333 & 0.905 & 3 & Estonia & 0.474 & 1.09 \\ 
  4 & Czech Republic & 0.229 & 1.041 & 4 & Czech Republic & 0.221 & 1.253 \\ 
  5 & Australia & 0.199 & 1.062 & 5 & Australia & 0.208 & 1.294 \\ 
  6 & Ireland & 0.185 & 0.928 & 6 & Poland & 0.177 & 1.152 \\ 
  7 & Switzerland & 0.15 & 0.991 & 7 & Switzerland & 0.158 & 1.195 \\
\hline
\end{tabular}
\caption{Comparison of rankings of country-wise average latent skill levels between proposed method and RMSD method for science domain.
} 
\label{tab: science with variance}
\end{center}
\end{table}

\begin{landscape}
    
\begin{table}[!htbp]
\begin{center}
\fontsize{10}{10}\selectfont
\begin{tabular}{c|c|c|c|c|c|c|c|c|c|c|c}
 \hline
Rank& RMSD 0.1 &  $\hat{\mu}_k$ & $\hat{\sigma}_k$ &  Rank &RMSD 0.15 & $\hat{\mu}_k$ & $\hat{\sigma}_k$ & Rank & Baseline & $\hat{\mu}_k$ & $\hat{\sigma}_k$ \\
 \hline
1 & Japan & 0.552 & 1.163 & 1 & Japan & 0.551 & 1.151 & 1 & Japan & 0.609 & 0.902 \\ 
  2 & Korea & 0.435 & 1.25 & 2 & Korea & 0.397 & 1.238 & 2 & Korea & 0.51 & 0.966 \\ 
  3 & Estonia & 0.364 & 1.106 & 3 & Estonia & 0.325 & 1.104 & 3 & Estonia & 0.459 & 0.851 \\ 
  4 & Czech Republic & 0.082 & 1.236 & 4 & Czech Republic & 0.042 & 1.242 & 4 & Australia & 0.23 & 0.998 \\ 
  5 & Australia & 0.069 & 1.292 & 5 & Australia & 0.038 & 1.287 & 5 & Czech Republic & 0.229 & 0.97 \\ 
  6 & Poland & 0.055 & 1.155 & 6 & Poland & 0.018 & 1.147 & 6 & Poland & 0.218 & 0.904 \\ 
  7 & Ireland & 0.014 & 1.1 & 7 & Ireland & -0.023 & 1.097 & 7 & Ireland & 0.196 & 0.855 \\
\hline
\hline
\end{tabular}
\caption{Comparison of rankings of country-wise average latent skill levels of RMSD methods with thresholds 0.1 and 0.15 and baseline methods for science domain.} 
\label{tab: rmsd 0.1 0.15 science}
\end{center}
\end{table}
\end{landscape}

\begin{table}[!h]
\centering
\fontsize{10}{10}\selectfont
\begin{tabular}{cc|cc}
  \hline
  $j$ & $\hat{a}_j$ & $j$ & $\hat{a}_j$ \\ 
  \hline
1   & 0.52 & 53 & 1.67 \\ 
  2   & 0.81 & 54 & 1.85 \\ 
  3   & 0.66 & 55 & 1.25 \\ 
  4   & 0.86 & 56 & 1.16 \\ 
  5   & 1.32 & 57 & 1.53 \\ 
  6   & 0.36 & 58 & 1.56 \\ 
  7   & 0.78 & 59 & 0.49 \\ 
  8   & 1.04 & 60 & 1.40 \\ 
  9   & 0.59 & 61 & 0.64 \\ 
  10   & 1.00 & 62 & 1.89 \\ 
  11   & 1.61 & 63 & 0.97 \\ 
  12   & 1.20 & 64 & 1.65 \\ 
  13   & 0.97 & 65 & 0.74 \\ 
  14   & 0.70 & 66 & 1.61 \\ 
  15   & 0.69 & 67 & 1.70 \\ 
  16   & 1.34 & 68 & 1.49 \\ 
  17   & 0.70 & 69 & 1.59 \\ 
  18   & 1.48 & 70 & 1.04 \\ 
  19   & 1.01 & 71 & 0.89 \\ 
  20   & 0.83 & 72 & 1.41 \\ 
  21   & 0.59 & 73 & 1.44 \\ 
  22   & 0.45 & 74 & 1.11 \\ 
  23  & 0.79 & 75 & 1.17 \\ 
  24   & 0.67 & 76 & 0.82 \\ 
  25  & 1.15 & 77 & 1.14 \\ 
  26   & 1.17 & 78 & 1.43 \\ 
  27  & 1.21 & 79 & 1.13 \\ 
  28   & 1.68 & 80 & 1.68 \\ 
  29   & 1.61 & 81 & 0.48 \\ 
  30   & 1.15 & 82 & 0.88 \\ 
  31   & 1.16 & 83 & 1.25 \\ 
  32   & 1.88 & 84 & 1.15 \\ 
  33   & 1.00 & 85 & 1.42 \\ 
  34   & 1.44 & 86 & 0.66 \\ 
  35   & 0.83 & 87 & 0.75 \\ 
  36  & 1.66 & 88 & 1.04 \\ 
  37  & 1.00 & 89 & 1.28 \\ 
  38   & 0.60 & 90 & 2.25 \\ 
  39   & 0.88 & 91 & 1.91 \\ 
  40   & 1.29 & 92 & 1.77 \\ 
  41   & 1.32 & 93 & 1.15 \\ 
  42  & 2.14 & 94 & 0.82 \\ 
  43   & 2.00 & 95 & 1.34 \\ 
  44   & 1.57 & 96 & 0.76 \\ 
  45  & 1.36 & 97 & 1.77 \\ 
  46   & 0.80 & 98 & 0.94 \\ 
  47   & 1.24 & 99 & 0.59 \\ 
  48   & 1.53 & 100 & 1.13 \\ 
  49   & 0.74 & 101 & 1.49 \\ 
  50   & 0.98 & 102 & 0.95 \\ 
  51   & 1.33 & 103 & 0.82 \\ 
  52   & 1.22 & 104 & 1.15 \\ 
   \hline
\end{tabular}~\quad\quad\quad\begin{tabular}{cc|cc}
  \hline
  $j$ & $\hat{d}_j$ & $j$ & $\hat{d}_j$ \\ 
  \hline
1   & 0.04 & 53 & -0.13 \\ 
  2   & -1.18 & 54 & 1.90 \\ 
  3   & 0.06 & 55 & -0.06 \\ 
  4   & -0.67 & 56 & 0.88 \\ 
  5   & -2.51 & 57 & -1.72 \\ 
  6   & 0.38 & 58 & -2.91 \\ 
  7   & -0.01 & 59 & -0.63 \\ 
  8   & 0.97 & 60 & 0.56 \\ 
  9   & 0.84 & 61 & -0.35 \\ 
  10   & 0.34 & 62 & 0.10 \\ 
  11   & 0.74 & 63 & -0.10 \\ 
  12   & 0.30 & 64 & -1.10 \\ 
  13   & -1.39 & 65 & 0.21 \\ 
  14   & 2.05 & 66 & -1.83 \\ 
  15   & -0.24 & 67 & -1.60 \\ 
  16   & -0.14 & 68 & -2.29 \\ 
  17   & 0.43 & 69 & 0.12 \\ 
  18   & -0.97 & 70 & 1.55 \\ 
  19   & -0.43 & 71 & -0.64 \\ 
  20   & 0.80 & 72 & 0.03 \\ 
  21   & -0.40 & 73 & 1.64 \\ 
  22   & -0.31 & 74 & -0.63 \\ 
  23   & -0.50 & 75 & -0.30 \\ 
  24   & 0.68 & 76 & 0.53 \\ 
  25   & 0.02 & 77 & 0.14 \\ 
  26   & -0.84 & 78 & -1.54 \\ 
  27   & 0.35 & 79 & 1.80 \\ 
  28   & 1.27 & 80 & -0.61 \\ 
  29   & -0.37 & 81 & -1.46 \\ 
  30   & 1.90 & 82 & -0.47 \\ 
  31   & -0.76 & 83 & -1.51 \\ 
  32   & 0.16 & 84 & 0.39 \\ 
  33   & 1.28 & 85 & 0.00 \\ 
  34   & 0.78 & 86 & 1.00 \\ 
  35   & 1.06 & 87 & -0.77 \\ 
  36   & 1.74 & 88 & -0.24 \\ 
  37   & 0.15 & 89 & -1.18 \\ 
  38   & -0.51 & 90 & 1.59 \\ 
  39   & 1.23 & 91 & 0.26 \\ 
  40   & 0.64 & 92 & -1.61 \\ 
  41   & 0.99 & 93 & -1.53 \\ 
  42   & -0.79 & 94 & -0.38 \\ 
  43   & 1.45 & 95 & -0.00 \\ 
  44   & -0.86 & 96 & 0.07 \\ 
  45   & 0.92 & 97 & -0.99 \\ 
  46   & 0.14 & 98 & -0.54 \\ 
  47   & 0.37 & 99 & 0.44 \\ 
  48   & 2.09 & 100 & -1.18 \\ 
  49   & -0.62 & 101 & 2.00 \\ 
  50   & -0.89 & 102 & 0.43 \\ 
  51   & 1.31 & 103 & -0.03 \\ 
  52   & 2.15 & 104 & 0.72 \\ 
   \hline
\end{tabular}
\caption{Estimated $\hat{a}_j$ and $\hat{d}_j$ for Model~\eqref{eq:irf-dif} on the science data.}
\label{tab:science a d}
\end{table}

\begin{landscape}
{\small
\begin{longtable}{c|cccccccccccccccccc}
\caption{Estimated $\hat{\gamma}_{jk}$ parameters for $k=1,...,18$ for Model \eqref{eq:irf-dif} on the science data.} \label{tab:science gamma 1} \\
 & 1 & 2 & 3 & 4 & 5 & 6 & 7 & 8 & 9 & 10 & 11 & 12 & 13 & 14 & 15 & 16 & 17 & 18 \\ 
  \hline
1 & -0.17 & 0.20 & -0.06 & -0.10 & -0.52 & -0.52 & -0.68 & -0.06 & 0.11 & 0.35 & 0.30 & -0.30 & 0.23 & -0.48 & 0.30 & -0.36 & 0.45 & 0.05 \\ 
  2 & -0.08 & 0.38 & 0.08 & -0.02 & -1.16 & -0.47 & -0.49 & -0.05 & -0.04 & 0.33 & -0.53 & 0.21 & -0.46 & -0.00 & 0.37 & 0.23 & -0.49 & 0.24 \\ 
  3 & 0.17 & -0.09 & -0.22 & 0.15 & -0.16 & -0.29 & -0.14 & 0.09 & 0.20 & 0.11 & 0.13 & -0.21 & 0.10 & -0.53 & -0.46 & -0.41 & 0.21 & 0.05 \\ 
  4 & -0.88 & 0.29 & -0.00 & -0.55 & 0.23 & -0.02 & 0.08 & 0.21 & 0.04 & 0.26 & -0.12 & 0.13 & 0.29 & 0.18 & -0.20 & -0.13 & -0.75 & 0.03 \\ 
  5 & -0.03 & -0.35 & -0.47 & 0.13 & 0.35 & 0.24 & 0.05 & 0.06 & -0.46 & 0.17 & -0.10 & -0.44 & -0.24 & 0.32 & -0.27 & 0.02 & 0.00 & -0.24 \\ 
  6 & 0.57 & -0.10 & -0.20 & 0.27 & -0.17 & -0.71 & -0.45 & -0.17 & 0.00 & -0.12 & -0.30 & 0.06 & -0.13 & -0.38 & -0.21 & 0.12 & 0.34 & -0.39 \\
\end{longtable}
}
\end{landscape}

\begin{landscape}
    
    {\small
\begin{longtable}{c|ccccccccccccccccccc}
\caption{Estimated $\hat{\gamma}_{jk}$ parameters for $k=19,...,37$ for Model \eqref{eq:irf-dif} on the science data.}\label{tab:science gamma 2} \\
  \hline
 & 19 & 20 & 21 & 22 & 23 & 24 & 25 & 26 & 27 & 28 & 29 & 30 & 31 & 32 & 33 & 34 & 35 & 36 & 37 \\ 
  \hline
1 & 0.30 & 0.79 & 0.68 & 0.21 & 0.36 & -1.27 & -0.28 & 0.00 & 0.16 & 0.52 & 0.67 & -0.20 & 0.42 & -0.09 & -0.25 & -0.10 & -0.32 & 0.31 & -0.03 \\ 
  2 & 0.46 & -1.00 & 0.66 & 1.11 & 0.52 & -0.43 & -0.13 & -0.17 & 0.11 & 0.12 & -1.27 & 0.09 & 0.02 & -0.16 & -0.60 & 0.02 & 0.46 & 0.24 & -0.32 \\ 
  3 & -0.10 & -0.64 & -0.83 & 0.06 & 0.27 & -0.80 & -0.11 & 0.15 & 0.28 & 0.40 & 0.09 & -0.18 & 0.31 & -0.16 & 0.16 & -0.14 & -0.63 & -0.00 & 0.29 \\ 
  4 & 0.39 & -0.40 & -0.11 & 0.42 & 0.37 & 0.26 & -0.40 & -0.88 & 0.54 & 0.06 & -0.11 & -0.75 & -0.03 & -0.09 & 0.00 & 0.12 & -0.47 & -0.93 & -0.79 \\ 
  5 & -0.23 & -0.40 & -0.67 & 0.01 & 0.16 & 0.05 & -0.78 & 0.13 & 0.12 & 0.25 & -0.43 & 0.21 & -0.29 & 0.00 & 0.32 & -0.18 & -0.24 & -0.00 & 0.08 \\ 
  6 & 0.18 & 0.42 & 0.36 & -0.31 & 0.04 & -0.16 & 0.30 & 0.10 & 0.78 & -0.33 & 0.52 & -0.23 & 0.27 & 0.04 & 0.39 & -0.06 & -0.04 & 0.06 & 0.49 \\
\end{longtable}
}
\end{landscape}

\subsection{Additional results for Reading Analysis}
\label{sec-real-additional-reading}

\begin{figure}[!h]
    \centering
    \includegraphics[width=0.3\linewidth]{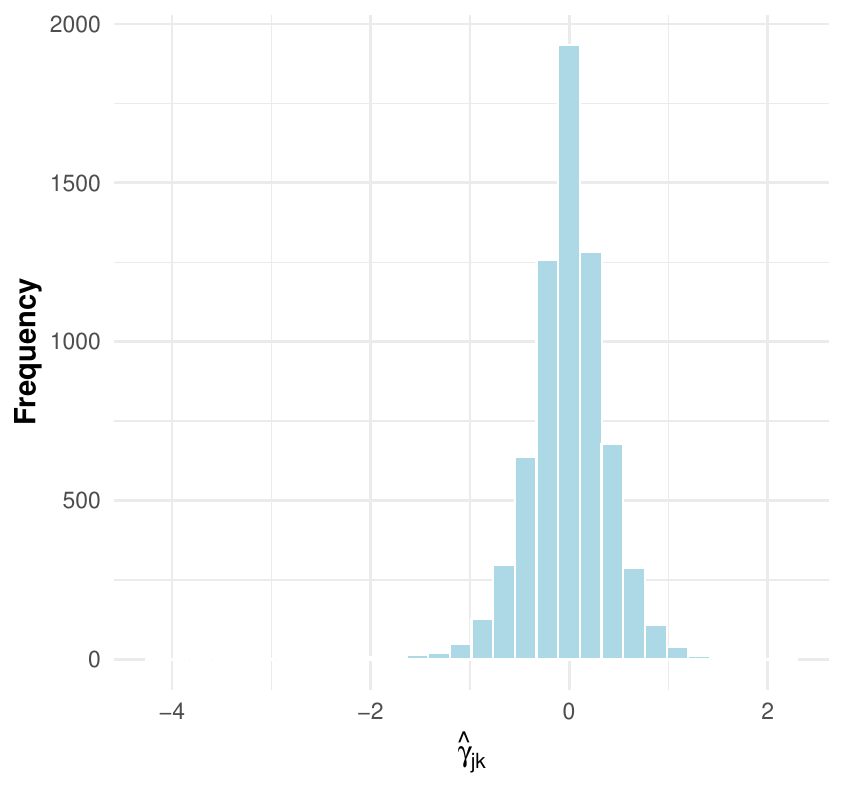}
    \caption{Histogram of estimated $\hat{\gamma}_{jk}$s for the reading domain.}
    \label{fig:reading-hist}
\end{figure}

The estimated mean parameters $\hat{\mu}_k$ and variance parameters $\hat{\sigma}_k$ from all methods are given in Tables~\ref{tab: reading with variance}--\ref{tab: rmsd 0.1 0.15 reading}.
Table~\ref{tab: rmsd 0.1 0.15 reading} also summarizes the country rankings obtained by the RMSD methods with thresholds of 0.10 and 0.15 and by the baseline method.  Table~\ref{tab:reading a d} reports the estimated $\hat{a}_j$ and $\hat{d}_j$ parameters from the proposed method for the reading data. In addition, Tables~\ref{tab:reading gamma 1}--\ref{tab:reading gamma 2} provide the estimated $\hat{\gamma}_{jk}$ values for the reading data. We also examine the estimated DIF parameters $\hat{\gamma}_{jk}$ and find that many of them are close to zero, as shown by the histogram in Figure~\ref{fig:reading-hist}.

In addition, we explored the differences in means and identified several interesting patterns, which serve as valuable complements to the country rankings. For instance, our proposed method estimates the difference in means $\hat{\mu}_k$ between the United Kingdom and the United States as $0.159 - 0.112 = 0.047$, whereas the RMSD method gives $0.485 - 0.484 = 0.001$. This indicates that our method estimates a relatively larger performance difference between the two countries, while the RMSD method considers them to be at nearly the same level. Another illustrative case is the comparison between Germany and Finland: our method estimates the difference as $0.044 - (-0.004) = 0.048$, whereas the RMSD method gives a difference of $0.4024 - 0.4020 = 0.0004$. Hence, our method indicates a more noticeable performance difference between these countries, whereas the RMSD method treats them as nearly identical. Moreover, we use a parametric bootstrap procedure to conduct hypothesis tests on whether the differences in means equal zero, and the resulting p-values are extremely small, indicating a significant difference in means between countries.

\begin{table}[!h]
\begin{center}
\fontsize{8}{8}\selectfont
\renewcommand{\arraystretch}{0.9}
\begin{tabular}{c|c|c|c|c|c|c|c}
 \hline
Rank& Proposed&  $\hat{\mu}_k$ & $\hat{\sigma}_k$ &  Rank &RMSD& $\hat{\mu}_k$ & $\hat{\sigma}_k$ \\
 \hline
 1 & Korea & 0.570 & 1.046 & 1 & Korea & 0.768 & 1.043 \\ 
  2 & Ireland & 0.455 & 0.943 & 2 & Ireland & 0.719 & 0.959 \\ 
  3 & Japan & 0.412 & 0.94 & 3 & Estonia & 0.685 & 0.956 \\ 
  4 & Estonia & 0.355 & 0.921 & 4 & Japan & 0.672 & 0.982 \\ 
  5 & Australia & 0.294 & 1.129 & 5 & Australia & 0.613 & 1.179 \\ 
  6 & Czech Republic & 0.276 & 1.092 & 6 & Czech Republic & 0.587 & 1.109 \\ 
  7 & New Zealand & 0.195 & 1.163 & 7 & New Zealand & 0.583 & 1.196 \\
\hline
\end{tabular}
\caption{Comparison of rankings of country-wise average latent skill levels between proposed method and RMSD method for reading domain.} 
\label{tab: reading with variance}
\end{center}
\end{table}

\begin{landscape}
    
\begin{table}[!htbp]
\begin{center}
\fontsize{10}{10}\selectfont
\begin{tabular}{c|c|c|c|c|c|c|c|c|c|c|c}
 \hline
Rank& RMSD 0.1 &  $\hat{\mu}_k$ & $\hat{\sigma}_k$ &  Rank &RMSD 0.15 & $\hat{\mu}_k$ & $\hat{\sigma}_k$ & Rank & Baseline & $\hat{\mu}_k$ & $\hat{\sigma}_k$ \\
 \hline
1 & Korea & 0.735 & 1.049 & 1 & Korea & 0.741 & 1.06 & 1 & Korea & 0.974 & 0.976 \\ 
  2 & Ireland & 0.58 & 0.942 & 2 & Japan & 0.529 & 1.01 & 2 & Japan & 0.76 & 0.905 \\ 
  3 & Japan & 0.556 & 0.99 & 3 & Estonia & 0.515 & 0.944 & 3 & Ireland & 0.751 & 0.839 \\ 
  4 & Estonia & 0.554 & 0.943 & 4 & Ireland & 0.513 & 0.932 & 4 & Estonia & 0.731 & 0.84 \\ 
  5 & Australia & 0.488 & 1.163 & 5 & Australia & 0.434 & 1.163 & 5 & Australia & 0.673 & 1.044 \\ 
  6 & Czech Republic & 0.445 & 1.081 & 6 & Czech Republic & 0.412 & 1.083 & 6 & Czech Republic & 0.653 & 0.98 \\ 
  7 & New Zealand & 0.443 & 1.185 & 7 & New Zealand & 0.389 & 1.174 & 7 & New Zealand & 0.632 & 1.053 \\
\hline
\end{tabular}
\caption{Comparison of rankings of country-wise average latent skill levels of RMSD methods with thresholds 0.1 and 0.15 and baseline methods for reading domain.} 
\label{tab: rmsd 0.1 0.15 reading}
\end{center}
\end{table}
\end{landscape}

\begin{table}[ht]
\centering
\fontsize{10}{10}\selectfont
\begin{tabular}{cc|cc|cc}
  \hline
  $j$ & $\hat{a}_j$ & $j$ & $\hat{a}_j$ & $j$ & $\hat{a}_j$ \\ 
  \hline
1   & 0.67 & 62 & 1.47 & 123 & 1.03 \\ 
  2   & 1.18 & 63 & 1.23 & 124 & 0.96 \\ 
  3   & 1.72 & 64 & 1.48 & 125 & 1.16 \\ 
  4   & 1.12 & 65 & 1.48 & 126 & 1.16 \\ 
  5   & 0.62 & 66 & 1.67 & 127 & 0.88 \\ 
  6   & 1.69 & 67 & 1.20 & 128 & 1.10 \\ 
  7   & 1.13 & 68 & 0.98 & 129 & 0.99 \\
\hline
\end{tabular}~\quad\quad\quad \begin{tabular}{cc|cc|cc}
  \hline
  $j$ & $\hat{d}_j$ & $j$ & $\hat{d}_j$ & $j$ & $\hat{d}_j$ \\ 
  \hline
1   & 0.40 & 62 & 1.07 & 123 & -0.24 \\ 
  2   & 0.52 & 63 & 1.05 & 124 & 1.30 \\ 
  3   & 2.59 & 64 & -0.16 & 125 & 0.46 \\ 
  4   & 0.88 & 65 & 1.45 & 126 & 2.01 \\ 
  5   & 0.42 & 66 & -0.32 & 127 & 1.29 \\ 
  6   & 1.05 & 67 & -0.75 & 128 & 1.64 \\ 
  7   & 0.38 & 68 & 1.53 & 129 & -0.77 \\
\hline
\end{tabular}
\caption{Estimated $\hat{a}_j$ and $\hat{d}_j$ for Model~\eqref{eq:irf-dif} on the reading data.}
\label{tab:reading a d}
\end{table}

\begin{landscape}
{\small
\begin{longtable}{c|cccccccccccccccccc}
\caption{Estimated $\hat{\gamma}_{jk}$ parameters for $k=1,...,18$ for Model \eqref{eq:irf-dif} on the reading data.} \label{tab:reading gamma 1} \\
 & 1 & 2 & 3 & 4 & 5 & 6 & 7 & 8 & 9 & 10 & 11 & 12 & 13 & 14 & 15 & 16 & 17 & 18 \\ 
  \hline
1 & 0.00 & -0.16 & 0.10 & 0.16 & 0.05 & -0.10 & -0.09 & 0.26 & -0.12 & -0.05 & -0.21 & 0.05 & 0.02 & -0.00 & 0.50 & 0.15 & 0.06 & -0.09 \\ 
  2 & -0.68 & 0.29 & 0.48 & -0.32 & 0.58 & 0.66 & 0.41 & -0.24 & -1.09 & -0.52 & 0.38 & 0.71 & 0.68 & 0.90 & -0.00 & -0.04 & -0.61 & 0.72 \\ 
  3 & -0.53 & 0.07 & -0.08 & -0.46 & 0.34 & 0.39 & 0.41 & 0.22 & -0.46 & 0.01 & -0.00 & 0.10 & 0.21 & 0.11 & 0.16 & -0.10 & -0.51 & 0.11 \\ 
  4 & -0.85 & 0.38 & 0.28 & -0.55 & -0.15 & 0.08 & 0.21 & 0.44 & -0.06 & -0.91 & 0.04 & 0.16 & 0.46 & 0.03 & 0.30 & -0.08 & -0.54 & -0.50 \\ 
  5 & -0.23 & 0.20 & 0.20 & 0.03 & -0.04 & 0.19 & 0.21 & 0.30 & -0.49 & 0.21 & 0.23 & -0.10 & 0.09 & -0.07 & -0.70 & -0.13 & 0.13 & 0.22 \\ 
  6 & -0.29 & -0.16 & -0.09 & -0.19 & -0.48 & -0.73 & 0.32 & -0.31 & -0.05 & 0.46 & 0.78 & -0.03 & 0.27 & 0.68 & 1.42 & 1.01 & -0.03 & -0.00 \\
\end{longtable}
}
\end{landscape}

\begin{landscape}{\small
\begin{longtable}{c|ccccccccccccccccccc}
\caption{Estimated $\hat{\gamma}_{jk}$ parameters for $k=19,...,37$ for Model \eqref{eq:irf-dif} on the reading data.}\label{tab:reading gamma 2} \\
  \hline
 & 19 & 20 & 21 & 22 & 23 & 24 & 25 & 26 & 27 & 28 & 29 & 30 & 31 & 32 & 33 & 34 & 35 & 36 & 37 \\ 
  \hline
1 & 0.33 & 0.19 & 0.04 & 0.27 & -0.18 & -0.26 & -0.05 & 0.30 & -0.17 & -0.05 & -0.17 & 0.13 & -0.22 & -0.00 & 0.28 & 0.15 & -0.00 & 0.04 & -0.21 \\ 
  2 & 0.84 & -0.08 & -0.06 & 0.41 & -0.23 & -0.00 & 0.40 & -0.39 & -0.41 & -0.01 & 0.05 & -0.03 & 0.18 & 0.26 & -0.19 & 0.49 & 0.99 & -0.74 & -0.31 \\ 
  3 & 0.12 & -0.62 & -0.57 & 0.51 & 0.53 & -0.17 & 0.51 & -0.53 & -0.32 & 0.08 & -0.13 & 0.22 & -0.09 & -0.20 & -0.02 & 0.32 & -0.17 & -0.72 & -0.51 \\ 
  4 & 0.66 & -0.81 & -0.36 & 0.34 & 0.07 & -0.11 & -0.02 & -0.77 & -0.23 & -0.15 & 0.00 & 0.33 & 0.21 & 0.28 & -0.16 & 0.50 & 0.16 & -0.81 & -0.74 \\ 
  5 & -0.15 & 0.11 & -0.56 & 0.04 & -0.30 & 0.17 & 0.00 & -0.14 & 0.19 & -0.53 & -0.04 & -0.12 & 0.22 & 0.04 & -0.06 & 0.11 & -0.09 & -0.11 & -0.15 \\ 
  6 & -1.15 & -0.26 & -0.51 & 1.16 & 0.71 & -0.39 & 0.46 & -0.63 & 0.44 & 0.50 & 0.40 & -0.00 & 0.75 & 0.21 & 0.12 & -0.17 & -1.99 & 0.02 & -0.81 \\
\end{longtable}
}
\end{landscape}

\bibliographystyle{apalike}
\bibliography{Bibliography-MM-MC}
\end{appendix}
\end{document}